\documentclass{ametsoc}

\modulolinenumbers[200]
\nolinenumbers

\journal{jcli}

\bibpunct{(}{)}{;}{a}{}{,}

\DeclareMathOperator{\stdev}{stdev}

\title{Indo-Pacific variability on seasonal to multidecadal timescales. \\ Part I: Intrinsic SST modes in models and observations}

\authors{Joanna Slawinska\correspondingauthor{Joanna Slawinska, Center for Environmental Prediction, Rutgers, The State University of New Jersey, 14 College Farm Road, New Brunswick, NJ, 08901}} 

\affiliation{Center for Environmental Prediction, Rutgers, The State University of New Jersey, \\ New Brunswick, New Jersey, USA} 

\email{joanna.slawinska@envsci.rutgers.edu}

\extraauthor{Dimitrios Giannakis}

\extraaffil{Courant Institute of Mathematical Sciences, New York University, New York, New York, USA}

\abstract{The variability of Indo-Pacific SST on seasonal to multidecadal timescales is investigated using a recently introduced technique called nonlinear Laplacian spectral analysis (NLSA). Through this technique, drawbacks associated with ad hoc pre-filtering of the input data are avoided, enabling recovery of low-frequency and intermittent modes not previously accessible via classical approaches. Here, a multiscale hierarchy of spatiotemporal modes is identified for Indo-Pacific SST in millennial control runs of CCSM4 and CM3 and in HadISST data. On interannual timescales, a mode with spatiotemporal patterns corresponding to the fundamental component of ENSO emerges, along with ENSO-modulated annual modes consistent with combination mode theory. The ENSO combination modes also feature prominent activity in the Indian Ocean, explaining significant fraction of the SST variance in regions associated with the Indian Ocean dipole. A pattern resembling the tropospheric biennial oscillation emerges in addition to ENSO and the associated combination modes. On multidecadal timescales, the dominant NLSA mode in the model data is predominantly active in the western tropical Pacific. The interdecadal Pacific oscillation also emerges as a distinct NLSA mode, though with smaller explained variance than the western Pacific multidecadal mode. Analogous modes on interannual and decadal timescales are also identified in HadISST data for the industrial era, as well as in model data of comparable timespan, though decadal modes are either absent or of degraded quality in these datasets.}

\begin{document}

\maketitle

\section{\label{secIntro}Introduction}

The internal variability of the Indo-Pacific Ocean is of wide scientific and societal interest as it has been found to be associated with a plethora of phenomena characterizing the Earth's climate. Among these phenomena is the El Ni\~no-Southern Oscillation (ENSO)---the dominant mode of interannual variability of the coupled climate system \citep[]{Bjerknes69,Philander90,NeelinEtAl98,WangPicaut04,SarachikCane10,Trenberth13,WangEtAl17}. The corresponding warm El Ni\~no and cold La Ni\~na events occur predominantly over the tropical Pacific Ocean, and are sustained by ocean-atmosphere interactions producing a positive feedback, known as Bjerknes feedback \citep[][]{Bjerknes69,Wyrtki75,CaneZebiak85}, whereby a positive SST anomaly in the eastern equatorial Pacific induces anomalous westerly surface winds which reinforce the original SST anomaly through anomalous downwelling  in the eastern Pacific. Negative-feedback mechanisms, leading to a self-sustained interannual oscillation \citep[][]{CaneZebiak85}, have been extensively studied through conceptual models emphasizing equatorial wave propagation \citep[][]{SuarezSchopf88,BattistiHirst89,PicautEtAl97,WeisbergWang97}, heat recharge/discharge \citep[][]{Jin97,Jin97b}, or aspects of both \citep[][]{Wang01,GrahamEtAl15}. Such models are part of a model hierarchy for ENSO that includes low-order deterministic and stochastic models \citep[][]{ThompsonBattisti01,TimmermannEtAl03,Kleeman08,GehneEtAl14,ThualEtAl16}, intermediate-complexity models \citep[][]{ZebiakCane87,SuarezSchopf88,JinNeelin93,KleemanEtAl99}, and coupled climate models \citep[][]{NeelinEtAl94,DeserEtAl12b,ChoiEtAl13,BellengerEtAl14}.

Individual El Ni\~no/La Ni\~na events occur every 3 to 8 years and are locked to the seasonal cycle \citep[][]{RasmussonCarpenter82}, developing early in the boreal summer and peaking during the following winter. Among many studies on the phase synchronization of ENSO with the annual cycle, quadratic nonlinearities in the coupled atmosphere-ocean system have been proposed as a mechanism for producing so-called ENSO combination modes with spectral peaks at the sum and difference of the ENSO and annual cycle frequencies \citep[][]{SteinEtAl11,SteinEtAl14,StueckerEtAl13,StueckerEtAl15,RenEtAl16}. As a result of this coupling, the atmospheric response of ENSO combination modes exhibits a southward migration of surface zonal winds taking place in boreal winter \citep[][]{McGregorEtAl12}, which is generally associated with the seasonally-locked termination of ENSO events \citep[][]{Vecchi06,LengaigneEtAl06}. Another important feature of ENSO, which is a hallmark of nonlinear dynamics, is that asymmetries exist between the El Ni\~no and La Ni\~na phases, characterized by positively skewed SST anomalies due to strong El Ni\~no events \citep[][]{BurgersStephenson99,AnJin04,WellerCai13}.  

The tropospheric biennial oscillation \citep[TBO;][]{Meehl87,Meehl93,Meehl94,Meehl97,LiEtAl02,LiEtAl06} and the Indian Ocean dipole \citep[IOD;][]{SajiEtAl99, WebsterEtAl99} are other prominent modes of interannual variability  confined to the Indo-Pacific which are also seasonally locked and driven by strong atmosphere--ocean coupling. The TBO describes biennial variability of the Asian-Australian monsoon in a process where a weak Australian monsoon in boreal winter with anomalous easterlies and negative SST anomalies over the Maritime Continent is followed by a strong Indian monsoon \citep[][]{Li13} with positive SST anomalies over the Maritime Continent and a westward shift of Walker circulation, which is followed in turn by a strong Australian monsoon (and opposite-sign SST circulation anomalies) in the subsequent boreal winter \citep[e.g.,][]{Meehl93,Meehl97,MeehlArblaster02,WuKirtman04,LiEtAl06}. On the other hand, the IOD emerges in late spring, peaks in October, and quickly diminishes afterwards.  The IOD is usually measured by means of the IOD index, which is defined as the difference in SST anomalies between the western (50$^\circ$E--70$^\circ$E, 10$^\circ$S--10$^\circ$N) and eastern (90$^\circ$E--110$^\circ$E, 10$^\circ$S--0$^\circ$) parts of the Indian Ocean \citep[][]{SajiEtAl99}. A positive IOD phase is associated with a negative SST anomaly west of Sumatra, which is reinforced through Bjerknes feedback by coupling with correspondingly weaker convection and westerlies.  

Although the IOD was discovered over 15 years ago and has received considerable attention by the climate community, it is still not fully understood. Several studies \citep[e.g.,][and references therein]{AnnamalaiEtAl04} report that the IOD is correlated with the ENSO and have argued that the former is forced by the latter. Others \citep[e.g.,][]{YamagataEtAl03, AshokEtAl03, BeheraEtAl06} claim that the IOD is an intrinsic mode of the Indian Ocean and  pointed to cases of strong IOD events emerging in the absence of El Ni\~no/La Ni\~na events. Establishing the independence of the highly correlated IOD and ENSO events through classical methods such as empirical orthogonal function (EOF) analysis is far from straightforward, and despite extensive research on IOD dynamics and climatic impacts \citep[e.g.,][]{AshokEtAl07,ManatsaEtAl08, CaiEtAl12, CaiEtAl13, PeplerEtAl14}, there is no consensus on whether it is an intrinsic mode of Indian Ocean variability or a regional manifestation of ENSO. Therefore, alternative ways of reexamining its nature should still be sought. Similarly, the origin and nature of the TBO is a debated topic with studies finding that it is an ENSO-slaved phenomenon \citep[][]{Goswami95}, others that it can exist independently of ENSO forcing \citep[][]{LiEtAl06}, and yet others questioning its existence altogether \citep[][]{StueckerEtAl15b}.

Recently, \citet{Dommenget11} and \citet{ZhaoNigam15} undertook an effort to verify previously applied methodologies and argued that upon careful analysis no IOD pattern emerges independently of ENSO. Moreover, \citet{IzumoEtAl10, IzumoEtAl14} and \citet{YuanEtAl11,YuanEtAl13} reported that the IOD is a good predictor of ENSO more than one year ahead. If true, this would potentially improve the predictability of ENSO beyond the so-called ``spring barrier'' \citep[][]{WebsterHoyos10}; a well recognized scientific challenge and at the same time an issue of significant importance to economies around the world. Arguably, however, until the relation between the IOD and ENSO is correctly established, progress in recognizing the impact of these phenomena on other parts of the climate system will be hampered. 

Beyond prediction of interannual fluctuations of the Earth climate system, prognosing decadal to centennial variability also constitutes a grand challenge for climate science and potentially entails benefits to economies worldwide. Most studies on the decadal variability of the Indo-Pacific basin have focused on the Pacific Decadal Oscillation \citep[][]{MantuaEtAl97} or the Interdecadal Pacific Oscillation \citep[IPO;][]{PowerEtAl99}, which are generally regarded as the same phenomenon. The PDO is traditionally defined over the North Pacific ($20^\circ$N--$65^\circ$N, $120^\circ$E--110$^\circ$W) through the first EOF of seasonally detrended monthly data. The IPO is usually defined as the leading EOF of low-pass filtered SST anomalies over the whole Pacific basin. It is often described as ENSO-like decadal variability due to the similarity of its spatial pattern to ENSO, although the two patterns differ in their relative activity in the tropical and North Pacific Oceans.

A major challenge in research on decadal variability is the relatively short observational record from which it is difficult to robustly establish modes operating at decadal timescales and to project their activity into the future. \citet{HanEtAl14} examined a longer observational record and showed that the IPO does not explain in itself sea level rise over certain decades. Based on this finding, they hypothesized the existence of another decadal mode. They argued that this mode acts mainly over the tropical Indian Ocean (as opposed to the IPO which is active in the Pacific Ocean), but did not address the question of its origin. \citet{EnglandEtAl14} analyzed the most recent hiatus and associated it with exceptionally strong trade winds (also considered responsible for sea-level rise by \citet{HanEtAl14}) and increased heat uptake by the ocean. Notably, they observed that the magnitude of trade-wind strengthening could not be explained solely by the IPO, suggesting that other factors, e.g., the internal variability of the Indian Ocean, should play a role. Without further explanation of the origin of trade-wind enhancement, a heat budget analysis of the modeled hiatus was performed based on a simulation with prescribed values of trade winds (as models fail to produce as strong winds as observed; see \citet{LuoEtAl12}). Several studies  \citep[][]{KarnauskasEtAl09, SolomonNewman12, SeagerEtAl15} also stress the need to consider modes of Indo-Pacific internal variability other than the IPO to build a a more complete explanation of the recent climatic record. In particular, the second EOF of SST \citep[][]{Timmermann03, YehKirtman04, SunYu09, OgataEtAl13}, the spatial pattern of which consists mainly of an equatorial SST zonal dipole, have been associated sometimes with observed decadal fluctuations. 

In this work, presented as a two-part series, we reexamine the existence and significance of the internal variability of the Indo-Pacific Ocean using a recent data analysis technique called nonlinear Laplacian spectral analysis \citep[NLSA;][hereafter, GM]{GiannakisMajda11c,GiannakisMajda12a,GiannakisMajda13,GiannakisMajda14}. NLSA combines ideas from delay-coordinate embeddings of dynamical systems and kernel methods for machine learning to extract spatiotemporal modes of variability from high-dimensional time series. Unlike classical linear techniques such as EOF analysis, which typically require filtering or detrending of the data to isolate the timescale of interest, NLSA is able to simultaneously extract modes of multiple timescales with no ad hoc preprocessing of the data. The NLSA modes are computed from the eigenfunctions of a discrete Laplace-Beltrami operator on the nonlinear manifold sampled by the data. These eigenfunctions can be thought of as nonlinear analogs of the principal components (PCs) in EOF analysis, and capture intrinsic timescales associated with the dynamical system generating the data \citep[][]{BerryEtAl13,Giannakis16}. As a result, NLSA is a useful tool for objectively recovering quasiperiodic patterns such as the dominant oscillations in the Indo-Pacific system from high-dimensional time series. Among other applications, NLSA and related kernel algorithms have been used to recover and predict low-frequency modes of variability in the North Pacific Ocean \citep[][]{GiannakisMajda12b, GiannakisMajda12c,BushukEtAl14,ComeauEtAl16}, including the PDO and the North Pacific gyre oscillation \citep[][]{DiLorenzoEtAl08}.

In this paper, we study NLSA modes of Indo-Pacific SST variability in control integrations of coupled climate models and in observational data. Our analyzed model data is monthly averaged output from millennial control integrations of the Community Climate System Model version 4 \citep[CCSM4;][]{GentEtAl11,DeserEtAl12b} and the Geophysical Fluid Dynamics Laboratory coupled climate model CM3 \citep[][]{GriffiesEtAl11}. As observational data, we study monthly-averaged HadISST data \citep[][]{RaynerEtAl03} spanning the industrial or satellite eras. Applied to these datasets, NLSA yields a hierarchy of modes spanning seasonal to multidecadal timescales. These modes include the annual cycle and its harmonics, and a family of aperiodic modes. In particular, the dominant patterns at the interannual timescale correspond to ENSO and the annual modulations of these patterns as predicted by combination mode theory. In spatiotemporal reconstructions, the family of combination modes extracted by NLSA displays strong activity in the eastern Pacific cold tongue, as well as the eastern equatorial Indian Ocean. Notably, the ENSO combination modes explain $ \sim 35\% $ of the SST variance over the eastern Indian Ocean region used to define IOD indices \citep[][]{SajiEtAl99}.  

In CCSM4, two pairs of modes, one representing the fundamental component of the TBO and the other being a combination mode between the TBO and the annual cycle, emerge independently of the ENSO and ENSO-A modes. The SST anomalies associated with the TBO modes are generally an order of magnitude weaker than the ENSO anomalies, so they are more sensitive to the timespan of available data for analysis. Thus, while analogs of these modes are also detected in the CM3 dataset (which is shorter than CCSM4 by 500 years) and industrial-era HadISST, they have noisier time series mixing annual with biennial signals. We were not able to successfully recover TBO modes from satellite-era HadISST data.    

In the decadal scale, the dominant mode recovered by NLSA from model data is characterized by a pronounced anomaly in the tropical equatorial West Pacific, with anomalies of opposite sign  identified west of Tasmania and over the eastern Pacific. This pattern, which we refer to here as the west Pacific multidecadal mode (WPMM), resembles more closely the second EOF of low-pass filtered SST as opposed to the IPO. As with the TBO, we find that the WPMM is best recovered from the CCSM4 data, where it is robust under changes of NLSA parameters and the choice of spatial domain. However, due to its multidecadal nature and small explained variance, it requires datasets spanning several centuries in order for it to be robustly detected. In particular, while modes resembling the WPMM from CCSM4 were recovered from both CM3 and industrial-era HadISST, these modes have noisier temporal patterns mixing decadal and higher-frequency timescales. Similar issues were encountered in experiments with $ \sim 150 $-year portions of the CCSM4 dataset, suggesting that the poorer quality of the WPMM in CM3 and HadISST is at least partly caused by the time series length as opposed to absence of that mode. Besides the WPMM, modes representing the IPO are recovered in the CCSM4, CM3, and industrial-era HadISST datasets, though the temporal patterns of these modes are also influenced by higher-frequency variability. 

In the second part of this work \citep[][hereafter Part~II]{GiannakisSlawinska16}, we study various linkages and climatic impacts of the modes identified in this paper. There, we establish that the ENSO and ENSO combination modes recovered by NLSA from SST data have physically meaningful associated surface wind and thermocline depth patterns characterizing the seasonally-locked  ENSO lifecycle  \citep[][]{StueckerEtAl13,StueckerEtAl15,McGregorEtAl12}, and inducing couplings between the Pacific and Indian Oceans with IOD-like spatial patterns. Similarly, we find that the surface wind and precipitation patterns associated with the TBO and TBO combination modes from CCSM4 are consistent with the seasonally-synchronized biennial oscillation mechanism involving the Asian-Australian monsoon proposed by \citet[][]{MeehlArblaster02}. In Part II, we also show that despite its relatively low explained variance, the WPMM has a significant dynamical role (at least in the CCSM4 data, where that mode is robustly recovered) as indicated by its clear sign-dependent modulating relationships with the interannual and biennial modes carrying most of the variance. In particular, the cold phase of the WPMM is associated with anomalous westerlies in the equatorial Indian Ocean and the central Pacific, as well as a deep thermocline in the eastern Pacific cold tongue consistent with strong ENSO and TBO variability. The WPMM is also found to correlate strongly with decadal precipitation over Australia in CCSM4.

The plan of this paper is as follows. Section~\ref{secNLSA} contains a high-level description of NLSA algorithms. In section~\ref{secDataset}, we describe the datasets studied in this work. In section~\ref{secNLSAModes}, we present and discuss the modes of Indo-Pacific SST recovered by NLSA from the CCSM4 dataset. Section~\ref{secSensitivity} contains a sensitivity analysis of these results on the choice of NLSA parameters, spatial domain, temporal extent of the data, and the presence of noise. In sections~\ref{secGFDL} and~\ref{secObsModes}, we present and discuss the corresponding Indo-Pacific SST modes in the CM3 and HadISST data, respectively. The paper ends in section~\ref{secSummary} with concluding remarks. A comparison of the NLSA modes recovered from CCSM4 with those obtained via SSA is included in an Appendix. Movies illustrating the dynamical evolution of the NLSA modes are provided as online supplementary material. 

\section{\label{secNLSA}Nonlinear Laplacian spectral analysis algorithms} 

NLSA (GM) is a nonlinear time series analysis technique designed to extract intrinsic spatiotemporal modes of variability from high-dimensional data generated by dynamical systems. In this section, we present an overview of the technique, referring the reader to the references cited above and \citet[][]{Giannakis15} for more thorough discussions.  

Let $ \pmb{x} = \{ x_0, x_1, x_2, \ldots, x_n \} $ be a multivariate time series consisting of $ d $-dimensional samples $ x_i $ taken uniformly at times $ t_i = i \, \delta t $. In sections~\ref{secDataset}--\ref{secObsModes}  ahead, the $ x_i $ will be vectors consisting of monthly averaged SST gridpoint values in the Indo-Pacific sector. In NLSA, the samples are viewed as the values of an observable defined on the phase space of an abstract dynamical system (here, the Earth's climate system), and the goal is to decompose $ \pmb{ x } $ into a collection of temporal and spatial patterns with clear timescale separation and without prefiltering the data. In the Indo-Pacific SST application studied here these patterns include the seasonal cycle, interannual oscillations such as ENSO, decadal oscillations such as the WPMM, and also modulated patterns representing interactions of these basic modes. 

Similarly to classical eigendecomposition techniques such as EOF analysis, singular spectrum analysis \citep[SSA;][]{GhilEtAl02}, and the extended EOF (EEOF) analysis (which is equivalent to SSA), NLSA extracts these patterns through the eigenvectors of a linear operator defined on the dataset. However, instead of the global covariance operator utilized in (E)EOFs and SSA, NLSA employs a Laplace-Beltrami operator defined on the nonlinear manifold sampled by the data and approximated through kernel techniques \citep[][]{BelkinNiyogi03,CoifmanLafon06}. The eigenfunctions of this operator, which can be thought of as nonlinear analogs to principal components (PCs), are used for dimension reduction and spatiotemporal pattern extraction. A key feature of the Laplace-Beltrami operator in NLSA is that it is constructed from time-lagged sequences of data as opposed to individual snapshots ($x_i$). As has been shown theoretically \citep[][]{BerryEtAl13}, the eigenfunctions from time-lagged embedded data separate the observed signal into intrinsic timescales associated with stable Lyapunov directions of the dynamics, such as the quasiperiodic oscillations which are the focus of this work.    

\subsection{\label{secEmbedding}Time-lagged embedding}
Time-lagged embedding, which is also familiar from SSA and EEOFs, involves mapping each snapshot $ x_i $ into a time lagged sequence $ X_i = ( x_i, x_{i-1}, \ldots, x_{i-{q+1}} ) \in \mathbb{ R }^{qd} $, where $ q $ is the number of lags. This technique was originally introduced in state-space reconstruction methods for dynamical systems \citep[][]{PackardEtAl80,Takens81,BroomheadKing86,SauerEtAl91}, where it was established that for sufficiently large $ q $, the dataset $ \pmb{X} = \{ X_1, X_2, \ldots, X_{N} \} $, $ N = n -q +1 $, in lagged embedding space lies with high probability on a manifold $ \mathcal{ M } $ which is equivalent (diffeomorphic) to the attractor of the dynamical system generating the data, even if the samples $ x_i $ do not preserve the attractor (i.e., the observations are incomplete, as is frequently the case in atmosphere ocean science). 

Time-lagged embedding modifies the topology of the data, making it more Markovian if the observations are incomplete, but it also modifies its geometry. That is, inner products in embedding space and the corresponding distances, $ \lVert X_i - X_j \rVert = \sqrt{ ( X_i - X_j )^\mathrm{ T } ( X_i - X_j ) } $, between pairs of lagged embedded vectors depend not only on the snapshots $ x_i $ and $ x_j $, but also on the dynamical trajectories that the system took to arrive at these snapshots. As a result, distances in lagged embedding space induce a Riemannian geometry on $ \mathcal{ M } $ that depends on the dynamical system generating the data. NLSA employs kernel methods for nonlinear data analysis \citep[][]{BelkinNiyogi03,CoifmanLafon06} to empirically approximate differential operators that depend on the induced geometry, and it performs dimension reduction and spatiotemporal pattern extraction through the eigenfunctions of these operators, as follows.    

\subsection{\label{secEigenfunctions}Laplace-Beltrami eigenfunctions}

Let $ \xi_i = X_i - X_{i-1} $  be the time tendency of state $ X_i $, and let $ \cos \theta_{ij} = \xi_i^\mathrm{ T } ( X_i - X_j ) / ( \lVert \xi_i \rVert \lVert X_i - X_j \rVert ) $ be the cosine of the angle between $ \xi_i $ and the relative displacement vector $X_i - X_j $ between states $ X_i $ and $ X_j $. Fixing the parameters $ \epsilon > 0 $ and $ \zeta \in [ 0, 1 ) $, we define the kernel function
\begin{equation}
  \label{eqKCone}
  K( X_i , X_j ) = \exp\left( - \frac{ \lVert X_i - X_j \rVert^2 }{ \epsilon \lVert \xi_i \rVert \lVert \xi_j \rVert } \sqrt{ ( 1 - \zeta \cos^2 \theta_{ij} )( 1 - \zeta \cos^2 \theta_{ji} ) } \right).
\end{equation}
This quantity describes an anisotropic pairwise measure of similarity between states $ X_i $ and $ X_j $ that favors states which are mutually aligned to the dynamical flow (i.e., states with $ \cos^2 \theta_{ij} $ and $ \cos^2\theta_{ji}\approx 1$). Moreover, $ K $ is exponentially decaying with a rate of decay (bandwidth) controlled by $ \epsilon $. The kernel in~\eqref{eqKCone} was introduced in \citet[][]{Giannakis15}, and includes the kernel employed in the earlier work of GM as the special case $ \zeta = 0 $. The family of kernels in~\eqref{eqKCone} can also be viewed as a generalization of radial Gaussian kernels, $ K( x_i, x_j ) = e^{-\lVert x_i - x_j \rVert^2 / \epsilon } $ \citep[which are popular for nonlinear dimension reduction; e.g.,][]{BelkinNiyogi03,CoifmanLafon06}, through the incorporation of delay embeddings and the $ \lVert \xi_i \rVert $ and $\cos \theta_{ij} $ terms that depend on the time tendencies of the data.  We use the symbol $ \pmb{ K } $ to represent the $ N \times N $ symmetric matrix with elements $ K_{ij} = K( X_i, X_j ) $ computed from a kernel $ K $. 

Consider now an arbitrary scalar-valued function $ f $ on $ \mathcal{ M } $. Associated with $ f $ is the time series (temporal pattern) $ \{ f_1, f_2, \ldots, f_N \} $ with $ f_i = f( X_i ) $, which we can represent by an $ N $-dimensional column vector $ \pmb{ f } = ( f_1, f_2, \ldots, f_N )^\mathrm{T} $. As the number of samples $ N $ grows, the matrix-vector product $ \pmb{ K } \pmb{ f } = \pmb{ g } $ approximates the action of an integral operator on $ \mathcal{ M } $. Due to the exponential decay of the kernel, as the bandwidth parameter $ \epsilon $ tends to zero, the component $ g_i $ of $ \pmb{ g } $ depends on the structure of $ f $ and the properties of the kernel in a small neighborhood of $ X_i $ in a manner that was elucidated in a series of works on harmonic analysis and machine learning \citep[][]{BelkinNiyogi03,CoifmanLafon06,Singer06,VonLuxburgEtAl08,BerrySauer15}. Here, we follow the procedure introduced by \citet[][]{CoifmanLafon06}, which involves normalizing $ \pmb{ K } $ to create a Markov matrix $ \pmb{ P } $ through the sequence of operations,
\begin{equation}
  \label{eqDM}
  \tilde{\pmb{K}} = \pmb{K} \pmb{Q}^{-1}, \quad \pmb{P}=\pmb{D}^{-1}\tilde{\pmb{K}},
\end{equation}
where $ \pmb{Q} $ and $ \pmb{D} $ are diagonal matrices with nonzero entries given by the column vectors $ \pmb{q} = \pmb{K} \pmb{1} $ and $ \pmb{d} = \tilde{\pmb{K}} \pmb{1} $, respectively, and $ \pmb{1} $ is the $ N $-dimensional column vector with elements all equal to 1. By construction, $ \pmb{ P } $ is similar to the symmetric matrix $ \pmb{S} = \pmb{D}^{1/2} \pmb{P} \pmb{D}^{-1/2} = \pmb{D}^{-1/2} \tilde{\pmb{K}} \pmb{D}^{-1/2} $, and therefore it possesses eigenvectors $ \{ \pmb{\phi}_0, \pmb{\phi}_1, \ldots, \pmb{\phi}_{N-1} \} $ which form a complete basis of $ \mathbb{ R }^N $.  One can check that the $ \pmb{\phi}_k $ are orthogonal with respect to the weighted inner product $ \pmb{ \phi}^\mathrm{T}_k \pmb{M} \pmb{\phi}_l = \delta_{kl} $, where $ \pmb{M} $ is the diagonal matrix with $ M_{ii} = D_{ii} / \sum_{j=1}^N D_{jj} $.  Moreover, because $ \pmb{P} $ is an irreducible right-stochastic matrix with $ \sum_{j=1}^N P_{ij}=1$, the eigenvalues $\lambda_k $  corresponding to $ \pmb{\phi}_k $ have the ordering $ 1 = \lambda_0 > \lambda_1 \geq \lambda_2 \geq \cdots \geq \lambda_{N-1} $, where the eigenvector corresponding to $ \lambda_0 $ is the constant eigenvector $ \pmb{\phi}_0 = \pmb{1} $. In NLSA, the orthogonal time series corresponding to the eigenfunctions are used as temporal modes of variability which are nonlinear analogs to the PC time series in EEOF analysis and SSA. 

\citet{CoifmanLafon06} established that in the limit of large data ($ N \to \infty$ and $ \epsilon \to 0$) and for the radial Gaussian kernel, the operator $ (\pmb{I}-\pmb{P}) / \epsilon $ converges to the Laplace-Beltrami operator associated with the Riemannian metric inherited by the manifold through its embedding in data space. In particular, the eigenvectors $ \pmb{\phi}_k = ( \phi_{1k}, \phi_{2k}, \ldots, \phi_{Nk} )^\mathrm{T} $ approximate the eigenfunctions $ \phi_k $ of the continuous operator at the sampled data points, i.e., $ \phi_{ik} \approx \phi_k( X_i ) $. An important property of the eigenfunctions \citep[][]{BelkinNiyogi03} is that they satisfy an optimality condition for the preservation of local distances in nonlinear dimension reduction maps of the form $ \Phi : \mathcal{ M } \mapsto \mathbb{ R }^m $ with $ \Phi( X_i ) = ( \phi_{i1}, \phi_{i2}, \ldots, \phi_{im} ) $. Moreover, the eigenvalues give rise to a measure of roughness for the corresponding eigenfunctions through the quantity $ 1 - \lambda_k $. In particular, the latter is proportional to the average gradient of $ \phi_k $ on $ \mathcal{ M } $ so that eigenfunctions of $ P $ with $ \lambda_k $ close to 1 correspond to ``large-scale'' weakly oscillatory functions. Note that $ \phi_k $ being a weakly oscillatory function on $ \mathcal{ M } $ does not imply that the time series $ \{ \phi_{1k}, \phi_{2k}, \ldots, \phi_{Nk} \} $ has low frequency---the timescales present in the time series are an outcome of both the geometrical structure of $ \phi_k $ on $ \mathcal{ M} $ and the sampling trajectory of the dynamics.  

 The analysis of \citet{CoifmanLafon06} was generalized by \citet{BerrySauer15}, who established that a general local kernel, not necessarily isotropic (e.g., the kernel in~\eqref{eqKCone}), can be used to approximate the Laplace-Beltrami operator of a Riemannian metric that depends on the leading moments of the kernel about the diagonal, $ X_i = X_j $. Thus, through a suitable choice of kernel one can construct dimension reduction coordinates with desirable properties for the data analysis task at hand. Here, our objective is to produce eigenfunctions with enhanced timescale separation, and the features of the kernel in~\eqref{eqKCone} which contribute to this skill are time-lagged embedding and the directional dependence of the kernel on the local dynamical flow. Specifically, \citet[][]{BerryEtAl13} theoretically established that time-lagged embedding significantly enhances the ability of the eigenfunctions to capture dynamically stable quasiperiodic patterns with timescale separation; a property that was experimentally observed in \citet[][]{GiannakisMajda11c,GiannakisMajda12a,GiannakisMajda12b,GiannakisMajda13}. Moreover, \citet{Giannakis15} established that in the limit $ \zeta \to 1 $ (where the directional dependence of the dynamical flow in the kernel is maximally significant) the eigenfunctions recover slow intrinsic timescales of the dynamical system generating the data. Both time-lagged embedding and the $ \zeta $-dependent terms contribute to the clean temporal character of the modes recovered in section \ref{secNLSAModes} ahead requiring no pre-filtering of the data.   
 
As a simple example illustrating the interpretation of the eigenvalues as measures of roughness, and the fact that the eigenvalues are generally not related to explained variance, consider a dataset $ \{ x_0, \ldots, x_n \} $ generated by a time-periodic process. That is, we have $ x_i = F( \theta_i ) $, where $ F $ is a one-to-one, periodic function of period $ 2 \pi $ taking values in $ \mathbb{ R }^d $, and $ \theta_i = \omega t_i $ for some angular frequency $ \omega $. Since $ F $ depends on a single periodic degree of freedom ($\theta_i$) and is one-to-one, the data $ x_i $ lie on a one-dimensional subset $ S $ of $ \mathbb{ R }^d $ which has the topology of a circle. On $ S $, we can define the Fourier functions $ \psi_k $, where $ k \in \{ 0, 1, 2, \ldots \} $ is an integer-valued wavenumber, $ \psi_k( \theta ) = \cos( k \theta / 2 ) $ when $ k $ is even, and $ \psi_k( \theta ) = \sin( ( k + 1 ) \theta / 2) $ when $ k $ is odd. The Fourier functions are eigenfunctions of the canonical Laplace-Beltrami operator on the circle, $ \upDelta  = - \partial^2 / \partial \theta^2 $, corresponding to the eigenvalues $ \mu_k = ( k / 2 )^2 $ and $ \mu_k = ( ( k + 1 ) / 2 )^2 $ for even and odd $ k $, respectively. Intuitively, Fourier functions with large wavenumber $ k $ are ``rough'' functions on the circle, and the fact that $ \mu_k $ scales as $ k^2 $ embodies that notion. In particular, the eigenvalues satisfy the relationship 
\begin{equation}
  \label{eqLambdaK} \mu_k = \frac{ \int_0^{2\pi} ( \partial \psi_k / \partial \theta )^2 \, d \theta } { \int_0^{2\pi} \psi^2_k \, d \theta }.
\end{equation}
For the family of kernels in~\eqref{eqKCone} and the diffusion maps normalization, the numerical eigenfunctions $ \pmb{\phi}_k $ converge  to $ \psi_k $ in the sense that in the limit of large data $ \phi_{ik} \to \psi_k( \theta_i ) $. Moreover, the numerical eigenvalues $ \lambda_k $ have the limiting behavior $( 1 - \lambda_k ) / \epsilon \to \mu_k $, which, taking into account~\eqref{eqLambdaK}, is consistent with the statement made earlier that $ 1 - \lambda_k $ is a measure of roughness proportional to the average gradient of $ \psi_k $ on the data manifold.

As an illustration that the leading eigenfunctions from NLSA can capture low-variance, yet dynamically important modes of variability, consider the same periodic example with $ d = 3 $ and $ F( 
\theta ) = ( F_1( \theta), F_2( \theta ), F_3( \theta ) ) = ( \cos \theta , \sin\theta, R \cos( l \theta ) ) $ with $ l \gg 1 $. This particular choice of $ F $ describes a ``wavy'' circle exhibiting oscillatory behavior in the $ F_3 $ direction. For $ R $ sufficiently large, that direction carries the majority of the signal's variance, and the leading EOF is $ ( 0, 0, 1 ) $. As a result, the leading PC is proportional to $F_3 $ and exhibits variability at the frequency $ l \omega $. On the other hand, the leading eigenfunctions from NLSA are not affected by the properties of $ F $ (which are extrinsic to the dynamical system), and approximate the lowest-wavenumber Fourier functions, $ \sin \theta $ and $ \cos \theta $, capturing the fundamental frequency $ \omega $. 

Of course, the datasets generated by the atmosphere ocean climate system are of significantly higher complexity than the one-dimensional periodic example discussed here, but the basic behavior discussed above generalizes to systems with multiple periodic and nonperiodic degrees of freedom. In such real-world settings, we identify physically important modes by selecting a conservatively large number of eigenfunctions (ordered in order of decreasing $ \lambda_k $) and manually examining the properties of the eigenfunction time series and the spatiotemporal reconstructions (performed as described in section~\ref{secNLSA}\ref{secReconstruction} below) of climatic variables of interest. Once a candidate set of patterns of interest has been identified, we perform various sensitivity tests to verify their robustness as described in section~\ref{secSensitivity} ahead, and then focus on the patterns that have been deemed robust. As stated earlier, the family of kernels in~\eqref{eqKCone} formulated in lagged embedding space is crucial in ensuring that individual eigenfunctions recover modes of variability with timescale separation and clear physical interpretability. An example, involving North Pacific mixed-layer temperature data, of the timescale mixing that can take place in the eigenfunctions  if no embedding is performed can be seen in Fig.~6 in \citet{GiannakisMajda13}. 

\subsection{\label{secReconstruction}Spatiotemporal reconstruction}

Consider a time series $ \pmb{y}= \{ y_0, y_1, \ldots, y_n \} $ with $ y_i \in \mathbb{ R }^{d'} $, sampled at the same times $ t_i $ as the input data $ \pmb{x} $ used in NLSA. Let also $ Y_i = ( y_i, y_{i-1}, \ldots, y_{i-q+1} ) $ be the corresponding time-lagged embedded data. In this paper, we will always set $ \pmb{ y } = \pmb{ x } $ (i.e., $ \pmb{y} $ will be Indo-Pacific SST data), but in Part~II $ \pmb{ y } $ will also represent other fields of interest such as Indo-Pacific surface winds. We reconstruct spatiotemporal patterns associated with $ \pmb{ y } $ from the eigenfunctions following the convolution approach used in EEOF analysis and SSA \citep[][]{GhilEtAl02}. Selecting an eigenfunction $ \pmb{\phi}_k $, we first compute the spatiotemporal pattern $ \pmb{Y}_k $ in lagged-embedding space given by $ \pmb{Y}_k =  \pmb{Y} \pmb{M} \pmb{\phi}_k \pmb{\phi}_k^\mathrm{T} $, where $ \pmb{ Y }_k $ is a $ (qd') \times N $ matrix. That pattern is then projected to a spatiotemporal pattern $ \pmb{y}_k $ in $ d' $-dimensional physical space by averaging along $ d \times q $ blocks of $ \pmb{Y}_k $ corresponding to the lagged windows; see, e.g., equation~[1] in~\citet[][]{GiannakisMajda12a} for an explicit formula. Note that $ \pmb{y}_k $ is a $ d \times n $ matrix, and the matrix element $  y_{k,ij} $ corresponds to the value of the reconstructed signal at the $ i $-th gridpoint in physical space at time $ t_j $. 

Given a collection of indices $ \kappa = \{ k_1, k_2, \ldots, k_r \} $, we define the reconstructed pattern $ \pmb{y}_\kappa $ as the sum $ \pmb{ y}_\kappa = \sum_{j=1}^r \pmb{y}_{k_j} $. In sections~\ref{secNLSAModes}--\ref{secObsModes} ahead, $ \kappa $ will frequently consist of pairs of in-quadrature (90$^\circ $ out of phase) eigenfunctions representing oscillatory phenomena (e.g., the annual cycle), and families of modes related by amplitude modulations (e.g., ENSO and its modulation by the annual cycle). For each such pattern $ \pmb{ y}_\kappa $ we  compute the time average $ \bar y_{\kappa,i} = \sum_{j=1}^n y_{\kappa,ij} / n $ at each gridpoint and the associated explained variance $ \sigma_{\kappa,i}^2  =\sum_{j=1}^n ( y_{\kappa,ij} - \bar y_{\kappa,i} )^2 / n $. We also compute a spatially aggregated average $ \sigma^2_\kappa = \sum_{i=1}^d \sigma_{\kappa,i}^2 $. Note that because the $ \pmb{y}_\kappa $ are not constrained to be orthogonal in time and/or space,  $ \sigma_{\kappa,i} $ and $ \sigma_\kappa $ do not add in quadrature.

If all of the eigenfunctions are used, i.e., $ \kappa = \{ 0, \ldots, N - 1 \} $, $ \pmb{y}_\kappa $ recovers the original data $ \{ y_1, \ldots, y_n \} $ exactly (note that the initial sample $ y_0 $ is not recovered since the corresponding $ x_0 $ sample is ``used up'' to compute the time tendency $ \xi_0 $ in the kernel in~\eqref{eqKCone}). Of course, in practice one performs dimension reduction using $ m \ll N $ eigenfunctions. These $ m $ eigenfunctions need not be consecutive, and in what follows we will employ a set of nonconsecutive eigenfunctions representing important annual, interannual, and decadal modes of variability of Indo-Pacific SST.  

\section{\label{secDataset}Dataset description}

\subsection{Model data}

We analyze 1300 years of monthly averaged SST data from a control integration of CCSM4 \citep[][]{ccsm4,GentEtAl11}. This control run is forced with preindustrial greenhouse gas levels, and has an ocean grid of approximately $ 1^\circ $ nominal resolution. In this configuration, CCMS4 is known to simulate realistic ENSO and Pacific decadal variability \citep[][]{DeserEtAl12b}. This work focuses on the Indo-Pacific domain, which we define here as the ocean gridpoints in the longitude-latitude box 28$^\circ$E--70$^\circ$W and 60$^\circ$S--20$^\circ$N. Working throughout with the model's native ocean grid, the spatial dimension (number of gridpoints) of the dataset is $ d = \text{44,471} $. We also analyze 800 years of monthly SST data from a preindustrial control integration of CM3 \citep{GriffiesEtAl11, gfdl-cm3} over the same Indo-Pacific domain. In this case, the nominal resolution is slightly lower corresponding to  24,612 gridpoints. Both datasets exhibit a small amount of drift that does not exceed 0.01 K per century.  Note that the data is not subjected to preprocessing such as removal of the seasonal cycle, low-pass filtering, or detrending to remove the drift prior to analysis via NLSA. As will be discussed below, the ability to capture amplitude modulation relationships (e.g., the modulation of the annual cycle by ENSO through combination modes) through NLSA relies crucially on temporal information which would be lost by prefiltering the data. 

\subsection{\label{secObsData}Observational data}

We also study HadISST data \citep[][]{RaynerEtAl03,HadISST13} over the same Indo-Pacific domain. This dataset consists of monthly averaged SST data on a uniform 1$^\circ$ latitude-longitude grid containing $ d = \text{20,960} $ gridpoints in our Indo-Pacific domain. We use either the full, 140-year HadISST dataset from January 1870 to December 2013, or the shorter, but more reliable, portion of that dataset from January 1979 to December 2009 spanning the satellite era (the latter dataset was truncated to 2009 for compatibility with the reanalysis data used in Part~II). Unlike the model data, these datasets exhibit a significant trend component which we do not remove as we found that NLSA naturally separates the statistically stationary and trend components of the input data into distinct eigenfunctions. We defer the study of such trend modes to future work.  

\section{\label{secNLSAModes}Spatiotemporal modes recovered from CCSM4 data}

We extract spatiotemporal modes of Indo-Pacific SST using the NLSA algorithm described in section~\ref{secNLSA} applied to the 1300 y CCSM4 control run. To induce timescale separation between multidecadal, interannual, and seasonal modes, we use a 20-year embedding window corresponding to $ q = 240 $ monthly lags. Note that the embedding window used in this work is significantly longer than the two-year window used by GM in studies of North Pacific SST variability via NLSA. Following \citet[][]{Giannakis15}, we work with the kernel in~\eqref{eqKCone} for the parameter values $ \epsilon = 1 $ and $ \zeta = 0.99 $. Using these parameter values, we compute the leading 200 Laplace-Beltrami eigenfunctions and their associated spatiotemporal patterns.  We will discuss the sensitivity of our results with respect to the NLSA parameters as well as the choice of spatial domain, timespan of the data, and the presence of noise in section~\ref{secSensitivity}.

In what follows and in Part II, out of the 200 computed eigenfunctions we focus primarily on an 18-element subset representing prominent modes of Indo-Pacific variability---the annual cycle and its harmonics, ENSO, the combination modes associated with the modulation of the annual cycle by ENSO, the WPMM, the IPO, the TBO, and the combination modes associated with the modulation of the annual cycle by the TBO. These modes were identified through the temporal power spectra of the eigenfunction time series and the corresponding spatiotemporal reconstructions. The global ordering of the corresponding eigenfunctions in the spectrum (ordered as described in section~\ref{secNLSA}\ref{secEigenfunctions}) is $ 1, \ldots, 12, 90, 93, 116, 117, 146, 147 $. For notational clarity, we denote these  eigenfunctions by $ \phi_1, \phi_2, \ldots,\phi_{18} $, respectively in accordance with their ordering within the selected subset. In addition to this subset, the NLSA spectrum contains several other ``secondary'' ENSO modes and their combinations with the annual cycle and its harmonics forming an ENSO frequency cascade \citep[][]{StueckerEtAl15}. In Part~II, we will examine the role of these modes in El Ni\~no-La Ni\~na asymmetries. 

Throughout this section, we frequently make reference to the amplitude and fractional variance explained by our modes. By amplitude we mean half of the difference between the maximum and minimum values of the SST anomalies at a given region, reconstructed for a given set of modes as described in section~\ref{secNLSA}\ref{secReconstruction} and visualized in the movies in the online supplementary material. Variances are also computed as described in section~\ref{secNLSA}\ref{secReconstruction} either globally, or at each gridpoint to construct spatial maps of explained variance. Depending on the context, we report fractional variances relative to the raw data, the raw data with the seasonal cycle subtracted, the reconstructed data using all modes $ \{ \phi_1, \ldots, \phi_{18} \} $, and the reconstructed data using all modes except from those representing the seasonal cycle. For conciseness, we sometimes refer to the raw or reconstructed explained variance computed after removal of the seasonal cycle as ``nonperiodic explained variance''.  We note again that since the reconstructed patterns are not constrained to be orthogonal (neither in time nor in space), our computed variances are not additive.

\subsection{\label{secPeriodic}Periodic modes} 

The first six eigenfunctions, $ \phi_1, \ldots, \phi_6 $, come in doubly degenerate pairs that represent the annual cycle and its first two harmonics. The corresponding time series and spatial composites are depicted in Fig.~\ref{figPeriodic}, and 150 years of reconstructed signal is visualized in movie 1 in the online supplementary material. The variance explained by these modes relative to the raw SST data is shown in Figs.~\ref{figVarRaw}(a--c). 
 
The temporal patterns in this family have the structure of in-quadrature sinusoidal waves of frequency 1, 2, and 3 y$^{-1}$, respectively for $ \{ \phi_1, \phi_2 \} $, $ \{ \phi_3, \phi_4 \} $, and $ \{ \phi_5, \phi_6 \} $. In the spatial domain, the annual pair $ \{ \phi_1,  \phi_2 \} $ describes fluctuations of predominantly  zonal character (Fig.~\ref{figPeriodic}(c)) with pronounced strength in the southern hemisphere midlatitudes and a relatively weak signal over tropics. The corresponding explained variance relative to the raw data reaches more than 60\% south of Tropic of Capricorn (see Fig.~\ref{figVarRaw}(a)). In particular, the amplitude of the reconstructed anomalies reaches up to 5 K in a zonal belt centered around the $ 35^\circ $S parallel, explaining up to 70\% of the raw variability east of Australia. The variance explained by the annual modes relative to the data reconstructed from the 16-eigenfunction family (not shown here) is even higher, reaching up to 90\% in that region.  Averaged over the whole Indo-Pacific basin, the variance associated with the annual modes is approximately 46\% and 58\% of the raw and reconstructed data, respectively, with negligible contribution from the tropics. 

The semiannual periodic pair $ \{ \phi_3, \phi_4 \} $ features a more complex spatial structure than the annual pair, with its activity taking place not just in the midlatitudes but also over the tropics (see Fig.~\ref{figPeriodic}(f) and movie~1 in the online supplementary material). The strongest reconstructed fluctuations for this mode occur in the western and northern parts of the Indian Ocean (especially the coastal waters of Somalia, the Arabian Sea, and the Bay of Bengal), the vicinity of the Indonesian Archipelago, and the eastern tropical Pacific, where anomaly amplitudes up to $\sim$1.5 K are observed. Averaged over the whole Indo-Pacific basin, the semiannual modes explain approximately 10\%  and 15\% of the raw and reconstructed variance, respectively. Regionally, the explained variance of these modes can be significantly higher, reaching 25--50\% of the raw variability over the tropical Indian Ocean and the Indonesian Archipelago (see Fig.~\ref{figVarRaw}(b)).

Similarly to the semiannual modes, the dominant spatial anomalies for the triennial modes $ \{ \phi_5, \phi_6 \} $ are concentrated  in the tropics, but generally feature smaller scales and weaker amplitudes (see Fig.~\ref{figPeriodic}(i) and movie~1 in the online supplementary material). Amplitudes up to $\sim 0.5 $ K are observed over the Gulf of Aden and the west part of Arabian Sea, where the resulting variance amounts to $\sim 20\%$ of the raw variability (see Fig.~\ref{figVarRaw}(c)). These anomalies diminish gradually towards the equator, eventually changing sign along the coast of Africa and between the Equator and Tropic of Capricorn. Significant anomalies are also found over the tropical Pacific, first emerging west of the Galapagos, and subsequently propagating westward. These anomalies reach amplitudes up to 0.5 K, and locally explain up to 20\% of the raw variance. Other notable anomalies associated with these modes occur within the equatorial belt over the Indian Ocean, as well as over waters located north of Australia, west of New Guinea, and south of Sulawesi. 

\subsection{\label{secInterannual}ENSO and ENSO combination modes}

ENSO and its associated combination modes are captured by several NLSA modes. In this paper, we focus on the mode families $ \{ \phi_7, \phi_8 \} $, $ \{ \phi_9, \phi_{10} \} $, and $ \{ \phi_{11}, \phi_{12}  \} $ which represent the fundamental component of ENSO and amplitude modulations of the annual cycle by ENSO, respectively. The spatial and temporal patterns associated with these families are displayed in Fig.~\ref{figInterannual}  and movies~2 and 3 in the online supplementary material. Spatial maps of the fractional variance explained by these modes, computed relative to the raw SST data, the raw SST data minus the seasonal cycle, and the signal reconstructed by the nonperiodic modes $ \{ \phi_7, \ldots, \phi_{14} \} $ in our selected mode family are shown in Figs.~\ref{figVarRaw}(d--f), ~\ref{figVarRawWithoutPer}(a--c), and~\ref{figVarRecWithoutPer}(a--c).

Modes $ \{\phi_7, \phi_8\} $ capture the dominant features of ENSO through an in-quadrature pair of amplitude-modulated waveforms (see Fig.~\ref{figInterannual}(b)). These waveforms have a carrier frequency of $ \nu_\text{ENSO} \approx (\text{4 y})^{-1} $ and a low-frequency modulating envelope evolving on decadal timescales.  The resulting Fourier spectrum (Fig.~\ref{figInterannual}(a)) is strongly peaked at the carrier frequency, but is broad and contains appreciable power on decadal to centennial timescales. In Part~II, we will make a connection between the amplitude envelope of this family and the WPMM.

Spatially (see Fig.~\ref{figInterannual}(c) and movie 2 in the online supplementary material),  the $ \{\phi_7,\phi_8 \} $ cycle initiates with positive SST anomalies emerging west of the Galapagos Islands, exceeding at times 1 K there. At the same time, a center of opposite-sign and slightly weaker anomalies develops over the South China Sea, as well as along the eastern and western coast of Australia. This phase lasts a few months and is associated with the westward expansion of the eastern tropical Pacific anomalies.  Moreover, an El Ni\~no-like triangular pattern develops as this anomaly weakens while spreading simultaneously south and north to the subtropical eastern Pacific. At the same time, positive anomalies up to $\sim 0.5 $ K develop over Indian Ocean waters north of 40$^\circ$S and south of an arc crossing through the Gulf of Aden and the Perth Basin. Also, negative anomalies propagate east of Australia and intensify over the southeast Pacific Ocean.   A double triangular pattern of opposite-sign anomalies eventually covers the majority of the subtropical Pacific waters east of the warm pool, and persists for several months before dissipating, ending one half of the ENSO cycle. 

Averaged in time, the explained variance of modes $ \{ \phi_7, \phi_8 \} $ reaches up to 20\% and  50\% of the raw (Fig.~\ref{figVarRaw}(d)) and reconstructed (not shown) variability of the equatorial regions of the Pacific Ocean. Averaged over the whole Indo-Pacific basin, these variances amount to 4.6\% and 5.7\%, respectively. When all but periodic modes are taken into account, the explained variance of the reconstructed data increases to 70\% in the equatorial Pacific (see Fig.~\ref{figVarRecWithoutPer}(a)) and to 40\% for the whole Indo-Pacific basin.  

The next two pairs of eigenfunctions, $ \{ \phi_9, \phi_{10} \} $ and $ \{ \phi_{11}, \phi_{12} \} $, shown in Figs.~\ref{figInterannual}(e, h), describe amplitude modulations of the annual cycle by the primary ENSO modes. In the Fourier domain, these modulations are evidenced by strong peaks in the spectra  in Figs.~\ref{figInterannual}(d, g) at the frequencies $ ( \mbox{1 y$^{-1}$ } )  - \nu_\text{ENSO} = 0.75 $ y$^{-1} $ and $  ( \mbox{1 y$^{-1}$ } ) + \nu_\text{ENSO} = 1.25 $ y$^{-1}$, respectively. Moreover, the amplitude envelopes of $ \{ \phi_9, \phi_{10} \}$ and $ \{ \phi_{11}, \phi_{12} \} $, computed via Hilbert transforms, correlates almost perfectly (correlation coefficient 0.99 at $ p $-value numerically equal to zero based on a $ t$-test with $(\text{1300 y}) \times \nu_\text{ENSO} - 2 = 323 $ degrees of freedom) with the corresponding amplitude envelope of $ \{ \phi_7, \phi_8 \} $. A more direct test for modulating behavior is to examine the relative phases between the complex numbers $ z = \phi_7+ \mathrm{i} \phi_8 $,  $ z_- = \phi_9 + \mathrm{i} \phi_{10} $, and  $ z_+ = \phi_{11} + \mathrm{i} \phi_{12} $. As shown in Fig.~\ref{figPhase}, the real and imaginary parts of $ z_- / z $ and $ z_+ / z_- $ are to a good approximation in-quadrature sinusoidal waves of frequency 1 y$^{-1}$ and 2 y$^{-1}$, respectively. This confirms that  $ \{ \phi_9, \phi_{10} \} $ and $ \{ \phi_{11}, \phi_{12} \} $ are modulations of $ \{ \phi_7, \phi_8 \} $ by the annual cycle, and that the peak frequency of $ \{ \phi_9, \phi_{10} \} $ lies between the frequencies of its parent signals (ENSO and the annual cycle), whereas the peak frequency of $ \{ \phi_{11}, \phi_{12} \} $ exceeds the frequencies of both of its parent signals. Hereafter, we denote the families $ \{ \phi_9, \phi_{10} \} $ and $ \{ \phi_{11}, \phi_{12} \} $ by ENSO-A1 and ENSO-A2, respectively, where ENSO-A stands for ``ENSO-annual''. 

The dominant frequencies of ENSO-A1 and ENSO-A2 are consistent with a frequency cascade attributed to quadratic nonlinearities in the coupled equations of motion for the ocean and atmosphere, leading to a synchronization between the seasonal cycle and ENSO \citep[][]{McGregorEtAl12,StueckerEtAl13,StueckerEtAl15,SteinEtAl14,RenEtAl16}. In previous works, these combination modes were either explicitly constructed through products of interannual ENSO and annual periodic signals \citep[][]{StueckerEtAl15}, through Hilbert transform techniques applied to ENSO indices \citep[][]{SteinEtAl11,SteinEtAl14}, or they were identified via EOF analysis of surface atmospheric circulation fields \citep[][]{McGregorEtAl12,StueckerEtAl13,RenEtAl16}. In the EOF-based analyses, ENSO and the combination modes at the $ ( \mbox{1 y$^{-1}$ } ) \pm \nu_\text{ENSO} $ frequencies (a total of six real-valued or four complex-valued signals) were represented as a single pair of in-quadrature PCs.  In  NLSA, the full ENSO, ENSO-A1 and ENSO-A2 families emerge naturally from Indo-Pacific SST model data. In sections~\ref{secGFDL} and \ref{secObsModes} we will see that these modes are also recovered from the CM3 and HadISST datasets, respectively, and in Part~II we will establish that the surface atmospheric circulation patterns due to these modes are consistent with the southward shift of zonal wind anomalies. The deterministic nature of the relationship between ENSO and ENSO-A1,2 has been proposed as source of ENSO termination predictability \citep[][]{StueckerEtAl13,StueckerEtAl15}. 

The spatiotemporal pattern associated with ENSO-A1 (shown in Fig.~\ref{figInterannual}(f) and movie~3 in the online supplementary material) consists primarily of SST anomalies emerging periodically along the coast of equatorial South America, strengthening up to $\sim 0.2 $ K in the vicinity of the Galapagos Islands, and subsequently propagating westward over the equatorial Pacific until dissipating around the date line. The termination of this equatorial wavelike is associated with simultaneous emergence of another cluster of SST anomalies located southwest of the coast of Sumatra. These anomalies have the same sign as the terminating wave, reach amplitudes up to $\sim 0.4 $ K, and persist west of the Sunda Strait for 6--7 months. In the meantime, a southwestward-propagating SST anomaly develops in the northeast Indian Ocean. This traveling disturbance appears to reflect upon reaching the east coast of Africa near the equator, and to cross the Indian Ocean for a second time until it merges with the cluster of anomalies west of the Sunda Strait (see, e.g., January--September of simulation year 1170 in movie~3 in the online supplementary material). After the decay of the latter anomalies, a pronounced disturbance develops along the north coast of Australia, and splits into a part that propagates slowly along the west Australian coast and a part that crosses the Pacific Ocean in a southeastward direction. Towards the end of this phase, new anomalies develop over the eastern tropical Pacific, and the cycle repeats. 

Overall, the vicinity of  the Galapagos Islands (and more broadly the eastern and central equatorial Pacific) and the Indian Ocean west of the Sunda Strait are regions where a significant amount of variability can be attributed to the ENSO-A1 family. In particular, the corresponding explained variance of the reconstructed data is up to 35\% west of the Sunda Strait and 13.5\% for the whole Indo-Pacific basin when all but periodic modes are taken into account (see Fig.~\ref{figVarRecWithoutPer}(b)). Moreover, as shown in Figs.~\ref{figVarRaw}(e) and~\ref{figVarRawWithoutPer}(b), these modes explain up to $\sim 8\%$ and $ \sim 12\%$ of the variance of the raw and nonperiodic signal west of the Sunda Strait, averaging to 1.6\% and 4\%, respectively, for the whole domain. Note that the region west of the Sunda Strait is typically associated with the IOD as defined conventionally through an index based on SST anomalies measured partially there \citep[][]{SajiEtAl99}. As pointed out in section~\ref{secIntro}, although the nature of the IOD and its relation to ENSO have been studied extensively over the last 15 years, controversies remain, and no prevailing consensus has been achieved yet in the community. Our results indicate that in CCSM4 (and also for HadISST; see section~\ref{secObsModes}), a significant portion of the SST fluctuations contributing to the variability typically associated with the IOD can be interpreted as a  deterministic modulation of the annual cycle by the fundamental ENSO modes $ \{ \phi_7, \phi_8 \} $. 

The spatiotemporal patterns associated with  ENSO-A2 bear several similarities with the patterns described above for ENSO-A1, but also have notable differences. First, the amplitude of the former is $\sim30\% $ lower than that of the latter for most of the regions where these two modes are both active. Furthermore, in the case of ENSO-A2, the traveling disturbance propagating west of the Galapagos is faster, has significantly finer spatial structure than in the case of ENSO-A2, and propagates further to the north and west of the equatorial Pacific before reflecting back over the Celebes Sea.  Compared to ENSO-A1, the anomalies west of the Sunda Strait associated with ENSO-A2 last for a significantly shorter time, and exhibit a stronger intensification after the arrival of the wave crossing twice the Indian Ocean (first in a southwest and then in a southeast direction of propagation). The temporal evolution of the anomalies over the South China Sea is also shifted in phase relative to the equatorial wave propagation.  Averaged over the whole domain, these modes explain 1.1\% of the raw variance, 2.9\% of the raw nonperiodic variance, and up to 10.4\% of the reconstructed nonperiodic variance. As with ENSO-A1, the explained variance by the ENSO-A2 modes is especially pronounced over the eastern Indian Ocean, reaching 9\% of the raw variance, 12\% of the raw nonperiodic variance, and 35\% and of the reconstructed  nonperiodic variance northeast of the Sunda Strait (see Figs.~\ref{figVarRaw}(f), \ref{figVarRawWithoutPer}(c), and~\ref{figVarRecWithoutPer}(c), respectively). Thus, these modes are also a significant contributor to SST anomalies over the region traditionally associated with the IOD. Similarly to ENSO termination, the deterministic nature of the relationship between ENSO and ENSO-A1,2 has potential implications to SST predictability in IOD regions.  

\subsection{\label{secBiennial}TBO and TBO combination modes}

The fundamental component of the TBO is captured by eigenfunctions $\{ \phi_{15}, \phi_{16} \}$. These modes form a degenerate in-quadrature oscillatory pair which, as shown in Fig.~\ref{figBienn}(a,b), is of predominantly biennial periodicity with a peak frequency $\nu_\text{TBO} = 0.5 $ y$^{-1} $. Spatially (see Fig.~\ref{figBienn}(c) and movie 1 in the supplementary material), these modes are active over the majority of the Indo-Pacific sector and explain approximately 0.25\%, 1.3\%, and 5\% of the raw, nonperiodic, and reconstructed nonperiodic variance, respectively. For the region northeast and east of New Guinea  (see Figs.~\ref{figVarRaw}(g), \ref{figVarRawWithoutPer}(d), and~\ref{figVarRecWithoutPer}(d)), approximately 2\%, 4\%, and 22\% of the corresponding variances can be attributed to this pair. 

The cycle represented by $ \{ \phi_{15}, \phi_{16} \} $ starts with positive anomalies developing west of the Galapagos Islands which intensifies up to 0.15 K over a six-month period.  In the meantime, an anomaly of the same sign and similar amplitude emerges along the west coast of Sumatra and Java and another one dissipates over the South China Sea. Also, weaker negative anomalies grow southeast of New Zealand and off the coast of Somalia. Six months later, positive anomalies west of the Galapagos propagate westwards, mainly to the north of equator, and negative anomalies southeast of New Zealand start to propagate east while weakening in the meantime. The westward propagation of the disturbances  leads to a fast warming northeast of New Guinea three months later. This wave subsequently propagates in a southeast direction and gradually decays. In the meantime, the positive anomalies west of Sumatra dissipate, while negative anomalies intensify over the South China Sea. A simultaneous development of negative anomalies west of Galapagos initiates the second half (year) of the cycle.

Similarly to ENSO, the TBO modes $ \{ \phi_{15}, \phi_{16} \} $ have associated combination modes, $ \{ \phi_{17}, \phi_{18} \} $, representing the phase locking of the TBO with the annual cycle. As shown in Fig.~\ref{figBienn}(d,e), the dominant frequency of this oscillatory pair is $ \nu_\text{TBO}+ \text{1 y$^{-1}$} \simeq \text{1.5 y$^{-1}$} $, and its amplitude envelope (extracted via a Hilbert transform) has a correlation coefficient of 0.87 with the corresponding amplitude envelope of the fundamental TBO modes (the $ p$-value of this correlation coefficient is numerically zero based on a $ t $-test with  with $(\text{1300 y}) \times \nu_\text{TBO} - 2 = 748 $ degrees of freedom). Moreover,  a similar amplitude and frequency analysis as that performed in section~\ref{secNLSAModes}\ref{secInterannual} and Fig.~\ref{figPhase} (not shown here), reveals that $ \{ \phi_{17}, \phi_{18} \} $ is indeed a pair of combination modes analogous to ENSO-A2. Hereafter, we denote this pair by TBO-A. In the spatial domain (Fig.~\ref{figBienn}(f)), the SST patterns associated with these modes have strong loadings in the equatorial Indo-Pacific and especially over the Maritime Continent. In Part~II, we will see that together with the fundamental TBO modes these patterns are consistent with the surface wind and precipitation patterns associated with biennial variability of the Indian-Australian Monsoon \citep[][]{MeehlArblaster02}.  
  
\subsection{\label{secDecadal}Decadal and multidecadal modes}

\subsubsection{West Pacific multidecadal mode} 

Mode $ \phi_{13} $, whose temporal pattern and frequency spectrum are depicted in Figs.~\ref{figDec}(a, b), represents the dominant multidecadal fluctuations of SST as extracted in the CCSM4 control run via NLSA. Eigenfunctions with the qualitative features of $\phi_{13} $ occur consistently and with negligible corruption from shorter timescales over a range of embedding windows greater than a decade and over a range of the other NLSA parameters. The spatial patterns associated with the positive phase of $ \phi_{13} $ (see Fig.~\ref{figDec}(c) and movie 2 in the online supplementary material) consist predominantly of a globular SST anomaly with amplitudes up  to $\sim 0.25 $ K, which is bounded by New Guinea from the south, the dateline from the east, and Borneo from the west. A weaker anomaly of the same sign develops over the southern Pacific east of the dateline, while anomalies of the opposite sign prevail at the same time in the vicinity of Tasmania, gradually weakening westward for over 100$^\circ$ of longitude. Weaker fluctuations emerge also over other regions. In particular, the western Indian Ocean and the subtropical central Pacific are anomalously warmer whereas negative anomalies can be found south of Ecuador, east and west of Australia, the Indonesian archipelago, the Bay of Bengal, and the South China Sea. 

The significant events associated with the WPMM are of aperiodic character and vary in spatial extent and duration. The spatial pattern developing during these events consists of the anomalies described above persisting over the course of individual episodes that can last for several decades.  For example, the reconstructed years 1060--1120 in movie~2 in the online supplementary material are associated with a single prolonged event with a positive anomaly over the warm pool. 

The WPMM explains  $\sim 1\%$ of the reconstructed variance  averaged over the whole Indo-Pacific basin, and $\sim 12\% $ if all but periodic modes are taken into account. As shown in Figs.~\ref{figVarRaw}(h) and~\ref{figVarRecWithoutPer}(e), a strong imprint of this mode is found over the equatorial Pacific northeast of New Guinea, where it explains locally over 10\% and 50\% of the nonperiodic raw and reconstructed variance, respectively (see Figs.~\ref{figVarRawWithoutPer}(e) and~\ref{figVarRecWithoutPer}(e)). Similar values of the nonperiodic SST variance localized south of the subtropics and west of Tasmania can be explained by this mode. 

In summary, the spatial characteristics of the WPMM partially resemble those of a west-east decadal dipole, identified often by the second EOF of low-pass filtered SST  \citep[][]{Timmermann03, YehKirtman04, OgataEtAl13}. While the physical relevance of this EOF has sometimes been questioned (e.g., due to preprocessing of the data by low-pass filtering), the fact that the WPMM is recovered here from unprocessed data suggests that, at least in CCSM4, it has a physically meaningful dynamical origin. In Part~II, this assertion will be supported through an analysis of surface wind, thermocline depth, and precipitation fields associated with the WPMM.  

\subsubsection{Interdecadal Pacific Oscillation}

Besides the WPMM, the NLSA spectrum contains another low-frequency mode, extracted via eigenfunction $ \phi_{14} $. As shown in Figs.~\ref{figDec}(d--f), this eigenfunction has a strong decadal component,  with individual events usually lasting more than a decade, but is also influenced significantly by higher-frequency (biennial) fluctuations. According to the composite in Fig.~\ref{figDec}(f) and the reconstructions in movie 2 in the online supplementary material, this mode is active mainly over  the Pacific basin. In particular, it generates a pronounced anomaly up to  $\sim 0.1$ K prevailing over a large triangle-shaped region with a vertex in the central Pacific Ocean and an eastern edge along the coastline of South America. An equally pronounced but negative anomaly occurs south of the last anomaly reaching as far as 60$^\circ$S latitude and as far as Tasmania in the west. Such spatial characteristics, in particular the ENSO-like pattern over the eastern part of the South Pacific Ocean, resemble the pattern associated usually with IPO. The explained variance from $ \phi_{14} $,  averaged over the Indo-Pacific basin (with the Indian Ocean playing a minor role in comparison to other basins), amounts to less than 1\% of the raw and reconstructed variability, and is slightly less than 7\% if all but periodic modes are taken into account. Note that  $\phi_{14}$  explains two times less variance than the WPMM identified by $ \phi_{13}$. 

\section{\label{secSensitivity}Sensitivity analysis}

In this section, we verify the robustness of the Indo-Pacific SST modes presented in section~\ref{secNLSAModes} with respect to the choice of NLSA parameters (section~\ref{secSensitivity}\ref{secNLSAParameters}), the choice of spatial domain (section~\ref{secSensitivity}\ref{secDomain}), and the presence of independent and identically distributed (i.i.d.) Gaussian noise in the data (section~\ref{secSensitivity}\ref{secNoise}). Additional details on the choice of NLSA parameters can be found in GM and in \citet{Giannakis15}. Technical results on the behavior of NLSA and related kernel algorithms in the presence of i.i.d.\ noise are discussed in \citet{Giannakis16}. 

\subsection{\label{secNLSAParameters}Choice of NLSA parameters}

The parameters of NLSA as described in section~\ref{secNLSA} are the number of embedding lags $ q $, the kernel bandwidth $ \epsilon $, and the cone kernel parameter $ \zeta $. For the class of kernels in~\eqref{eqKCone} featuring $ \xi $-dependent normalization factors the nominal value of $ \epsilon $ is 1 \citep[][]{Giannakis15}. The results in this paper and in Part~II were obtained using  this parameter value but we have verified that qualitatively similar results can be obtained for values $ \epsilon $ in the interval $ [ 1, 5 ] $. Theoretically, for maximal capability to recover slow timescales (in the present context, decadal modes), $ \zeta $ should be set arbitrarily close to 1, but not equal to or exceeding 1  \citep[][]{Giannakis15}. Throughout, we work with $ \zeta = 0.99 $, though we find that our results are not too sensitive on values of $ \zeta $ in the interval $ [ 0.95, 1 ] $.

To verify the robustness of our results against the number of embedding lags, we performed analyses with $ q $ in the interval $ [ 48, 480 ] $ corresponding to physical embedding window lengths $ \Delta t $ in the range  4--40 years. The salient features of the periodic, ENSO and ENSO combination modes studied here  are robust under these changes. On the other hand, in order to cleanly resolve the WPMM, IPO, TBO, and TBO combination modes we found that $ \Delta t \gtrsim 10 $ y embedding windows are generally required. As stated in the preamble of section~\ref{secNLSAModes}, our nominal choice, $ q = 240 $, $ \Delta t = 20 $ y, is significantly longer than the 2-year embedding windows used in previous studies of North Pacific SST variability via NLSA \citep[GM;][]{Giannakis15}. We believe that the longer embedding windows needed to achieve timescale separation in the Indo-Pacific system studied here is at least partly caused by the broadband character of ENSO. In particular, we observe that modes with qualitatively similar characteristics as the ENSO and ENSO-A modes described in section~\ref{secNLSAModes}\ref{secInterannual} are present in the NLSA spectrum for embedding windows as small as 4 years. 

As the embedding window length increases, the power spectral densities of the ENSO and ENSO combination modes become increasingly concentrated around the fundamental ENSO and ENSO-A peaks, and additional, ``secondary'' ENSO modes appear with frequency peaks adjacent to the dominant ENSO peak. This behavior, illustrated in Fig.~\ref{figSpec}, is generally consistent with the presence of an ENSO frequency cascade in the data, which is progressively separated by NLSA into its constituent frequencies with increasing $ q $. Such a frequency cascade is expected on general grounds due to dynamical nonlinearities in the coupled atmosphere-ocean system \citep[][]{StueckerEtAl15}, and as $ q $ increases NLSA is theoretically expected \citep[][]{BerryEtAl13,Giannakis16} to increasingly resolve these frequencies through individual eigenfunctions. Together, these facts may explain the behavior observed in Fig.~\ref{figSpec}. It is important to note that the dominant ENSO frequency $ \nu_\text{ENSO} $, the phase relationships between the ENSO and ENSO-A modes (depicted in Fig.~\ref{figPhase}), and the structure of the corresponding spatial SST patterns (see Fig.~\ref{figInterannual}) are robust under changes of $ q $ despite the emergence of new ENSO-like modes. Moreover, the decadal ENSO envelopes are also reasonably robust for $ \Delta t \geq 10 $ y.  While a detailed study of the secondary ENSO modes is beyond the scope of this work, in Part~II we will demonstrate that the collection of fundamental and secondary NLSA ENSO modes provides a realistic representation of El Ni\~no-La Ni\~na SST asymmetries. 

\subsection{\label{secDomain}Choice of spatial domain}

To verify the robustness of our results against changes of the spatial domains we have performed NLSA on SST data from the same CCSM4 control integration on two subregions of our Indo-Pacific domain, namely the tropical Pacific band 100$^\circ$E--110$^\circ$W and -20$^\circ$S--20$^\circ$N  and the tropical Indian Ocean band 30$^\circ$E--110$^\circ$E and 30$^\circ$S--30$^\circ$N. The modes recovered by NLSA on these subdomains are in good agreement with the family presented in section~\ref{secNLSA}. This agreement is particularly good in the case of the annual and interannual modes in the top of the spectrum, i.e., the annual cycle and its harmonics, ENSO and ENSO-A modes. WPMM patterns can also be detected from these subdomains,  although in the case of the tropical Pacific subdomain the WPMM pattern is corrupted (presumably due to the regional dominance of ENSO activity). 

In general, we do not detect qualitatively new patterns among the leading NLSA modes from the two subdomains, suggesting that the Indian and Pacific Oceans exhibit strong dynamical couplings. We will study this point further in Part~II using spatiotemporal reconstructions of atmospheric circulation fields. The agreement of the eigenfunctions from the different domains is also consistent with theoretical invariance properties of eigenfunctions obtained from the kernel in~\eqref{eqKCone} due to delay embeddings \citep[][]{BerryEtAl13,Giannakis16} and the directionally dependent cone kernel structure \citep[][]{Giannakis15}. Together, these properties imply that for data generated by a single dynamical system (in this case, the Indo-Pacific atmosphere-ocean system) and for sufficiently many delays $ q $, different observation functions (in particular, SST over the full Indo-Pacific domains or the Pacific and Indian Ocean subdomains) should generically lead to similar leading eigenfunctions.

\subsection{\label{secTimespan}Influence of the temporal span of the input data}

Another factor whose relevance in obtaining our results we examine here is the millennial length of the CCSM4 dataset. For a dataset of this timespan, all modes presented in section~\ref{secNLSAModes} are reasonably well-sampled. In particular, approximately 30 strong events (split approximately evenly between positive and negative phases) are captured for the WPMM, the slowest mode in our mode family. The duration of these events can easily exceed the timespan of the satellite era, and can even be comparable to the length of the industrial era (see, e.g., the positive WPMM event emerging around year 1060 in movie~2 and lasting for approximately 65 years), raising questions about the detectability of these modes in observational datasets of limited temporal coverage and potentially significant systematic biases.  

Here, we examine the sensitivity of the modes in section~\ref{secNLSAModes} to the timespan of the input data by splitting the original 1300 y CCSM4 time series into eight disjoint consecutive subperiods, each spanning 162.5 y, and performing NLSA on each batch with all parameters kept the same as in the original setup. Examining the resulting eight sets of eigenfunctions, we find that the annual, ENSO, and ENSO-A modes can be robustly captured from these smaller datasets, but the quality of the WPMM, IPO, TBO and TBO-A modes is significantly impacted. In particular, just two out of eight sets of eigenfunctions included discernible WPMM-like modes, namely subperiods 1 and 3. SST reconstructions based on these eigenfunctions produced features resembling the WPMM recovered from the full dataset, but the eigenfunction time series were found to be significantly influenced by high-frequency variability, possibly due to mixing with ENSO. These findings indicate that the possibility of recovering clean decadal modes analogous to the WPMM from observational data through the current formulation of NLSA is limited, and we will verify that this is the case in section~\ref{secObsModes}. Thus, long climate simulations, or perhaps paleoclimatic data, remain indispensable for recovering multidecadal modes of variability via this method.

\subsection{\label{secNoise}Noise robustness}

As shown in section 4, the NLSA Indo-Pacific SST modes extracted from millenial CCSM4 data exhibit a remarkably clear temporal evolution for both the variance-dominating modes such as ENSO and the ENSO-A modes, as well as for the low-amplitude WPMM and TBO. Besides the timespan of the data, another potential barrier in the detection of these modes in real-world data is measurement noise, which is inevitably part of any observational dataset. Intuitively, one would expect the noise of given strength to have more pronounced impact on retrieval of lower-amplitude modes such as the WPMM and TBO, as opposed to variance-dominating modes such as ENSO. 

Here, in order to assess the robustness of our extracted modes against observational noise, we examine the results of NLSA applied to millennial Indo-Pacific SST data corrupted by i.i.d.\ Gaussian noise. It can be shown \citep[][]{Giannakis16} that the presence of i.i.d.\ noise (not necessarily Gaussian) leads to a random bias in the pairwise distances $ \lVert X_i - X_j \rVert $ between the data in lagged embedding space, but, by the law of large numbers, as the number of lags $ q $ grows the bias becomes a constant with high probability. This statistically constant bias enters in the kernel matrix $ \pmb{ K }$, but is canceled in the normalization procedure in~\eqref{eqDM} for constructing the Markov matrix $ \pmb{ P } $ used for computing eigenfunctions. Thus, time-delay embedding naturally endows NLSA with robustness against i.i.d.\ measurement noise, and the larger $ q $ is the larger the noise tolerance becomes. This result also holds for noises with spatially nonuniform statistics and spatial dependencies, but  otherwise i.i.d.\ in time. Of course, in practical applications there are limits on how large $ q $ can be; e.g., due to computational cost and the fact that $ q $ must be significantly smaller than the number of samples $ n $ to avoid spurious correlations. 

In separate calculations, we have computed eigenfunctions using data corrupted by i.i.d.\ Gaussian noise of standard deviation $ \sigma $ in the range 0.5--3 K using $ q = 240 $ lags as in the noise-free case. (Note that the presence of a time-independent, nonzero noise mean is irrelevant for NLSA since such a term cancels from all pairwise data differences $ X_i - X_j $.) We found that  the WPMM is still present in the spectrum for $ \sigma = 0.5 $ K, but the quality of its temporal pattern is substantially degraded. Increasing the noise to 1 K results in the WPMM being indistinguishable among the leading 200 eigenfunctions. On the other hand, for the same number of lags, ENSO and ENSO-A eigenfunctions can be detected for noise standard deviations up to 3 K and 2 K, respectively. In agreement with theoretical expectations, we also found out that the noise tolerance for these modes increased by increasing $ q $. 

\section{\label{secGFDL}Comparison with CM3 dataset}

In this section, we compare the Indo-Pacific SST modes extracted from the 1300-year  CCSM4 dataset (section~\ref{secNLSAModes}) against modes extracted from the 800-year CM3 dataset. We analyzed this dataset in the same fashion as CCSM4, and the results were tested using a wide range of NLSA parameters. As different models have different physics representations and initial and boundary conditions, we do not obtain a perfect agreement between the two analyses. Qualitatively, however, the periodic, ENSO, and ENSO-A modes recovered from CCSM4 can also be identified in the CM3 control run. Modes resembling the WPMM, IPO, and TBO are also recovered from CM3, but their temporal patterns are noisier than their CCSM4 analogs. Figures~\ref{figGFDLPeriodic}--\ref{figGFDLBiennDec} show the temporal patterns, frequency spectra, and spatial composites for the CM3 modes corresponding to the CCSM4 modes in Figs.~\ref{figPeriodic}, \ref{figInterannual}, \ref{figBienn}, and~\ref{figDec}. Below, we outline the main similarities and differences between the two sets of modes.

The annual, semiannual, and triennial cycles from CM3 (Fig.~\ref{figGFDLPeriodic}) have qualitatively similar spatial features to their CCSM4 counterparts (Fig.~\ref{figPeriodic}), but are generally of higher amplitude. The CM3 annual modes prevail over the  same regions as the corresponding CCSM4 modes; that is, they are predominantly active in the midlatitudes and off the shore of South America, but have a slight spatial shift towards lower latitudes. The CM3 semiannual modes also exhibit several spatial similarities with those from CCSM4; e.g., in the regions west of the Galapagos, the Maritime Continent, the western Arabian Sea, and over the Southern Ocean.  The  CM3 triennial modes display more pronounced discrepancies from CCSM4, particularly over the northern and eastern Indian Ocean and the Maritime Continent, where the SST anomalies are frequently of opposite sign. 

Similarly to the periodic modes, the interannual modes are generally stronger in CM3 than in CCSM4. As shown in Fig.~\ref{figGFDLInterannual}(c), the dominant interannual spatiotemporal pattern from CM3 features an ENSO-like triangular anomaly that compares well to the corresponding pattern from CCSM4 (Fig.~\ref{figInterannual}(c)). However, the dominant ENSO timescale range of 2--3 y for CM3 is significantly shorter than the 3--5 y ENSO timescale characteristic of CCSM4. Pronounced spatial differences between the two patterns are also present. In particular, the pattern of westward-propagating anomalies originating in the Galapagos Islands and gradually weakening in the eastern Pacific progresses significantly faster in CM3 than in CCSM4. In other regions, such as the sector west of Australia, the interannual anomalies in CM3 are significantly weaker than those in CCSM4. Despite these differences, the dominant interannual modes from CM3 exhibit multidecadal fluctuations which are qualitatively similar to the behavior observed in CCSM4.

ENSO combination modes  also have a pronounced signature in the CM3 dataset (Figs.~\ref{figGFDLInterannual}(f) and~\ref{figGFDLInterannual}(i), respectively) that similarly to other modes is stronger than CCSM4. The first family of ENSO combination modes (analogous to the ENSO-A1 modes in section~\ref{secNLSAModes}\ref{secInterannual}) grows west of the Galapagos and subsequently weakens rapidly during its westward propagation. Also, the opposite-signed anomaly accompanying the Galapagos pattern develops east of Borneo and New Guinea, as opposed to the South China Sea as observed in CCSM4. The anomaly west of Australia is weaker, whereas the one along the west coastline of Sumatra and Java is more pronounced in CM3 than in CCSM4. The properties of the second set of combination modes (analogous to ENSO-A2 in section~\ref{secNLSAModes}\ref{secInterannual}) differ between two models in a similar fashion as the first annual modulation. Despite these differences, both the first and second annual modulations of the interannual mode are very pronounced west of the Sunda Strait in the CM3 dataset, and can be associated with the IOD. 

Biennial modes representing the TBO are also captured in the CM3 dataset (Figs.~\ref{figGFDLBiennDec}(a--c)), albeit with temporal patterns influenced  by annual modulation. In both models, the biennial modes generate anomalies over the eastern Pacific and west of the Sunda Strait, and these anomalies have similar amplitude but different regional extent. Also, the westward propagation associated with this mode is much less pronounced over the tropical Pacific in CM3, and the emergence of anomalies west of the Sunda Strait is less coincident to the dissipation of the equatorial wave. We did not recover TBO combination modes from CM3 analogous to the TBO-A modes in Fig.~\ref{figBienn}(d--f). 

A WPMM-like mode is also recovered in the CM3 dataset (Fig.~\ref{figGFDLBiennDec}(d,e)), but in this case the mode's temporal pattern is somewhat mixed with an annual signal. The corresponding spatial pattern (Fig.~\ref{figGFDLBiennDec}(f)) exhibits strong anomalies in the western Pacific, but these anomalies are not as spatially pronounced as in CCSM4 (Fig.~\ref{figDec}(f)). Spatial differences also occur elsewhere and in particular over the southern midlatitudes. The second CM3 decadal mode has a cleaner temporal pattern (Figs.~\ref{figGFDLBiennDec}(g, h)) in comparison to its CCSM4 counterpart. Its corresponding spatial composite (Fig.~\ref{figGFDLBiennDec}(i)) resembles the IPO, featuring an ENSO-like signal over the tropical eastern Pacific which is surrounded by same-sign but weaker anomalies prevailing over some parts of the Indian and Pacific Oceans.  

\section{\label{secObsModes}Spatiotemporal modes recovered from observational data}

In this section, we present the Indo-Pacific SST mode family recovered by NLSA from HadISST data spanning. As stated in section~\ref{secDataset}\ref{secObsData} we do not detrend the data. Moreover, we work with the same bandwidth and cone kernel parameters for NLSA  as with the model data ($ \epsilon = 1 $ and $ \zeta = 0.99 $), but we decrease the embedding window to $ \Delta t = 4 $ y, as we found that for longer embedding windows (including the 20 year embedding window employed earlier) the top part of the spectrum becomes dominated by trend modes and ``combination modes'' between trend and periodic modes. Since we are not confident about the behavior of NLSA for such long embedding windows compared to the total time series length and simultaneous presence of trends, in what follows we present results obtained using a 4 year embedding window.    

In what follows, we focus on the 140-year dataset spanning the industrial era (1870--2013). With the choice of parameters described above, we identify a 16-mode family, consisting of eigenfunctions 1, \ldots, 6, 11, 14, \ldots, 19, 25, 26, 29 (ordered as described in section~\ref{secNLSA}\ref{secEigenfunctions}), which we label $ \phi_1, \ldots, \phi_{16} $ as in sections~\ref{secNLSAModes} and~\ref{secGFDL}. This family consists of periodic modes representing the annual cycle and its first two harmonics ($\phi_1, \ldots, \phi_6 $), the fundamental component of ENSO together with the associated ENSO-A1 and ENSO-A2 combination modes ($  \phi_7, \ldots, \phi_{12} $), an IPO mode ($ \phi_{13}$), a pair of TBO modes  ($\phi_{14}, \phi_{15}$), and a WPMM mode ($\phi_{16}$). Besides these modes, the NLSA spectrum from HadISST data contains trend-like modes as well as ``combination modes'' representing modulations of the internal variability modes by the trend modes. We defer study of these modes to future work. Also, taking into account the sensitivity analysis in section~\ref{secSensitivity}\ref{secTimespan}, we express caution about the robustness of the IPO and WPMM recovered from the HadISST data, though we find that the properties of these modes are broadly consistent with those recovered from the model data. Similarly, our confidence on the TBO modes from HadISST is somewhat diminished since we will see that its frequency peaks are not well-resolved, but nevertheless that mode retains some of the qualitative features observed in its counterpart from the model data. In separate calculations, we have computed NLSA modes from the portion of the HadISST dataset spanning the satellite era, and found that the periodic, ENSO, ENSO-A modes can also be recovered from that shorter dataset, although the WPPM, IPO, and TBO are not successfully recovered.   

\subsection{Periodic modes}
  
Figure~\ref{figObsPeriodic} shows power spectra, time series, and SST composites associated with the annual, $ \{ \phi_1, \phi_2 \} $, semiannual,  $ \{ \phi_3, \phi_4\} $, and triennial, $ \{ \phi_5, \phi_6\}$, modes recovered from the industrial-era HadISST data. Comparing that figure to Figs.~\ref{figPeriodic} and~\ref{figGFDLPeriodic}, it is evident that the periodic modes from HadISST are in good agreement with their CCSM4 and CM3 counterparts. In particular, SST patterns from the annual modes are mainly active in the southern midlatitudes and exhibit weak zonal variations, whereas the patterns from the semiannual modes exhibit strong activity in the Indian Ocean as expected from Asian monsoon variability. The spatial patterns of the triennial modes are somewhat noisier in HadISST than in the models, but display the same centers of activity in the Indian Ocean  off the African and Australian coasts at $ \sim 10^\circ$S latitudes and in the eastern Pacific cold tongue region.  

\subsection{ENSO and ENSO combination modes}

Next, we examine the collection of interannual modes $ \phi_7, \ldots, \phi_{12} $, whose power spectra, time series, and SST composites are shown in Fig.~\ref{figObsInterannual}.  The dynamic evolution of reconstructed SST, surface wind, and thermocline depth patterns associated with these modes will be discussed in more detail in Part~II (along with supplementary movies). Among these modes, the pair  $ \{ \phi_5, \phi_6 \} $ (Fig.~\ref{figObsInterannual}(a--c)) represents the fundamental component of ENSO. The dominant frequency of this pair, $ \nu_\text{ENSO} = 0.25$ $y^{-1}$, compares well with the corresponding frequency peak identified in the corresponding mode from CCSM4 (Fig.~\ref{figInterannual}(a)), but the spectra of the HadISST modes (Fig.~\ref{figObsInterannual}(a)) are significantly broader. This discrepancy is at least partly caused by  the different embedding windows used in each case, for, as shown in Fig.~\ref{figSpec}, the spectra of the HadISST ENSO modes compare well with the spectra of the CCSM4 ENSO modes computed for a 4-year embedding window. A comparison of the SST composites in Fig.~\ref{figObsInterannual}(c) and Figs.~\ref{figInterannual}(c) and~\ref{figGFDLInterannual}(c) reveals that the SST patterns from HadISST and CCSM4/CM3 are in good agreement over the tropical Pacific, but notable differences can be observed in the Indian Ocean and the Maritime Continent north of Borneo. In particular, instead of the dipolar Indian Ocean SST anomaly patterns observed in CCSM4 and CM3, the HadISST composites exhibit SST anomalies which have the same sign as the main eastern Pacific ENSO anomaly,  and extend over the majority of the Indian Ocean and the Maritime Continent region to the North of Borneo. In Part~II, we will see that the reconstructed SST fields from the HadISST ENSO modes reproduce well the historical El Ni\~no/La Ni\~na events, as is also suggestive by the eigenfunction time series in Fig.~\ref{figObsInterannual}(b) (e.g., notice the strong activity in the second half of the 1990s associated with the 1997--1998 El Ni\~no and the 1998--2000 La Ni\~nas). 

Modes $ \{ \phi_9, \phi_{10} \} $ (Fig.~\ref{figObsInterannual}(d--f)) and $ \{ \phi_{11}, \phi_{12} \} $ (Fig.~\ref{figObsInterannual}(g--i)) have spectral peaks at the ENSO combination frequencies $ \nu_\text{ENSO} - \text{1 y$^{-1}$} $ and $ \nu_\text{ENSO} + \text{1 y}^{-1} $, respectively, and their amplitude envelopes correlate significantly with the envelopes of the fundamental ENSO modes (correlation coefficients 0.96 and 0.78, respectively, at $ p \approx 0 $ using a $ t $-test with $ (\text{140 y}) \times \nu_\text{ENSO} - 2 = 33 $ degrees of freedom).  Moreover, the amplitudes and phases of these modes have analogous relationships with the fundamental ENSO modes $ \{ \phi_7, \phi_8 \} $ as those shown in Fig.~\ref{figPhase} for the corresponding CCSM4 modes. Based on these facts, we conclude that the pairs $ \{ \phi_9, \phi_{10} \} $ and $ \{ \phi_{11}, \phi_{12} \} $ are combination modes between ENSO and the annual cycle; we denote these pairs ENSO-A1 and ENSO-A2 following the notation of sections~\ref{secNLSAModes}\ref{secInterannual} and~\ref{secGFDL}. The SST composites from these modes (Fig.~\ref{figObsInterannual}(f, i)) exhibit significant activity over the Indian Ocean  (and in particular the region west of the Sunda Strait) as in the case of the ENSO-A modes from the model data (Figs.~\ref{figInterannual} and~\ref{figGFDLInterannual}) do. However, the ENSO-A composites from HadISST differ in their prominent activity in the central Pacific (at $\sim 120^\circ$W longitudes). In particular, the ENSO-A1 composite (Fig.~\ref{figObsInterannual}(i)) features two prominent anomaly clusters of opposite sign which are not seen in the model-based composites (Figs.~\ref{figInterannual}(i) and~\ref{figGFDLInterannual}(i)). Despite these differences, the ENSO and ENSO-A combination modes from HadISST explain comparable variance in the Indian Ocean as their counterparts from CCSM4 and CM3. Indeed, as shown in the variance maps in Fig. 15, the ENSO-A modes families  from HadISST each explain $ \sim 10\%$ of the nonperiodic SST variance in the eastern equatorial Indian Ocean region off the Sunda Strait, which is comparable to the corresponding variances explained by the CCSM4 modes (see section~\ref{secNLSAModes}\ref{secInterannual}).      

\subsection{Biennial modes}

HadISST mode pair $ \{ \phi_{14}, \phi_{15} \} $, depicted in Fig.~\ref{figObsDecadal}(a), displays a cluster of SST anomalies in the Maritime Continent and off Australia's north coast, together with a cluster of opposite-sign anomalies in the eastern equatorial Pacific off the Galapagos Islands (Fig.~\ref{figObsInterannual}(l)). Moreover, the frequency spectra of these modes (Fig.~\ref{figObsInterannual}(j)) show evidence of spectral peaks at the TBO frequency $ \nu_\text{TBO} = 0.5 $ y$^{-1} $ and the corresponding combination frequency $ \nu_\text{TBO} + \text{1 y$^{-1}$} = 1.5$ y$^{-1} $ (though these peaks are of marginal statistical significance). Due to these spatial and spectral characteristics, we interpret modes $ \{ \phi_{14}, \phi_{15} \} $ as representations of the TBO that mix the fundamental TBO frequency and the TBO combination frequency with the annual cycle into a single pair of modes (cf.\ the TBO modes from CCSM4 presented in section~\ref{secNLSAModes}\ref{secBiennial}, which resolve these frequencies into distinct mode pairs). The shortness of the embedding window used to compute modes from the observational data may be contributing to the observed mixing of the TBO frequencies. 

\subsection{Decadal modes}

Figure~\ref{figObsDecadal}(d--i) displays the power spectra, time series, and SST composites associated with modes $ \phi_{16} $ and $ \phi_{13}$; the modes  recovered from the HadISST data that most closely resemble the IPO and WPMM modes, respectively. In  particular, Fig.~\ref{figObsDecadal}(e) shows that $ \phi_{16} $ describes a multidecadal oscillation (albeit with noticeable mixing with high-frequency variability), essentially completing a single cycle over the 140-year long HadISST dataset (the time average of the  $ \phi_{16} $ time series is close to zero). The corresponding SST composites (Fig.~\ref{figObsDecadal}(f)) feature a cluster of negative SST anomalies in the central tropical Pacific which broadly resembles the western Pacific anomaly cluster in the WPMM recovered from CCSM4 but shifted to the East. The $ \phi_{16} $  composite in Fig.~\ref{figObsDecadal}(f) also features a central-eastern Pacific dipole as seen in the WPMM composite from CCSM4, but instead of the positive SST anomalies developing off the southern and eastern coasts of Australia in CCSM4, the HadISST mode features a positive anomaly cluster in the south central Pacific. As stated earlier, due to the shortness of the available HadISST data we do not have particularly high confidence in that mode, though in separate calculations we have confirmed that it appears in the NLSA spectrum for different embedding window lengths in the range 4--20 y. Notably, time series of $ \phi_{16} $ correlates strongly with low-frequency ENSO envelope. Turning now to $ \phi_{13}$,  this mode is characterized by a red-noise-like power spectrum (Fig.16(g)) and an SST composite (Fig.~\ref{figObsDecadal}(i)) characteristic of the IPO. As with the IPO modes from CCSM4 and CM3, the time series of $ \phi_{13} $ is noticeably influenced by interannual variability. Nevertheless it exhibits a discernible decadal signal that broadly resembles IPO variability.

\section{\label{secSummary}Concluding remarks}

In this paper, we have studied the variability of Indo-Pacific SST in millenial control integrations of the CCSM4 and CM3 coupled climate models and HadISST observational data for the industrial and satellite eras using a recent data analysis technique called nonlinear Laplacian spectral analysis \citep[GM,][]{Giannakis15}. An advantage of NLSA over traditional eigendecomposition techniques such as EOF analysis and SSA is that it requires no prefiltering of the input data to isolate the timescales of interest. Instead, NLSA extracts a multiscale hierarchy of modes through the eigenfunctions of a single kernel operator designed  to extract intrinsic timescales of data generated by complex dynamical systems. Applied to Indo-Pacific SST data, NLSA recovers a hierarchy of physically meaningful modes spanning seasonal to multidecadal timescales. Here, we have focused on the spatiotemporal properties of a subset of this family that includes the annual cycle and its harmonics, ENSO, combination modes between ENSO and the annual cycle (denoted here as ENSO A-modes), representations of the TBO (together with its associated combination modes) and IPO, and a new low-frequency pattern which we refer to as west Pacific multidecadal mode (WPMM). This family was best recovered in the 1300-y long CCSM4 dataset, where it was also found to be robust under changes of NLSA parameters and spatial domain and addition of i.d.\ Gaussian noise. The periodic, ENSO, and ENSO-A modes were also robustly recovered in CM3 and HadISST data (the latter, both for the industrial and satellite eras), but the TBO, WPMM, and IPO modes were found to be of poorer quality (in the sense of timescale mixing) in CM3 and industrial-era HadISST, and were not detectable in satellite-era HadISST data. A similar degradation of quality took place in separate analyses of $\sim 150$-year portions of the CCSM4 data, suggesting that the poor quality of the TBO, WPMM, IPO modes in HadISST data may be due to the timespan of that dataset as opposed to absence of these modes in nature.   

We showed that the ENSO and ENSO-A modes are naturally recovered though NLSA from raw monthly Indo-Pacific SST data from both models and observations. In particular, the recovered temporal patterns have to a high accuracy the frequencies, phase relationships, and multiplicities expected theoretically from quadratic nonlinearities in the governing equations of the coupled atmosphere-ocean system \citep[][]{McGregorEtAl12,SteinEtAl11,SteinEtAl14,StueckerEtAl13,StueckerEtAl15,RenEtAl16}.  Moreover, we showed that the ENSO-A modes explain a significant portion of SST variability over the area traditionally used to define IOD indices  \citep[][]{SajiEtAl99}. Specifically, these modes explain $\sim 35\% $ of the raw variance in these  regions after removal of the seasonal cycle. These results suggest that a significant portion of IOD variability (and potentially predictability) can be associated with deterministic modulations of the annual cycle by ENSO. 

Similarly to ENSO, we found that the phase locking of the TBO to the seasonal cycle can be represented through a distinct family of oscillatory patterns and their associated combination modes (denoted TBO-A modes). However, possibly due to their low amplitude in comparison to ENSO, a clean identification of these patterns was only possible in the 1300-long CCSM4 data. In particular, even though we did recover biennial modes in CM3 and in industrial-era HadISST data, these modes mixed together the TBO and TBO-A frequencies. Thus, our results from CCSM4 are consistent with studies attributing a distinct dynamical origin to the TBO \citep[e.g.,][]{MeehlArblaster02,LiEtAl06,MeehlArblaster11}, but the challenges in detecting these modes in the other datasets studied here are also representative of difficulties in TBO detection from model and observational data faced by other studies \citep[][]{StueckerEtAl15b}.     

Another result stemming from this work has been the identification of the WPMM as a distinct mode of decadal Indo-Pacific SST variability. The dominant feature of this mode is a persistent cluster of SST anomalies in the western Pacific. This  pattern, which was most clearly observed in the CCSM4 data and for a wide range of NLSA parameters, is distinct from the eastern Pacific decadal SST anomalies typically associated with the IPO \citep[][]{PowerEtAl99,MeehlEtAl13}. The WPMM also features prominent activity in the southern Pacific and Indian oceans, though the details of the corresponding spatial patterns were found to differ between the CCSM4, CM3 and HadISST datasets. 

In summary, the family of NLSA modes presented here provides an attractive low-dimensional basis to characterize numerous features of the Indo-Pacific climate system. In Part~II, we will study the SST, surface wind, thermocline depth, and precipitation patterns represented by the ENSO and TBO modes and the associated coupling between the Pacific and Indian Oceans, as well as the relationship between ENSO and TBO activity and the WPMM in CCSM4 data. We will also demonstrate the impacts of the WPMM on decadal precipitation over Australia in CCSM4.  

\acknowledgments
J.\ Slawinska received support from the Center for Prototype Climate Modeling at NYU Abu Dhabi and NSF grant AGS-1430051.  The research of D.\ Giannakis is supported by ONR Grant N00014-14-0150, ONR MURI Grant 25-74200-F7112, and NSF Grant DMS-1521775. We thank Sulian Thual for stimulating conversations and three anonymous reviewers for comments that led to significant improvement of the paper. This research was partially carried out on the high performance computing resources at New York University Abu Dhabi. 

\appendix

\section*{Comparison with SSA}

For completeness, we have compared the NLSA spatiotemporal patterns described in section~\ref{secNLSAModes} with the corresponding patterns recovered through SSA \citep[][]{GhilEtAl02}  using the same Indo-Pacific SST dataset from CCSM4 and $ \Delta t = 20 $ y embedding window as in section~\ref{secNLSAModes}. Representative temporal patterns from SSA and the associated phase composites are shown in Fig.~\ref{figSSAa} and Fig.~\ref{figSSAb}, where the modes are ordered in order of decreasing singular value. The main commonalities and differences between the two approaches are as follows.
 
\begin{figure}[t]
\noindent\includegraphics[width=\linewidth]{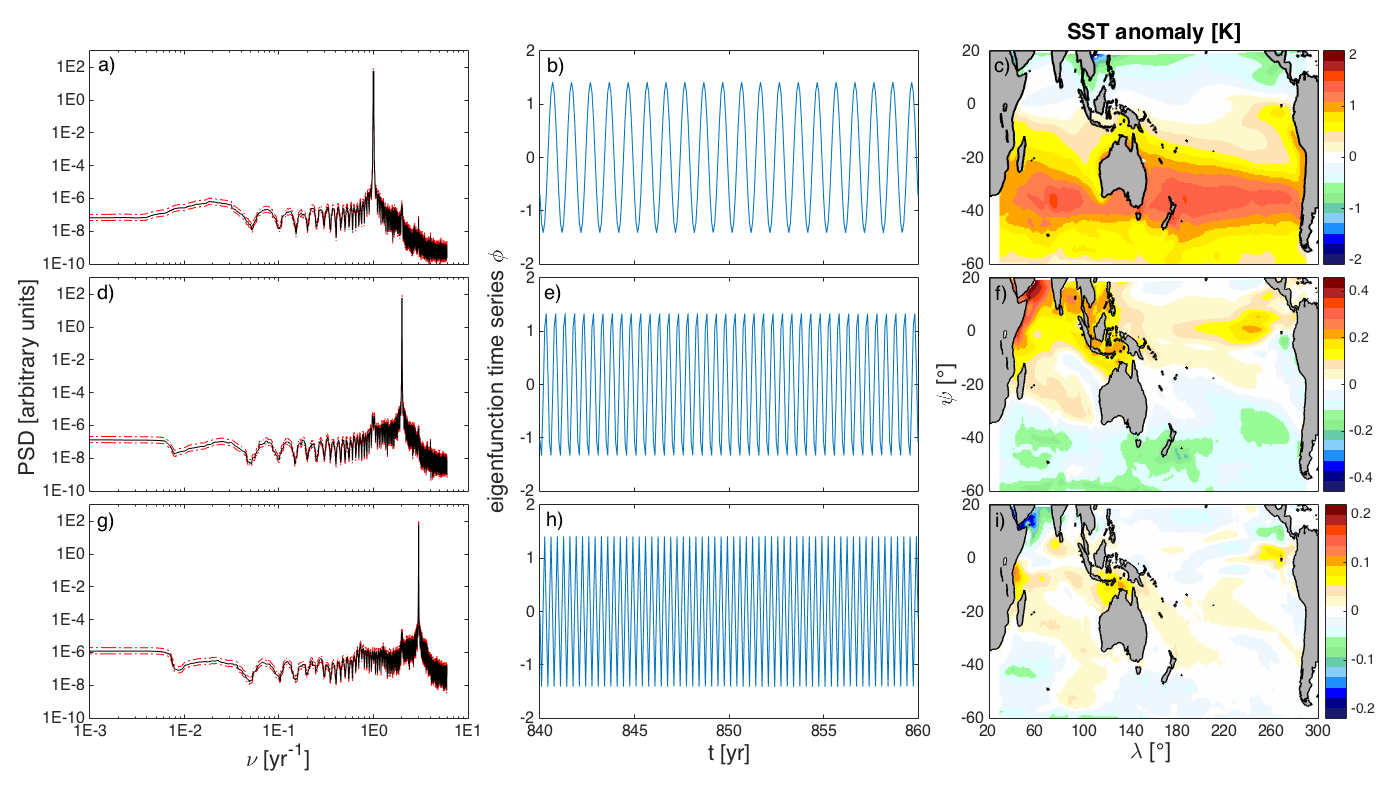}
\caption{Power spectral densities (a, d, g), 20-year portions of the eigenfunction time series  (b, e, h), and spatial composites (c, f, i) for NLSA modes $\phi_{1} $ (a--c), $ \phi_3 $ (d--f), and $ \phi_5 $  (g--i) extracted from the CCSM4 control integration. Each composite corresponds to data reconstructed for the corresponding eigenfunction as described in section~\ref{secNLSA}\ref{secReconstruction}, and subsequently averaged over the times for which the eigenfunction value $ \phi_i( t ) $ exceeds the standard deviation $ \stdev \phi_i $ of the eigenfunction time series. The modes shown here are the first modes in the families $\{ \phi_1, \phi_2 \} $, $  \{ \phi_3, \phi_4 \} $, and $ \{ \phi_5, \phi_6 \} $ representing the annual cycle and its first two harmonics, respectively. The second mode in each family is in quadrature with the first, and therefore has analogous spatial and temporal patterns. The power spectral densities were estimated via the multitaper method \citep[][]{GhilEtAl02,Thomson82} with time-bandwidth product $ p = 6 $ and $ K = 2p - 1 = 11 $ Slepian tapers. The effective half-bandwidth resolution for the $ N = \text{15,357} $ samples in the eigenfunction time series is $ \Delta \nu = p / ( N \delta t ) \approx 1/213 $ y$^{-1}$, where $ \delta t = 1 $ m is the sampling interval. Shown in red dashed lines are 95\% confidence intervals for the power spectrum computed using an $ F $-test with $ ( 2, 2 K - 2) $ degrees of freedom.}\label{figPeriodic}
\end{figure}

As in NLSA, the leading pairs of SSA modes are doubly-degenerate periodic modes associated with the annual cycle and its harmonics. The SSA spectrum contains two such pairs of annual and semiannual modes (not shown), whose spatial patterns are in good agreement with the corresponding NLSA patterns in Fig.~\ref{figPeriodic}. We note the SSA also recovers a pair of triennial modes (also not shown here) analogous to the NLSA modes $ \{ \phi_5, \phi_6 \} $, but the SSA triennial pair appears as the 22nd and 23rd modes in the spectrum. This is an example of the differences in mode ordering that can occur between NLSA (which orders the modes in order of increasing roughness on the data manifold) and SSA (which orders the modes in order of decreasing explained variance).

\begin{figure}[t]
\noindent\includegraphics[width=\linewidth]{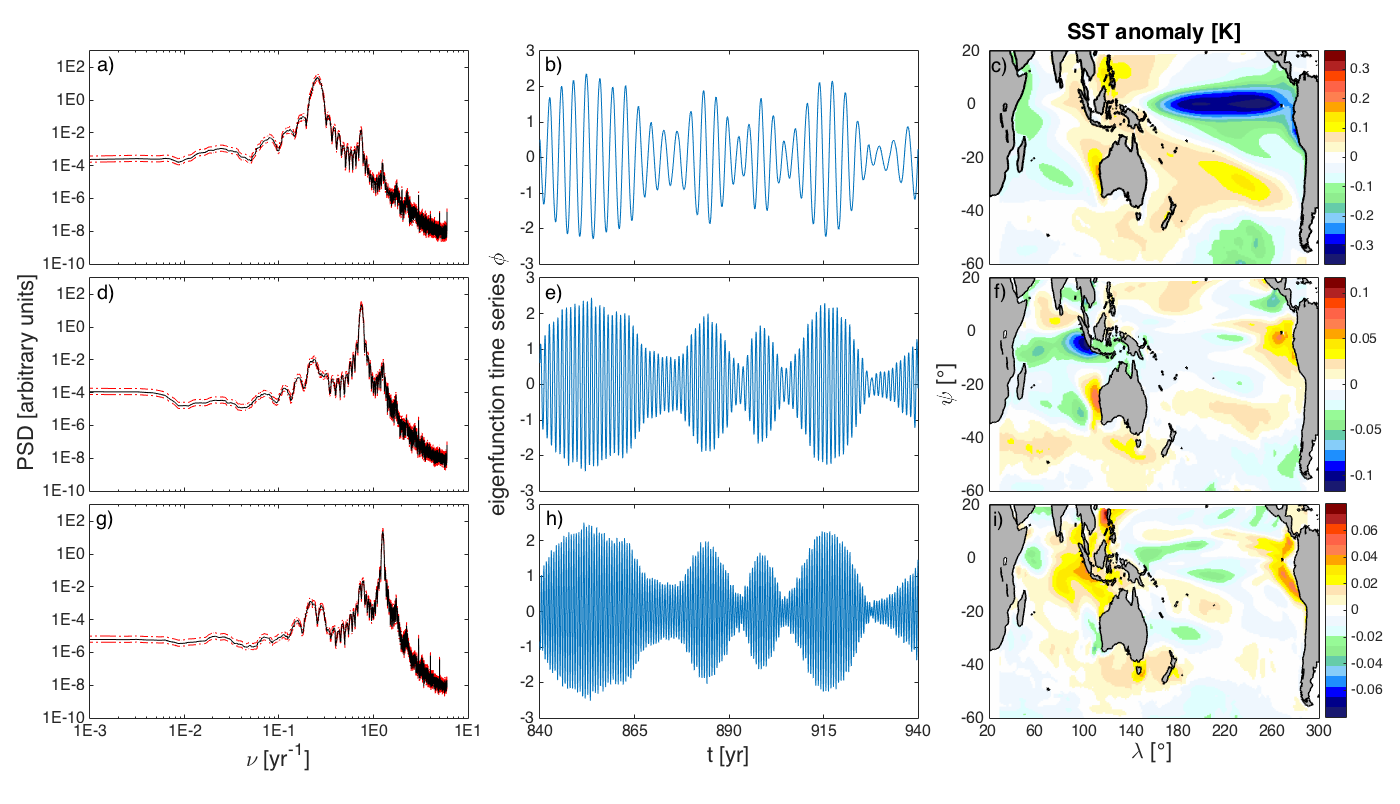}
\caption{Power spectral densities  (a, d, g), 100-year portions of the eigenfunction time series  (b, e, h), and spatial composites (c, f, i) for NLSA modes $\phi_7 $ (a--c), $ \phi_9 $ (d--f), and $ \phi_{11} $  (g--i) extracted from the CCSM4 control integration. The power spectral densities and composites were computed in the same manner as in Fig.~\ref{figPeriodic}. The modes shown here are the first modes in the families $\{ \phi_7, \phi_8 \} $, $ \{ \phi_9, \phi_{10} \} $, and $ \{ \phi_{11},  \phi_{12} \} $ representing the fundamental component of ENSO and the ENSO-A1 and ENSO-A2 combination modes, respectively. The second mode in each family is in quadrature with the first, and therefore has analogous spatial and temporal patterns.  Notice the strong activity exhibited by $ \phi_9 $ and $ \phi_{11} $ in the IOD region in panels (f, g).}\label{figInterannual}
\end{figure}

\begin{figure}[t]
\centering\includegraphics[width=\linewidth]{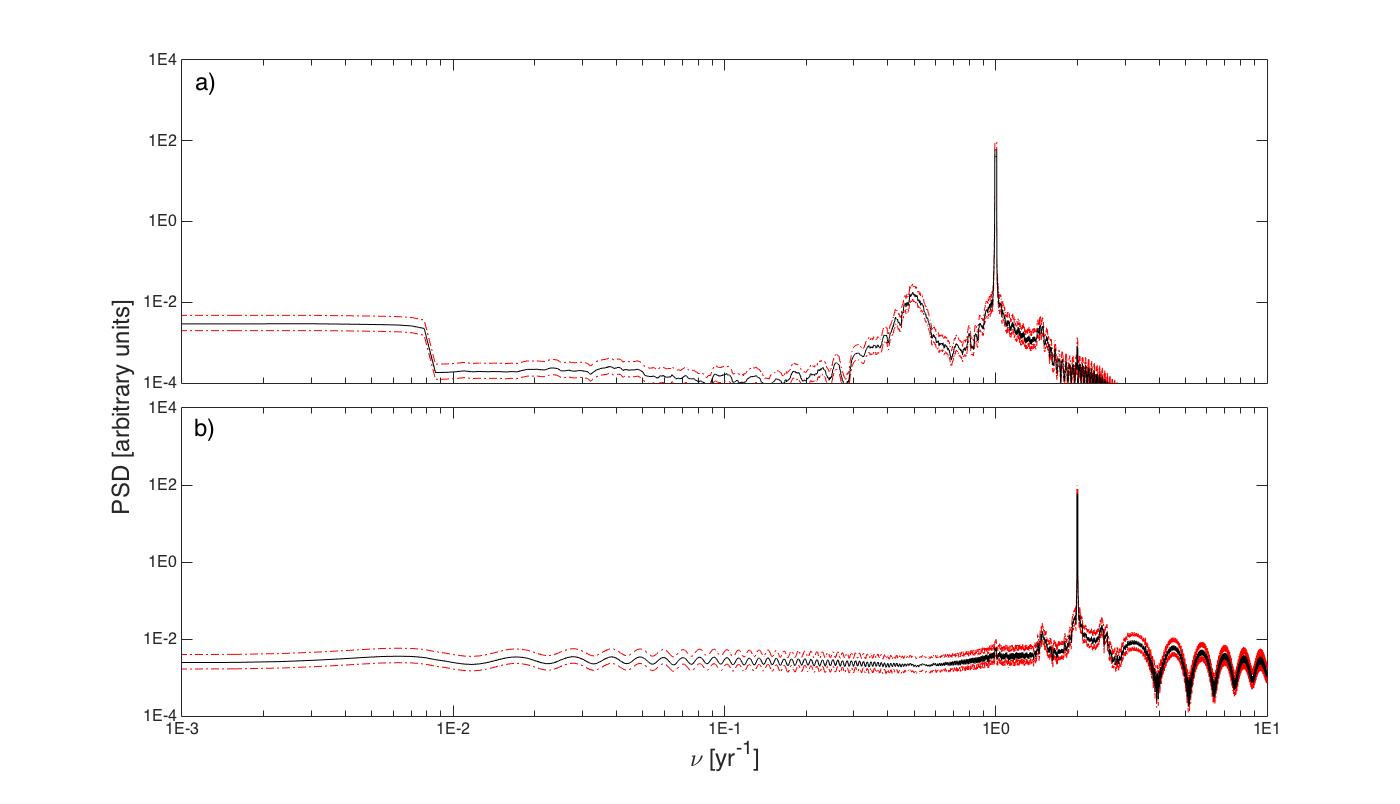}
\caption{Power spectral densities of the complex time series $ z_-/z $ and $ z_+/z_- $ ,  representing the amplitude and phase relationship between modes $\{\phi_9, \phi_{10}\}$ and $\{\phi_{7},\phi_{8}\}$, and  $\{\phi_{11}, \phi_{12}\}$ and $\{\phi_{9},\phi_{10}\}$, respectively. The power spectral densities were estimated in the same manner as in Fig.~\ref{figPeriodic}. The fact that $ z_{-}/z$ and $ z_{+}/z_{-} $ are near-perfect sinusoids with frequencies 1 y$^{-1} $ and 2 y$^{-1} $, respectively, shows that $ z_- $ and $ z_+$ are ENSO combination modes at the frequencies $ \nu_\text{ENSO} - \text{1 y$^{-1}$}$ and  $ \nu_\text{ENSO} + \text{1 y$^{-1}$} $, respectively.}\label{figPhase}
\end{figure}

The next pair of SSA modes (Fig.~\ref{figSSAa}(a--c)) represents the fundamental component of ENSO. As in the case of the NLSA ENSO modes in Fig.~\ref{figInterannual}(a--c), the SSA ENSO modes form a 90$^\circ$-out-of-phase oscillatory pair with an interannual dominant frequency of $\simeq 4$ y$^{-1}$ and the characteristic SST patterns of eastern Pacific El Ni\~no/La Ni\~na. Moreover, the SSA ENSO modes have a decadal envelope, which correlates almost perfectly with the corresponding envelope from NLSA (correlation coefficient 0.99; hereafter, unless otherwise stated  all amplitude envelopes are extracted via Hilbert transforms and the sample correlation coefficients have numerically zero $ p $-values based on $ t $-tests with $(\text{1300 y}) \times \nu_\text{ENSO} - 2 = 323 $ degrees of freedom).   

\begin{figure}[t]
\centering\includegraphics[width=\linewidth]{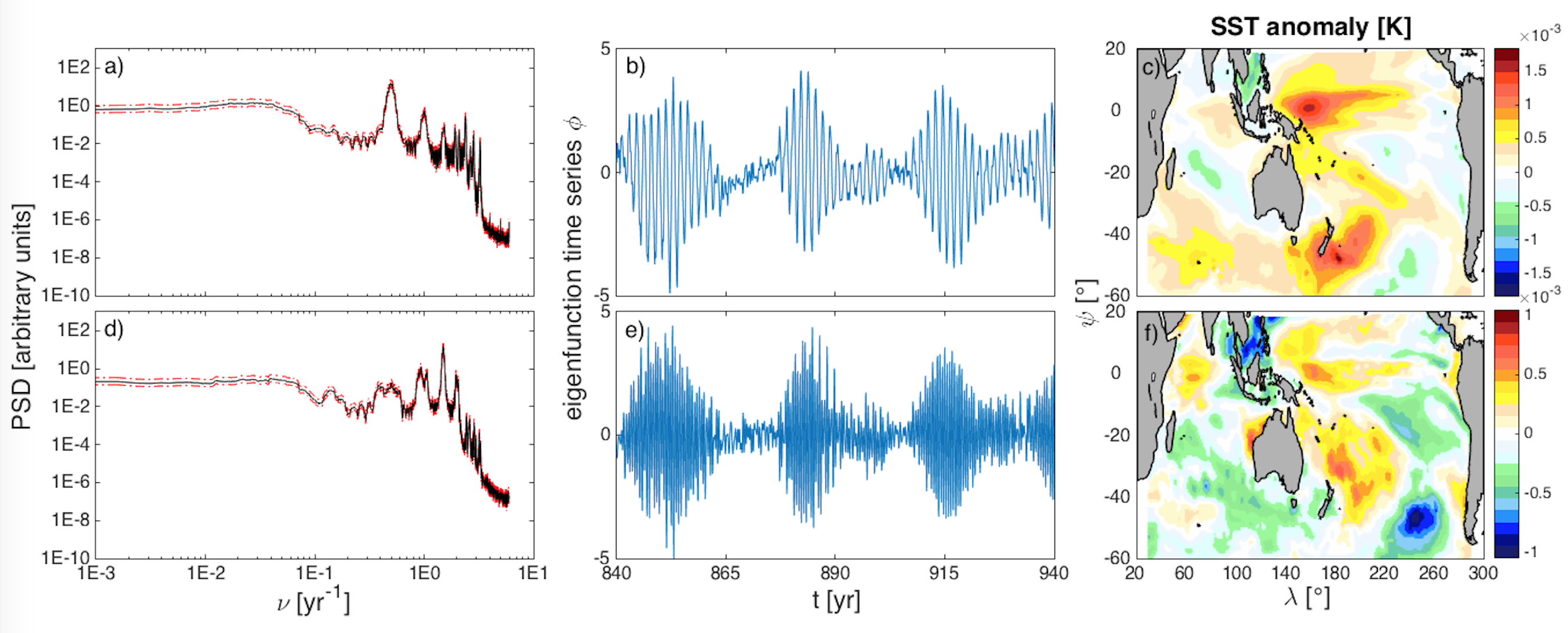}
  \caption{Power spectral densities (a, d), 100-year portions of the eigenfunction time series (b, e), and spatial composites (c, f) for NLSA modes $\phi_{15} $ (a--c), and $ \phi_{17} $ (d--f) extracted from the CCSM4 control integration. The power spectral densities and composites were computed in the same manner as in Fig.~\ref{figPeriodic}. The modes shown here are the first modes in the families $\{ \phi_{15}, \phi_{16} \} $ and $ \{ \phi_{17}, \phi_{18} \} $ representing the fundamental component of the TBO and the TBO-A combination modes, respectively. The second mode in each family is in quadrature with the first, and therefore has analogous spatial and temporal patterns.}
  \label{figBienn}
\end{figure}

Beyond the fundamental ENSO pair, the NLSA and SSA spectra differ significantly. In particular, the next several SSA modes have the structure of secondary ENSO modes as opposed to the ENSO-A combination modes recovered by NLSA. This is another example of the differences in mode ordering between SSA and NLSA. According to NLSA, the ENSO-A modes are represented by smoother functions on the data manifold than the secondary interannual modes, and therefore the former are leading the latter in the mode ordering, despite that the secondary interannual modes explain more variance than the ENSO-A modes. Two pairs of SSA modes that resemble the ENSO-A1 (Fig.~\ref{figInterannual}(d--f)) and ENSO-A2 (Fig.~\ref{figInterannual}(g--i)) modes are modes $ \{ 27, 28 \} $ (Fig.~\ref{figSSAa}(d--f)) and $ \{ 48, 49 \} $ (Fig.~\ref{figSSAa}(g--i)), respectively. These pairs exhibit the dominant frequencies which are consistent  with the theoretical ENSO-A frequencies. However, the envelopes of these modes do not correlate as strongly with the fundamental ENSO envelope as they do in NLSA. Specifically, the corresponding correlation coefficients are 0.75 and 0.94 respectively (recall that in case of NLSA, the corresponding values are greater than 0.99). In the spatial domain, SSA modes $ \{ 27, 28 \} $ and  $\{ 48, 49 \} $ are in reasonably good qualitative agreement with their NLSA counterparts in Fig.~\ref{figInterannual}(f); in particular, they exhibit strong activity in the eastern Indian Ocean region employed in IOD indices and southward-propagating anomalies off the west coast of Australia as discussed in section~\ref{secNLSAModes}\ref{secInterannual}. Besides the ENSO and ENSO-A modes, the SSA spectrum contains a pair of biennial modes, $ \{ 35, 36 \} $ (Fig.~\ref{figSSAa}(j--l)), which have significantly different spatiotemporal characteristics than the corresponding NLSA TBO modes (Fig.~\ref{figBienn}(a--c)). We did not find evidence of TBO-A combination modes in the SSA spectrum.

\begin{figure}[t]
\noindent\includegraphics[width=\linewidth]{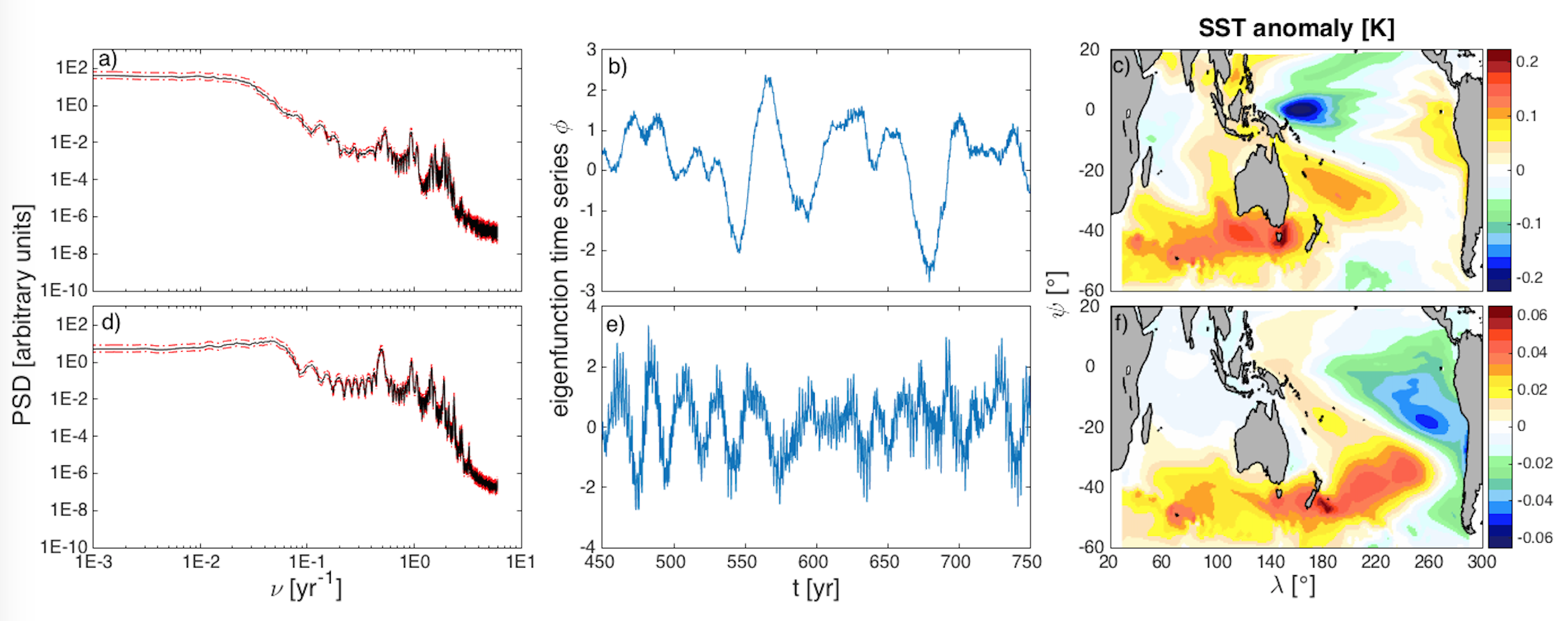}\\
\caption{Power spectral densities (a, d), 300-year portions of the eigenfunction time series  (b, e), and spatial composites (c, f) for NLSA modes $\phi_{13} $ (a--c) and $ \phi_{14} $ (d--f) extracted from the CCSM4 control integration. Modes $ \phi_{13} $ and $ \phi_{14} $ correspond to the WPMM and the IPO, respectively. The power spectral densities in (a, d) were computed in the same manner as those in Fig.~\ref{figPeriodic}. The composites in (f, i) were created by averaging the reconstructed data over periods for which $ \phi_i(t)>\max \phi_i - \stdev \phi_i $. For intermittent time series such as $ \phi_{13} $ and $ \phi_{14} $ with $ \max \phi_i \gg \stdev \phi_i $, the composites in (f, i) emphasize stronger events than composites created through the requirement $ \phi_i( t ) > \stdev \phi_i $.} \label{figDec}
\end{figure}

\begin{figure}[t]
\noindent\includegraphics[width=\linewidth]{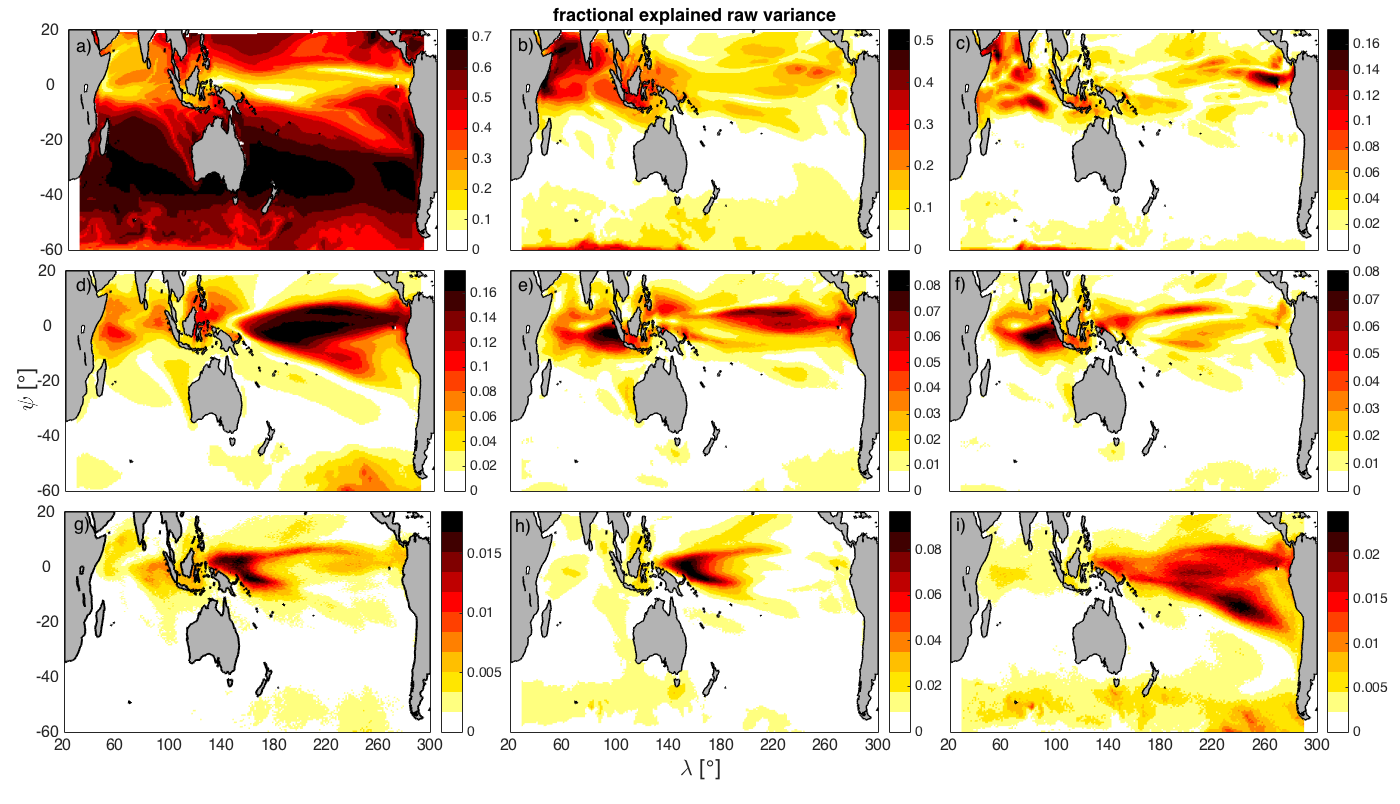}\\
\caption{Spatial map of time-averaged fraction of raw SST variance explained by NLSA modes $\{\phi_{1},\phi_{2}\}$ (a; annual modes), $\{\phi_{3},\phi_{4}\}$ (b; semiannual modes),  $\{\phi_{5},\phi_{6}\}$ (c; triennial modes),  $\{\phi_{7},\phi_{8}\}$ (d; fundamental ENSO modes),  $\{\phi_{9},\phi_{10}\}$ (e; ENSO-A1 modes),  $\{\phi_{11},\phi_{12}\}$ (f; ENSO-A2 modes), $\{\phi_{15},\phi_{16}\}$ (g; fundamental TBO modes),  $ \phi_{13}$ (h; WPMM), and $\phi_{14}$ (i; IPO), extracted from the CCSM4 dataset.}\label{figVarRaw}
\end{figure}

\begin{figure}[t]
\noindent\includegraphics[width=\linewidth]{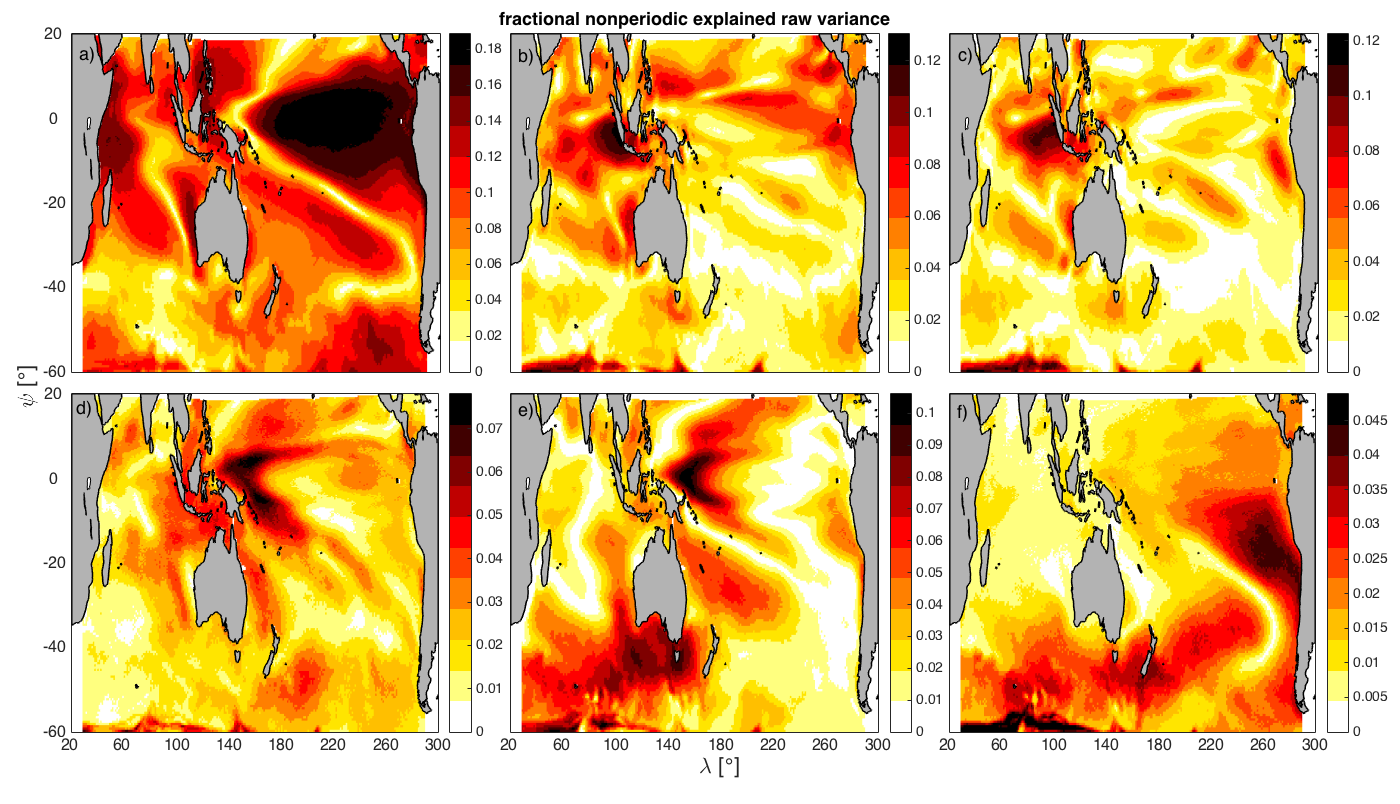}\\
\caption{Same as Fig.~\ref{figVarRaw}, but for the explained variance relative to the raw SST data after removal of the seasonal cycle.}\label{figVarRawWithoutPer}
\end{figure}

\begin{figure}[t]
\noindent\includegraphics[width=\linewidth]{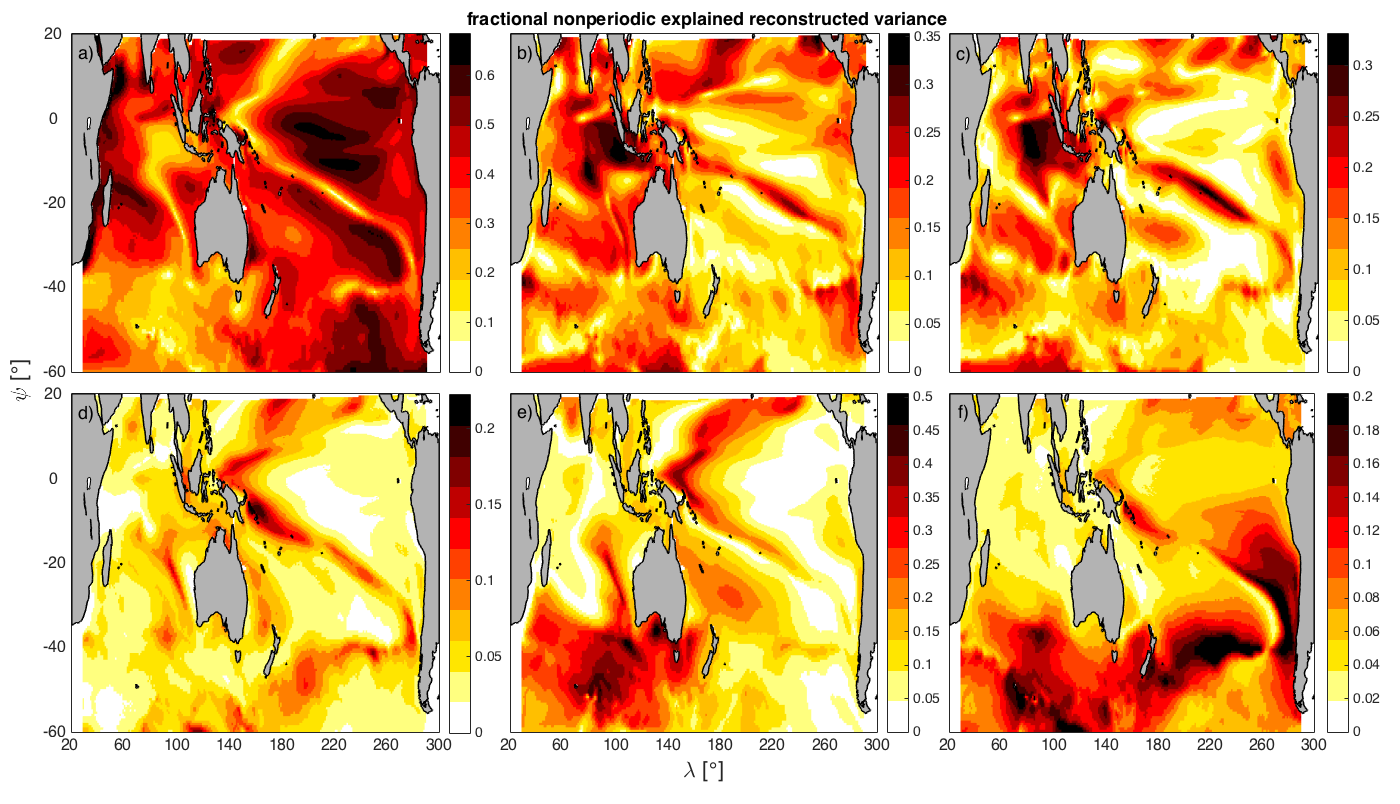}\\
\caption{Same as Fig.~\ref{figVarRaw}, but for the explained variance relative to the reconstructed nonperiodic SST data.}\label{figVarRecWithoutPer}
\end{figure}

\begin{figure}[t]
\noindent\includegraphics[width=\linewidth]{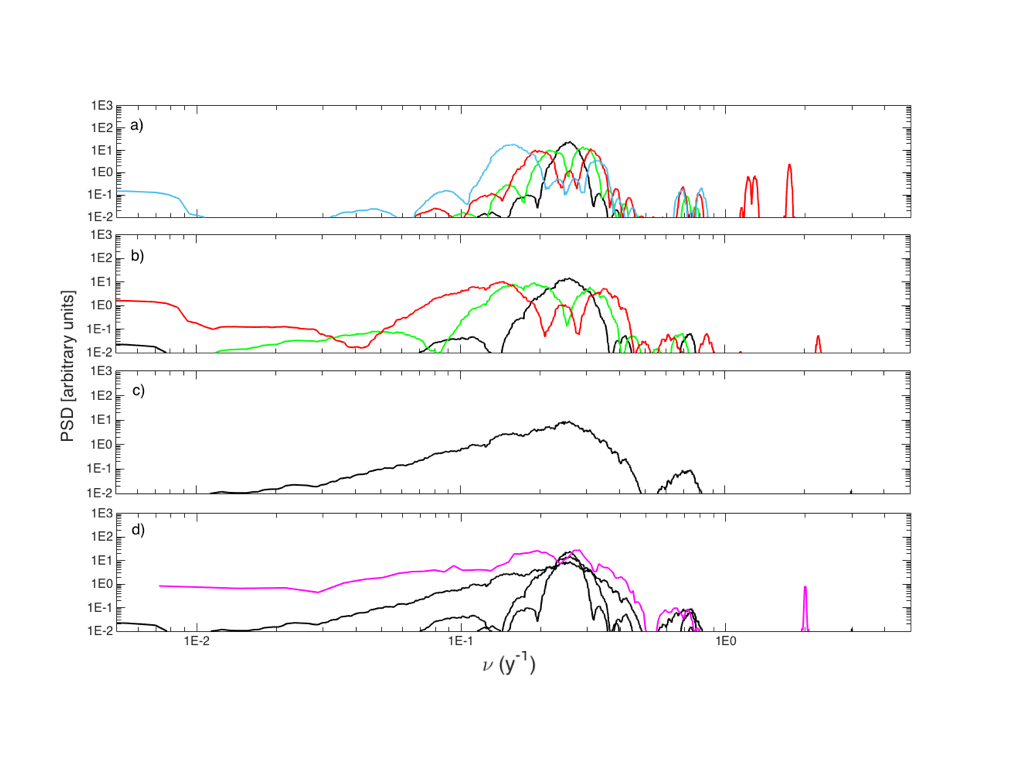}\\
  \caption{Power spectral densities for ENSO for different values of the embedding window length $ \Delta t $. (a--c) Fundamental (thick black lines) and secondary (thin colored lines) ENSO modes for (a) $ \Delta t = 20 $ y, (b) $ \Delta t = 10 $ y, and (c) $ \Delta t = 4 $ y obtained from the CCSM4 data. (d) Power spectral densities of the fundamental ENSO modes in (a--c) (black) and the fundamental ENSO modes from HadISST data in Fig.~\ref{figObsInterannual}(a) (red) shown together. The power spectral densities from the CCSM4 and HadISST data were estimated via the multitaper method as described in Figs.~\ref{figPeriodic} and~\ref{figObsPeriodic}, respectively.}
  \label{figSpec}
\end{figure}

\begin{figure}[t]
\noindent\includegraphics[width=\linewidth]{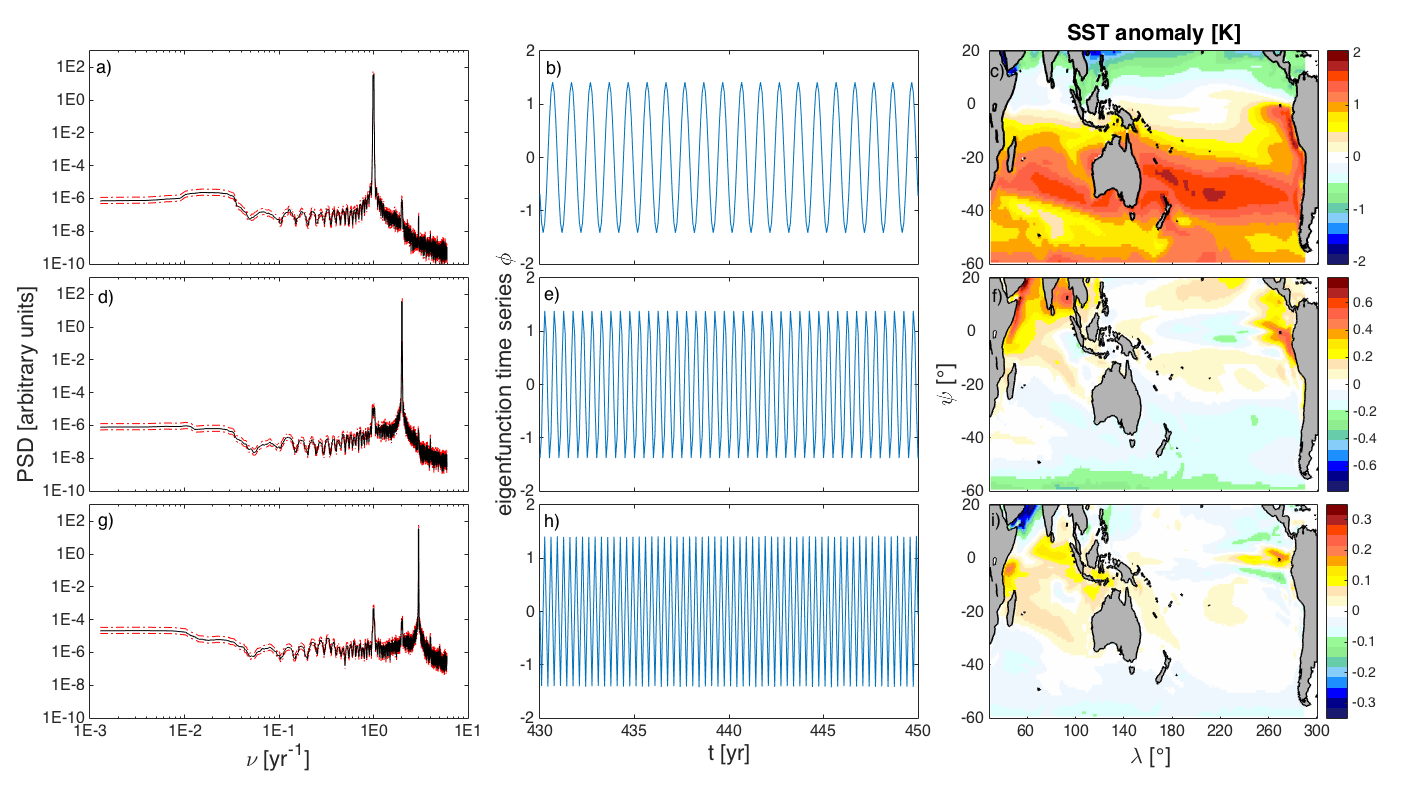}\\
\caption{Same as Fig.~\ref{figPeriodic}, but for the annual (a--c), semiannual (d--f), and triennial (g--i) modes of Indo-Pacific SST variability extracted with NLSA from 800 years of the CM3 control integration. In this case, the power spectral densities (computed via the multitaper method as in Fig.~\ref{figPeriodic}) have effective frequency resolution $ \Delta \nu = p / ( N \delta t ) \approx 1/130 $ y$^{-1}$ corresponding to the $ N = 9357 $ samples in the eigenfunction time series from CM3.}\label{figGFDLPeriodic}
\end{figure}

\begin{figure}[t]
\noindent\includegraphics[width=\linewidth]{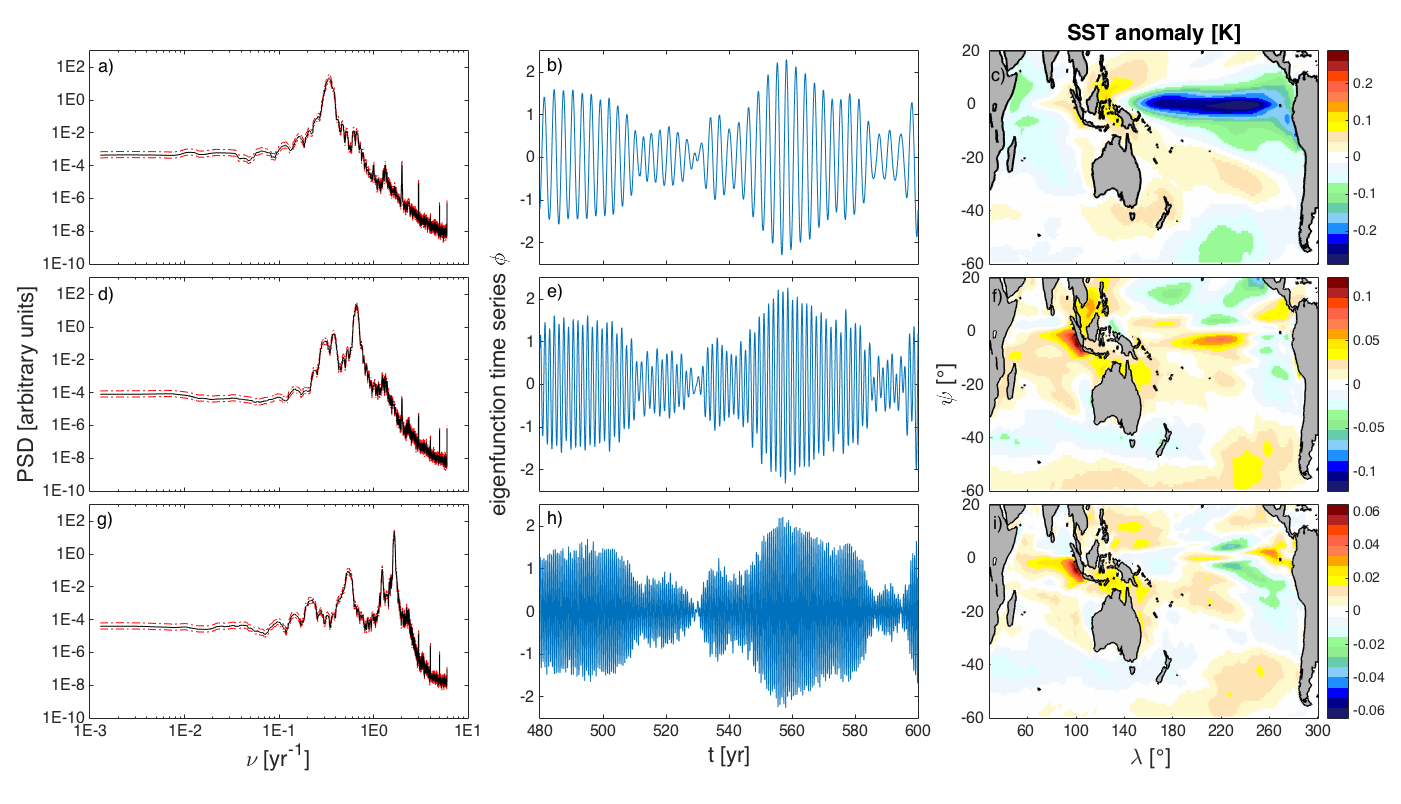}\\
\caption{Same as Fig.~\ref{figInterannual}, but for the ENSO (a--c), ENSO-A1 (d--f), and ENSO-A2 (g--i) modes extracted via NLSA from the CM3 dataset. The power spectral densities were estimated in the same manner as in Fig.~\ref{figGFDLPeriodic}.}\label{figGFDLInterannual}
\end{figure}

\begin{figure}[t]
\noindent\includegraphics[width=\linewidth]{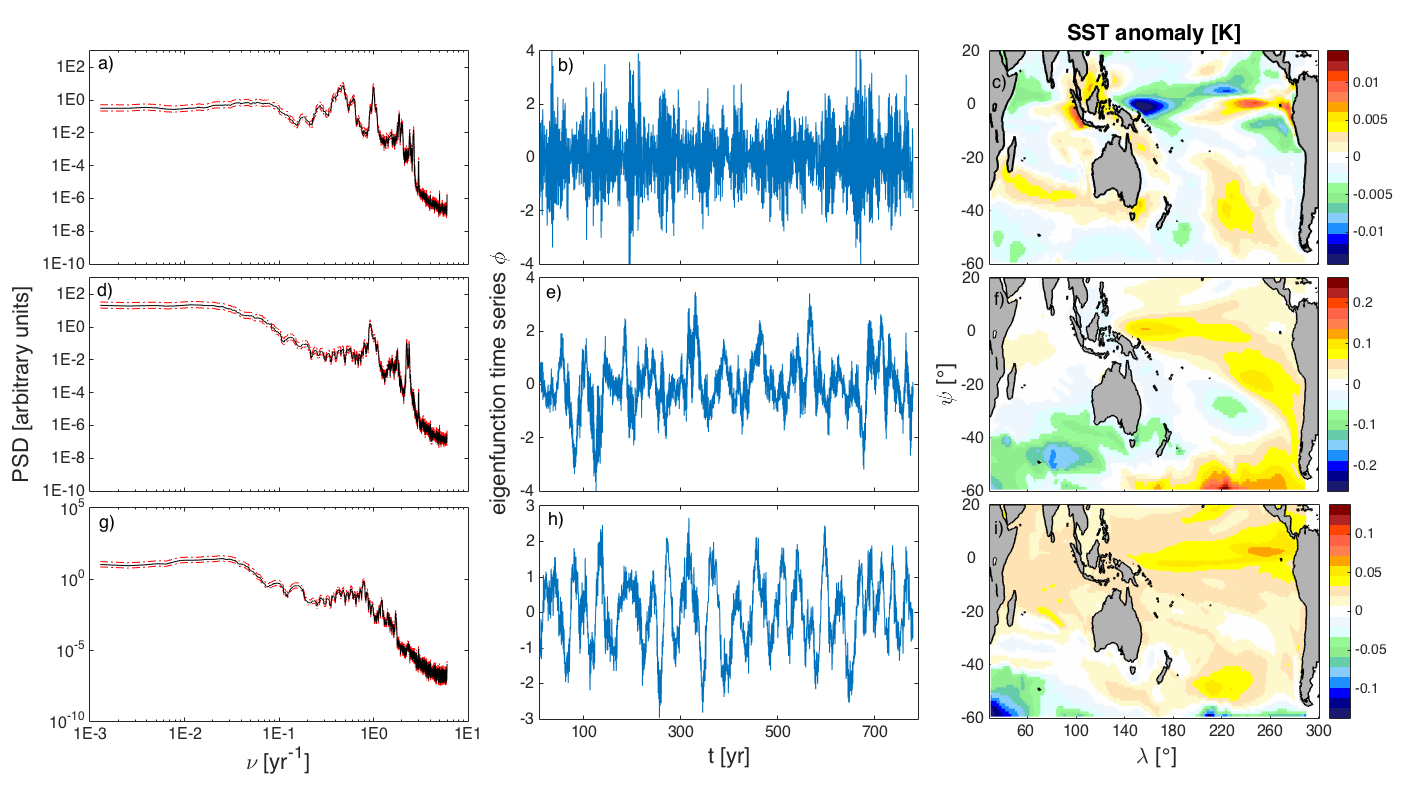}\\
\caption{Same as Fig.~\ref{figBienn}(a--c) and Fig.~\ref{figDec}, but for the TBO (a--c),  (d--f), and IPO  (g--i) modes extracted via NLSA from the CM3 dataset. The power spectral densities were estimated in the same manner as in Fig.~\ref{figGFDLPeriodic}.}\label{figGFDLBiennDec}
\end{figure}

\begin{figure}[t]
\noindent\includegraphics[width=\linewidth]{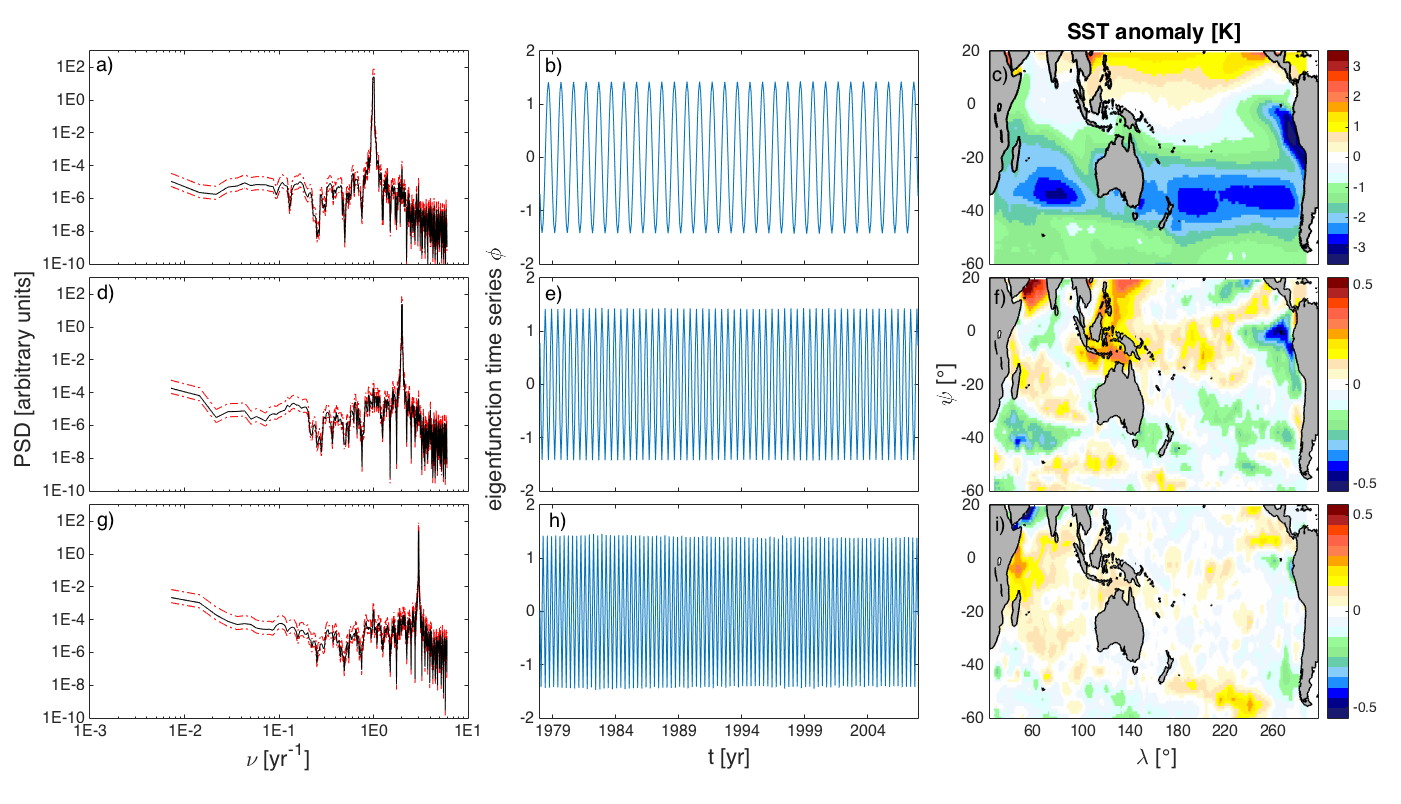}\\
  \caption{Same as Fig.~\ref{figPeriodic}, but for the annual (a--c), semiannual (d--f), and triennial (g--i) modes of Indo-Pacific SST variability extracted with NLSA from 140 years of HadISST data. In this case, the power spectral densities were computed via the multitaper method using a time-bandwidth product $ p = 2 $ and $ K = 2 p - 1 = 4 $ Slepian tapers. The effective frequency resolution is $ \Delta \nu = p / ( N \delta t ) \approx 1/69 $ y$^{-1}$, where $ N = 1665 $ is the number of samples in the eigenfunction time series from the HadISST data. Dashed red lines in the power spectral density plots indicate 95\% confidence levels computed using an $ F $-test with $ ( 2, 2 K - 2 ) $ degrees of freedom.}
  \label{figObsPeriodic}
\end{figure}

\begin{figure}[t]
\noindent\includegraphics[width=\linewidth]{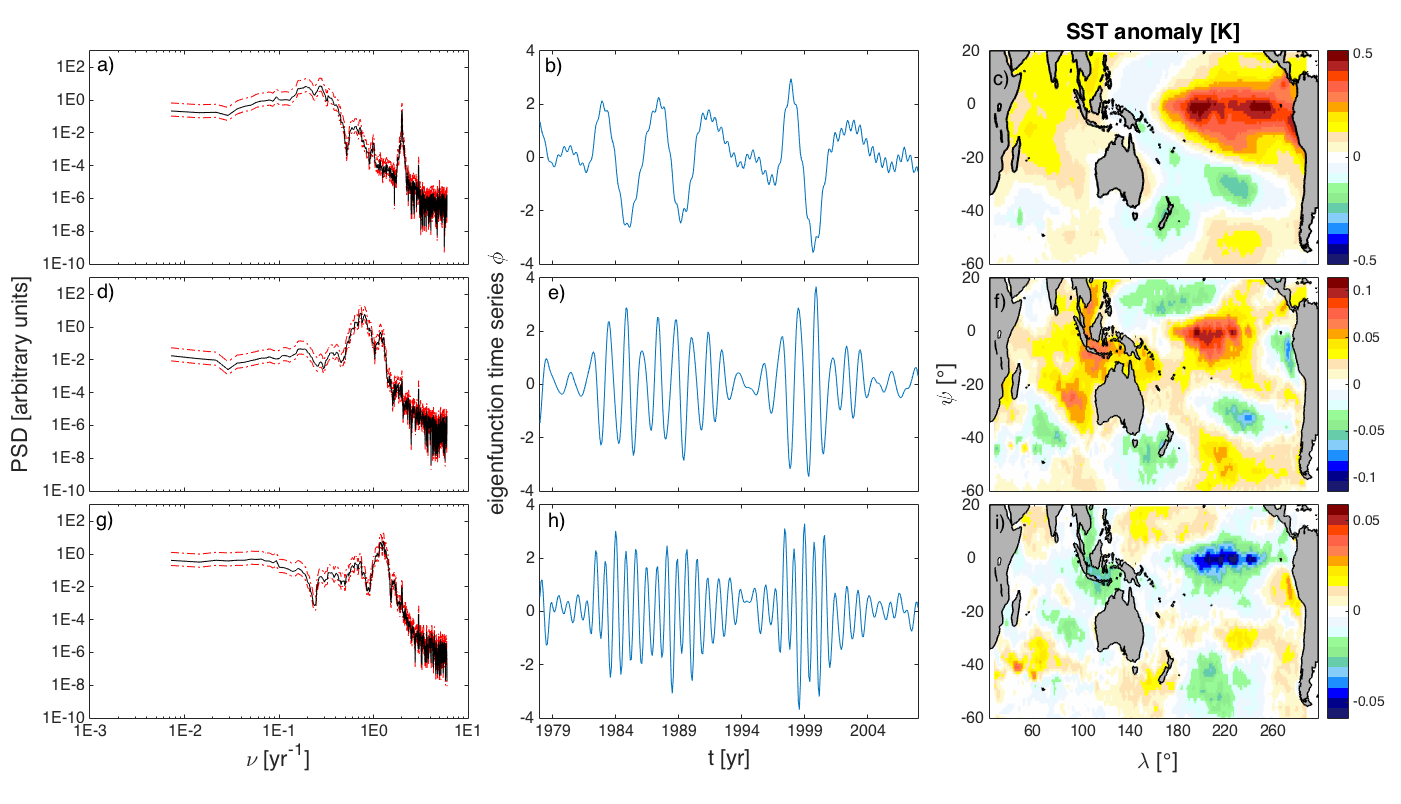}\\
  \caption{Same as Fig.~\ref{figInterannual}, but for the ENSO (a--c), ENSO-A1 (d--f), and ENSO-A2 (g--i) modes extracted via NLSA from the industrial-era HadISST dataset. The power spectral densities were estimated in the same manner as in Fig.~\ref{figObsPeriodic}.}
  \label{figObsInterannual}
\end{figure}

Next, consider the decadal modes. Figure~\ref{figSSAb}(a--c) displays the temporal pattern, frequency spectrum, and SST composite corresponding to the leading SSA decadal mode, mode~32. This mode has a red-noise-like frequency spectrum carrying the majority of its power on decadal timescales, though compared to the NLSA WPMM spectrum (Fig.~\ref{figDec}(d)) the frequency spectrum of the SSA mode decays less steeply and exhibits more power on shorter timescales. In the spatial domain, there is also a moderate agreement between SSA mode~32 and the WPMM in that both modes feature an decadal SST anomaly cluster in the equatorial western Pacific and coexisting with opposite-sign anomalies in the south and eastern Pacific, but in the case of SSA the amplitude is weaker.  Also, SSA mode~32 has weaker correlation (correlation coefficient 0.48) with the amplitude envelope of the SSA ENSO mode than the WPMM has with the amplitude envelope of the NLSA ENSO mode (correlation coefficient 0.63; see Part~II).

SSA modes~43 and~45 (Fig.~\ref{figSSAb}(d--i)) are also of decadal character. In the spatial domain, these modes describe an oscillation featuring a cluster of SST anomalies that originates in the equatorial eastern Pacific and propagates westward until it merges with an anomaly of the same sign that develops in the western equatorial Pacific. At the same time, another same-sign anomaly cluster emerges in the Southern Pacific east of the coast of New Zealand and travels towards the East. Overall, these motions produce an apparent basin-scale anticyclonic motion in the South Pacific Ocean. We were not able to identify a clear analog of this pattern in the NLSA spectrum. Besides modes~43--45, SSA contains additional decadal modes, but these modes are generally mixed with high-frequency ($\sim 1$ y$^{-1}$) oscillations and we do not discuss them here. We also note that as with NLSA there are trend-like modes in the SSA spectrum (not shown) that capture the small model drift present in the CCSM4 dataset.

In summary, applied to Indo-Pacific SST dataset from CCSM4, both SSA and NLSA are able to extract modes of adequate timescale separation, spanning seasonal to multidecadal timescales. The leading modes, in particular the annual and semiannual modes and the fundamental component of ENSO, are in good agreement between the two methods. Further down the spectrum, the two methods have significant differences. In particular, NLSA has higher skill in capturing the combination modes associated with the nonlinear interaction between ENSO and the annual cycle, and it also provides a better representation of the TBO and its associated combination modes. In the decadal scale, both SSA and NLSA produce clean decadal spatiotemporal patterns, although we were not able to find a direct correspondence between the decadal modes extracted via the two methods. In particular, we were not able to identify a decadal mode in the SSA spectrum having amplitude modulation relationships with ENSO of comparable strength as the WPMM does.
  
\bibliographystyle{ametsoc2014}

\begin{thebibliography}{114}
\providecommand{\natexlab}[1]{#1}
\providecommand{\url}[1]{\texttt{#1}}
\renewcommand{\UrlFont}{\rmfamily}
\providecommand{\urlprefix}{URL }
\expandafter\ifx\csname urlstyle\endcsname\relax
  \providecommand{\doi}[1]{doi:\discretionary{}{}{}#1}\else
  \providecommand{\doi}{doi:\discretionary{}{}{}\begingroup
  \urlstyle{rm}\Url}\fi
\providecommand{\eprint}[2][]{\url{#2}}

\bibitem[{An and Jin(2004)An, and Jin}]{AnJin04}
An, S.-I., and F.-F. Jin, 2004: Nonlinearity and asymmetry of {ENSO}.
  \textit{J. Climate}, \textbf{17}, 2399--2412,
  \doi{10.1175/1520-0442(2004)017<2399:NAAOE>2.0.CO;2}.

\bibitem[{Annamalai et~al.(2003)Annamalai, Murtugudde, Potemra, Xie, Liu,, and
  Wang}]{AnnamalaiEtAl04}
Annamalai, H., R.~Murtugudde, J.~Potemra, S.~P. Xie, P.~Liu, and B.~Wang, 2003:
  Coupled dynamics in the {I}ndian {O}cean: {S}pring initiation of the zonal
  mode. \textit{Deep-Sea Res.}, \textbf{50}, 2305--2330,
  \doi{10.1016/S0967-0645(03)00058-4}.

\bibitem[{Ashok et~al.(2003)Ashok, Guan,, and Yamagata}]{AshokEtAl03}
Ashok, K., Z.~Y. Guan, and T.~Yamagata, 2003: A look at the relationship
  between the {ENSO} and the {I}ndian {O}cean {D}ipole. \textit{J. Meteorol.
  Soc. Jpn.}, \textbf{81}, 41--56, \doi{10.2151/jmsj.81.41}.

\bibitem[{Ashok et~al.(2007)Ashok, Nakamura,, and Yamagata}]{AshokEtAl07}
Ashok, K., H.~Nakamura, and T.~Yamagata, 2007: Impacts of {ENSO} and {I}ndian
  {O}cean {D}ipole events on the {S}outhern {H}emisphere storm-track activity
  during austral winter. \textit{J. Climate}, \textbf{20}, 3147--3163,
  \doi{http://dx.doi.org/10.1175/JCLI4155.1}.

\bibitem[{Battisti and Hirst(1989)Battisti, and Hirst}]{BattistiHirst89}
Battisti, D.~S., and A.~C. Hirst, 1989: Interannual varibility in a tropical
  atmosphere--ocean model: {I}nfluence of the basic state, ocean geometry and
  nonlinearity. \textit{J. Atmos. Sci.}, \textbf{46~(12)}, 1687--1712,
  \doi{10.1175/1520-0469(1989)046<1687:IVIATA>2.0.CO;2}.

\bibitem[{Behera et~al.(2006)Behera, Luo, Masson, Rao, Sakuma,, and
  Yamagata}]{BeheraEtAl06}
Behera, S.~K., J.~J. Luo, S.~Masson, S.~A. Rao, H.~Sakuma, and T.~Yamagata,
  2006: A {CGCM} {S}tudy on the {I}nteraction between {IOD} and {ENSO}.
  \textit{J. Climate}, \textbf{19}, 1688--1705,
  \doi{http://dx.doi.org/10.1175/JCLI3797.1}.

\bibitem[{Belkin and Niyogi(2003)Belkin, and Niyogi}]{BelkinNiyogi03}
Belkin, M., and P.~Niyogi, 2003: Laplacian eigenmaps for dimensionality
  reduction and data representation. \textit{Neural Comput.}, \textbf{15},
  1373--1396, \doi{10.1162/089976603321780317}.

\bibitem[{Bellenger et~al.(2014)Bellenger, Guilyardi, Leloup, Lengaigne,, and
  Vialard}]{BellengerEtAl14}
Bellenger, H., E.~Guilyardi, J.~Leloup, M.~Lengaigne, and J.~Vialard, 2014:
  {ENSO} representation in climate models: from {CMIP3} to {CMIP5}.
  \textit{Climate Dyn.}, \textbf{42}, 1999--2018,
  \doi{10.1007/s00382-013-1783-z}.

\bibitem[{Berry et~al.(2013)Berry, Cressman, Greguric~Ferencek,, and
  Sauer}]{BerryEtAl13}
Berry, T., R.~Cressman, Z.~Greguric~Ferencek, and T.~Sauer, 2013: Time-scale
  separation from diffusion-mapped delay coordinates. \textit{SIAM J. Appl.
  Dyn. Sys.}, \textbf{12}, 618--649, \doi{10.1137/12088183X}.

\bibitem[{Berry and Sauer(2015)Berry, and Sauer}]{BerrySauer15}
Berry, T., and T.~Sauer, 2015: Local kernels and the geometric structure of
  data. \textit{Appl. Comput. Harmon. Anal.}, \doi{10.1016/j.acha.2015.03.002}.

\bibitem[{Bjerknes(1969)}]{Bjerknes69}
Bjerknes, J., 1969: Atmospheric teleconnections from the equatorial {P}acific.
  \textit{Mon. Wea. Rev.}, \textbf{97~(3)}, 163--172,
  \doi{10.1175/1520-0493(1969)097<0163:ATFTEP>2.3.CO;2}.

\bibitem[{Broomhead and King(1986)Broomhead, and King}]{BroomheadKing86}
Broomhead, D.~S., and G.~P. King, 1986: Extracting qualitative dynamics from
  experimental data. \textit{Phys. D}, \textbf{20~(2--3)}, 217--236,
  \doi{10.1016/0167-2789(86)90031-x}.

\bibitem[{Burgers and Stephenson(1999)Burgers, and
  Stephenson}]{BurgersStephenson99}
Burgers, G., and D.~B. Stephenson, 1999: The ``normality'' of {E}l {N}i\~no.
  \textit{Geophys. Res. Lett.}, \textbf{26~(8)}, 1027--1030,
  \doi{10.1175/1520-0442(2004)017<2399:NAAOE>2.0.CO;2}.

\bibitem[{Bushuk et~al.(2014)Bushuk, Giannakis,, and Majda}]{BushukEtAl14}
Bushuk, M., D.~Giannakis, and A.~J. Majda, 2014: Reemergence mechanisms for
  {N}orth {P}acific sea ice revealed through nonlinear {L}aplacian spectral
  analysis. \textit{J.\ Climate}, \textbf{27}, 6265--6287,
  \doi{10.1175/jcli-d-13-00256.}

\bibitem[{Cai et~al.(2012)Cai, Rensch, Cowan,, and Hendon}]{CaiEtAl12}
Cai, W., P.~Rensch, T.~Cowan, and H.~H. Hendon, 2012: An asymmetry in the {IOD}
  and {ENSO} teleconnection pathway and its impact on {A}ustralian climate.
  \textit{J. Climate}, \textbf{25}, 6518--6329,
  \doi{10.1175/JCLI-D-11-00501.1}.

\bibitem[{Cai et~al.(2013)Cai, Zheng, Weller, Collins, Cowan, Lengaigne, Yu,,
  and Yamagata}]{CaiEtAl13}
Cai, W., X.-T. Zheng, E.~Weller, M.~Collins, T.~Cowan, M.~Lengaigne, W.~Yu, and
  T.~Yamagata, 2013: Projected response of the {I}ndian {O}cean {D}ipole to
  greenhouse warming. \textit{Nat. Geosci.}, \textbf{6}, 999--1007,
  \doi{10.1038/ngeo2009}.

\bibitem[{Cane and Zebiak(1985)Cane, and Zebiak}]{CaneZebiak85}
Cane, M.~A., and S.~E. Zebiak, 1985: A theory for {E}l {N}i\~no and the
  {S}outhern {O}scillation. \textit{Science}, \textbf{228~(4703)}, 1085--1087,
  \doi{10.1126/science.228.4703.1085}.

\bibitem[{CCSM(2010)}]{ccsm4}
CCSM, 2010: {C}ommunity {C}limate {S}ystem {M}odel {V}ersion 4 {(CCSM4)} data.
  Earth System Grid at the National Center for Atmospheric Research, {URL}:
  https://www.earthsystemgrid.org/dataset/ucar.cgd.ccsm4.joc.b40.1850.track1.1deg.006.html,
  last accessed August 2016.

\bibitem[{Choi et~al.(2013)Choi, An, Yeh,, and Yu}]{ChoiEtAl13}
Choi, J., S.-I. An, S.-W. Yeh, and J.-Y. Yu, 2013: {ENSO}-{L}ike and
  {ENSO}-{I}nduced {T}ropical {P}acific {D}ecadal {V}ariability in {CGCM}s.
  \textit{J. Climate}, \textbf{26}, 1485--1501.

\bibitem[{Coifman and Lafon(2006)Coifman, and Lafon}]{CoifmanLafon06}
Coifman, R.~R., and S.~Lafon, 2006: Diffusion maps. \textit{Appl. Comput.
  Harmon. Anal.}, \textbf{21}, 5--30, \doi{10.1016/j.acha.2006.04.006}.

\bibitem[{Comeau et~al.(2016)Comeau, Zhao, Giannakis,, and
  Majda}]{ComeauEtAl16}
Comeau, D., Z.~Zhao, D.~Giannakis, and A.~J. Majda, 2016: Data-driven
  prediction strategies for low-frequency patterns of {N}orth {P}acific climate
  variability. \textit{Climate. Dyn}, \doi{10.1007/s00382-016-3177-5}, in
  press.

\bibitem[{Deser et~al.(2012)Deser, Phillips, Tomas, Okumura, Alexander,
  Capotondi,, and Scott}]{DeserEtAl12b}
Deser, C., A.~S. Phillips, R.~A. Tomas, Y.~M. Okumura, M.~A. Alexander,
  A.~Capotondi, and J.~D. Scott, 2012: {ENSO} and {P}acific decadal variability
  in the {C}ommunity {C}limate {S}ystem {M}odel {V}ersion 4. \textit{J.
  Climate}, \textbf{25}, 2622--2651, \doi{10.1175/JCLI-D-11-00301.1}.

\bibitem[{Di~Lorenzo et~al.(2008)}]{DiLorenzoEtAl08}
Di~Lorenzo, E., and Coauthors, 2008: {N}orth {P}acific {G}yre {O}scillation
  links ocean climate and ecosystem change. \textit{Geophys. Res. Lett.},
  \textbf{35}, L08\,607, \doi{10.1029/2007gl032838}.

\bibitem[{Dommenget(2011)}]{Dommenget11}
Dommenget, D., 2011: An objective analysis of the observed spatial structure of
  the tropical {I}ndian {O}cean {SST} variability. \textit{Climate Dyn.},
  \textbf{36}, 2129--2145, \doi{10.1007/s00382-010-0787-1}.

\bibitem[{England et~al.(2014)}]{EnglandEtAl14}
England, M.~H., and Coauthors, 2014: Recent intensification of wind-driven
  circulation in the {P}acific and the ongoing warming hiatus. \textit{Nat.
  Clim. Change}, \textbf{4}, 222--227, \doi{10.1038/nclimate2106}.

\bibitem[{Gehne et~al.(2014)Gehne, Kleeman,, and Trenberth}]{GehneEtAl14}
Gehne, M., R.~Kleeman, and K.~E. Trenberth, 2014: Irregularity and decadal
  variation in {ENSO}: a simplified model based on {P}rincipal {O}scillation
  {P}atterns. \textit{Climate Dynamics}, \textbf{43~(12)}, 3327--3350,
  \doi{10.1007/s00382-014-2108-6}.

\bibitem[{Gent et~al.(2011)}]{GentEtAl11}
Gent, P.~R., and Coauthors, 2011: The {C}ommunity {C}limate {S}ystem {M}odel
  version 4. \textit{J. Climate}, 4973--4991, \doi{10.1175/2011jcli4083.1}.

\bibitem[{GFDL(2012)}]{gfdl-cm3}
GFDL, 2012: {C}oupled {P}hysical {M}odel ({CM3}) data. Geophysical FLuid
  Dynamics Laboratory, {URL}: http://nomads.gfdl.noaa.gov/, accessed January
  2014.

\bibitem[{Ghil et~al.(2002)}]{GhilEtAl02}
Ghil, M., and Coauthors, 2002: Advanced spectral methods for climatic time
  series. \textit{Rev. Geophys.}, \textbf{40}, 1003,
  \doi{10.1029/2000rg000092}.

\bibitem[{Giannakis(2015)}]{Giannakis15}
Giannakis, D., 2015: Dynamics-adapted cone kernels. \textit{SIAM J. Appl. Dyn.
  Sys.}, \textbf{14~(2)}, 556--608, \doi{10.1137/140954544}.

\bibitem[{Giannakis(2016)}]{Giannakis16}
Giannakis, D., 2016: Data-driven spectral decomposition and forecasting of
  ergodic dynamical systems. \eprint{1507.02338}.

\bibitem[{Giannakis and Majda(2011)Giannakis, and Majda}]{GiannakisMajda11c}
Giannakis, D., and A.~J. Majda, 2011: Time series reconstruction via machine
  learning: Revealing decadal variability and intermittency in the {N}orth
  {P}acific sector of a coupled climate model. \textit{Conference on
  Intelligent Data Understanding 2011}, Mountain View, California.

\bibitem[{Giannakis and Majda(2012{\natexlab{a}})Giannakis, and
  Majda}]{GiannakisMajda12b}
Giannakis, D., and A.~J. Majda, 2012{\natexlab{a}}: Comparing low-frequency and
  intermittent variability in comprehensive climate models through nonlinear
  {L}aplacian spectral analysis. \textit{Geophys. Res. Lett.}, \textbf{39},
  L10\,710, \doi{10.1029/2012GL051575}.

\bibitem[{Giannakis and Majda(2012{\natexlab{b}})Giannakis, and
  Majda}]{GiannakisMajda12c}
Giannakis, D., and A.~J. Majda, 2012{\natexlab{b}}: Limits of predictability in
  the {N}orth {P}acific sector of a comprehensive climate model.
  \textit{Geophys. Res. Lett.}, \textbf{39}, L24\,602,
  \doi{10.1029/2012gl054273}.

\bibitem[{Giannakis and Majda(2012{\natexlab{c}})Giannakis, and
  Majda}]{GiannakisMajda12a}
Giannakis, D., and A.~J. Majda, 2012{\natexlab{c}}: Nonlinear {L}aplacian
  spectral analysis for time series with intermittency and low-frequency
  variability. \textit{Proc. Natl. Acad. Sci.}, \textbf{109~(7)}, 2222--2227,
  \doi{10.1073/pnas.1118984109}.

\bibitem[{Giannakis and Majda(2013)Giannakis, and Majda}]{GiannakisMajda13}
Giannakis, D., and A.~J. Majda, 2013: Nonlinear {L}aplacian spectral analysis:
  Capturing intermittent and low-frequency spatiotemporal patterns in
  high-dimensional data. \textit{Stat. Anal. Data Min.}, \textbf{6~(3)},
  180--194, \doi{10.1002/sam.11171}.

\bibitem[{Giannakis and Majda(2014)Giannakis, and Majda}]{GiannakisMajda14}
Giannakis, D., and A.~J. Majda, 2014: Data-driven methods for dynamical
  systems: Quantifying predictability and extracting spatiotemporal patterns.
  \textit{Mathematical and Computational Modeling: With Applications in
  Engineering and the Natural and Social Sciences}, R.~Melnik, Ed., Wiley,
  Hoboken, 288.

\bibitem[{Giannakis and Slawinska(2016)Giannakis, and
  Slawinska}]{GiannakisSlawinska16}
Giannakis, D., and J.~Slawinska, 2016: Indo-{P}acific variability on seasonal
  to multidecadal timescales. {P}art {II}: {M}ultiscale atmosphere-ocean
  linkages. \textit{J. Climate}, submitted.

\bibitem[{Goswami(1995)}]{Goswami95}
Goswami, B.~N., 1995: A multiscale interaction model for the origin of the
  tropspheric {QBO}. \textit{J. Climate}, \textbf{8}, 524--534,
  \doi{10.1175/1520-0442(1995)008<0524:AMIMFT>2.0.CO;2}.

\bibitem[{Graham et~al.(2015)Graham, Brown, Wittenberg,, and
  Holbrook}]{GrahamEtAl15}
Graham, F.~S., J.~N. Brown, A.~T. Wittenberg, and N.~J. Holbrook, 2015:
  Reassessing conceptual models of {ENSO}. \textit{J. Climate},
  \textbf{28~(23)}, 9121--9142, \doi{10.1175/JCLI-D-14-00812.1}.

\bibitem[{Griffies et~al.(2011)}]{GriffiesEtAl11}
Griffies, S.~M., and Coauthors, 2011: The {GFDL} {CM3} coupled climate model:
  {C}haracteristics of the ocean and sea ice simulations. \textit{J. Climate},
  \textbf{24}, 3520--3544, \doi{10.1175/2011JCLI3964.1}.

\bibitem[{HadISST(2013)}]{HadISST13}
HadISST, 2013: Hadley {C}entre {S}ea {I}ce and {S}ea {S}urface {T}emperature
  ({HadISST1}) data. Met Office Hadley Centre, {URL}:
  http://www.metoffice.gov.uk/hadobs/hadisst/data/download.html, accessed June
  2015.

\bibitem[{Han et~al.(2014)}]{HanEtAl14}
Han, W., and Coauthors, 2014: Intensification of decadal and multi-decadal sea
  level variability in the western tropical {P}acific during recent decades.
  \textit{Climate Dyn.}, \textbf{4}, 1357--1379,
  \doi{10.1007/s00382-013-1951-1}.

\bibitem[{Izumo et~al.(2014)Izumo, Lengaigne, Vialard, Luo, Yamagata,, and
  Madec}]{IzumoEtAl14}
Izumo, T., M.~Lengaigne, J.~Vialard, J.-J. Luo, T.~Yamagata, and G.~Madec,
  2014: Influence of the {I}ndian {O}cean {D}ipole and {P}acific recharge on
  the following year {E}l {N}i\~no: interdecadal robustness. \textit{Climate
  Dyn.}, \textbf{42}, 291--310, \doi{10.1007/s00382-012-1628-1}.

\bibitem[{Izumo et~al.(2010)}]{IzumoEtAl10}
Izumo, T., and Coauthors, 2010: Influence of the state of the {I}ndian {O}cean
  {D}ipole on the following years {E}l {N}i\~no. \textit{Nature Geosci},
  \textbf{3}, 168--172, \doi{10.1038/ngeo760}.

\bibitem[{Jin(1997{\natexlab{a}})}]{Jin97}
Jin, F.-F., 1997{\natexlab{a}}: An equatorial ocean recharge paradigm for
  {ENSO}. {P}art {I}: {C}onceptual model. \textit{J. Atmos. Sci.}, \textbf{54},
  811--829, \doi{10.1175/1520-0469(1997)054<0811:AEORPF>2.0.CO;2}.

\bibitem[{Jin(1997{\natexlab{b}})}]{Jin97b}
Jin, F.-F., 1997{\natexlab{b}}: An equatorial ocean recharge paradigm for
  {ENSO}. {P}art {II}: {A} striped-down coupled model. \textit{J. Atmos. Sci.},
  \textbf{54}, 830--847, \doi{10.1175/1520-0469(1997)054<0830:AEORPF>2.0.CO;2}.

\bibitem[{Jin and Neelin(1993)Jin, and Neelin}]{JinNeelin93}
Jin, F.-F., and J.~D. Neelin, 1993: Modes of interannual tropical
  ocean--atmosphere interaction---a unified view. {P}art {I}: {N}umerical
  results. \textit{J. Atmos. Sci.}, \textbf{50~(21)}, 3477--3503,
  \doi{10.1175/1520-0469(1993)050<3477:MOITOI>2.0.CO;2}.

\bibitem[{Karnauskas et~al.(2009)Karnauskas, Seager, Kaplan, Kushnir,, and
  Cane}]{KarnauskasEtAl09}
Karnauskas, K.~B., R.~Seager, A.~Kaplan, Y.~Kushnir, and M.~A. Cane, 2009:
  Observed {S}trengthening of the {Z}onal {S}ea {S}urface {T}emperature
  {G}radient across the {E}quatorial {P}acific {O}cean. \textit{J. Climate},
  \textbf{22~(16)}, 4316--4321, \doi{10.1175/2009JCLI2936.1}.

\bibitem[{{Kleeman}(2008)}]{Kleeman08}
{Kleeman}, R., 2008: {Stochastic theories for the irregularity of {ENSO}}.
  \textit{Philosophical Transactions of the Royal Society of London Series A},
  \textbf{366}, 2509--2524, \doi{10.1098/rsta.2008.0048}.

\bibitem[{Kleeman et~al.(1999)Kleeman, McCreary,, and Klinger}]{KleemanEtAl99}
Kleeman, R., J.~P. McCreary, Jr., and B.~A. Klinger, 1999: A mechanism for
  generating {ENSO} and {D}ecadal variability. \textit{Geophys. Res. Lett.},
  \textbf{26}, 1743--1746, \doi{10.1029/1999gl900352}.

\bibitem[{Lengaigne et~al.(2006)Lengaigne, Boulanger, Menkes,, and
  Spencer}]{LengaigneEtAl06}
Lengaigne, M., J.-P. Boulanger, C.~Menkes, and H.~Spencer, 2006: Influence of
  the seasonal cycle on the termination of {E}l {N}i\~no events in a {C}oupled
  {G}eneral {C}irculation {M}odel. \textit{J. Climate}, \textbf{19},
  1850--1868, \doi{dx.doi.org/10.1175/JCLI3706.1}.

\bibitem[{Li(2013)}]{Li13}
Li, T., 2013: \textit{Monsoon Climate Variabilities}, 27--51. American
  Geophysical Union, \doi{10.1029/2008GM000782},
  \urlprefix\url{http://dx.doi.org/10.1029/2008GM000782}.

\bibitem[{Li et~al.(2006)Li, Liu, Fu, Wang,, and Meehl}]{LiEtAl06}
Li, T., P.~Liu, X.~Fu, B.~Wang, and G.~A. Meehl, 2006: Spatiotemporal
  {S}tructures and {M}echanisms of the {T}ropospheric {B}iennial {O}scillation
  in the {I}ndo-{P}acific {W}arm {O}cean {R}egions. \textit{J. Climate},
  \textbf{19}, 3070--3087, \doi{http://dx.doi.org/10.1175/JCLI3736.1}.

\bibitem[{Li and Zhang(2002)Li, and Zhang}]{LiEtAl02}
Li, T., and Y.~Zhang, 2002: Processes that determine the quasi-biennial and
  lower-frequency variability of the south asian monsoon. \textit{Journal of
  the Meteorological Society of Japan. Ser. II}, \textbf{80~(5)}, 1149--1163,
  \doi{10.2151/jmsj.80.1149}.

\bibitem[{Luo et~al.(2012)Luo, Sasaki,, and Masumoto}]{LuoEtAl12}
Luo, J.~J., W.~Sasaki, and Y.~Masumoto, 2012: Indian {O}cean warming modulates
  {P}acific climate change. \textit{Proc. Natl. Acad. Sci.}, \textbf{109},
  18\,701--18\,706, \doi{10.1073/pnas.1210239109}.

\bibitem[{Manatsa et~al.(2008)Manatsa, Chingombe,, and
  Matarira}]{ManatsaEtAl08}
Manatsa, D., W.~Chingombe, and C.~H. Matarira, 2008: The impact of the positive
  {I}ndian {O}cean dipole on {Z}imbabwe droughts. \textit{Int. J. Climatol.},
  \textbf{28}, 2011--2029, \doi{10.1002/joc.1695}.

\bibitem[{Mantua et~al.(1997)Mantua, Hare, Zhang, Wallace,, and
  Francis}]{MantuaEtAl97}
Mantua, N.~J., S.~R. Hare, Y.~Zhang, J.~M. Wallace, and R.~Francis, 1997: A
  {P}acific {I}nterdecadal {C}limate {O}scillation with {I}mpacts on {S}almon
  {P}roduction. \textit{Bull. Amer. Meteor. Soc.}, \textbf{78}, 1069--1079,
  \doi{http://dx.doi.org/10.1175/1520-0477(1997)078<1069:APICOW>2.0.CO;2}.

\bibitem[{McGregor et~al.(2012)McGregor, Timmermann, Schneider, Stuecker,, and
  England}]{McGregorEtAl12}
McGregor, S., A.~Timmermann, N.~Schneider, M.~F. Stuecker, and M.~F. England,
  2012: The effect of the {S}outh {P}acific convergence zone on the termination
  of {E}l {N}i\~no events and the meridional asymmetry of {ENSO}. \textit{J.
  Climate}, \textbf{25}, 5566--5586, \doi{10.1175/JCLI-D-11-00332.1}.

\bibitem[{Meehl(1987)}]{Meehl87}
Meehl, G.~A., 1987: The annual cycle and interannual variability in the
  tropical {P}acific and {I}ndian {O}cean regions. \textit{Mon. Wea. Rev.},
  \textbf{115}, 27--50, \doi{10.1175/1520-0493(1987)115<0027:TACAIV>2.0.CO;2}.

\bibitem[{Meehl(1993)}]{Meehl93}
Meehl, G.~A., 1993: A {C}oupled {A}ir-{S}ea {B}iennial {M}echanism in the
  {T}ropical {I}ndian and {P}acific {R}egions: {R}ole of the {O}cean.
  \textit{J. Climate}, \textbf{6}, 31--41,
  \doi{10.1175/1520-0442(1993)006<0031:ACASBM>2.0.CO;2}.

\bibitem[{Meehl(1994)}]{Meehl94}
Meehl, G.~A., 1994: Coupled land-ocean-atmosphere processes and {S}outh {A}sian
  {M}onsoon variability. \textit{Science}, \textbf{266}, 263--267,
  \doi{10.1126/science.266.5183.263}.

\bibitem[{Meehl(1997)}]{Meehl97}
Meehl, G.~A., 1997: The {S}outh {A}sian {M}onsoon and the {T}ropospheric
  {B}iennial {O}scillation. \textit{J. Climate}, \textbf{10}, 1921--1943,
  \doi{10.1175/1520-0442(1997)010<1921:TSAMAT>2.0.CO;2}.

\bibitem[{Meehl and Arblaster(2002)Meehl, and Arblaster}]{MeehlArblaster02}
Meehl, G.~A., and J.~M. Arblaster, 2002: The tropospheric biennial oscillation
  and {A}sian--{A}ustralian monsoon rainfall. \textit{J. Climate}, \textbf{15},
  722--744, \doi{10.1175/1520-0442(2002)015<0722:TTBOAA>2.0.CO;2}.

\bibitem[{Meehl and Arblaster(2011)Meehl, and Arblaster}]{MeehlArblaster11}
Meehl, G.~A., and J.~M. Arblaster, 2011: Decadal {V}ariability of
  {A}sian-{A}ustralian {M}onsoon-{ENSO}-{TBO} {R}elationships. \textit{J.
  Climate}, \textbf{24}, 4925--4940, \doi{10.1175/2011JCLI4015.1}.

\bibitem[{Meehl et~al.(2013)Meehl, Hu, Arblaster, Fasullo,, and
  Trenberth}]{MeehlEtAl13}
Meehl, G.~A., A.~Hu, J.~M. Arblaster, J.~Fasullo, and K.~E. Trenberth, 2013:
  Externally {F}orced and {I}nternally {G}enerated {D}ecadal {C}limate
  {V}ariability {A}ssociated with the {I}nterdecadal {P}acific {O}scillation.
  \textit{J. Climate}, \textbf{26}, 7298--7310,
  \doi{10.1175/JCLI-D-12-00548.1}.

\bibitem[{Neelin et~al.(1998)Neelin, Battisti, Hirst, Jin, Wakata, Yamagata,,
  and Zebiak}]{NeelinEtAl98}
Neelin, J.~D., D.~S. Battisti, A.~C. Hirst, F.-F. Jin, Y.~Wakata, T.~Yamagata,
  and S.~E. Zebiak, 1998: {ENSO} theory. \textit{J. Geophys. Res.},
  \textbf{103~(C7)}, 14\,261–--14\,290, \doi{10.1029/97JC03424}.

\bibitem[{Neelin et~al.(1994)Neelin, Latif,, and Jin}]{NeelinEtAl94}
Neelin, J.~D., M.~Latif, and F.-F. Jin, 1994: Dynamics of coupled
  ocean-atmosphere models: {T}he tropical problems. \textit{Annu. Rev. Fluid
  Mech.}, \textbf{26}, 617--659, \doi{10.1146/annurev.fl.26.010194.003153}.

\bibitem[{Ogata et~al.(2013)Ogata, Xie, Wittenberg,, and Sun}]{OgataEtAl13}
Ogata, T., S.-P. Xie, A.~Wittenberg, and D.-Z. Sun, 2013: Interdecadal
  {A}mplitude {M}odulation of {E}l {N}i\~no {S}outhern {O}scillation and {I}ts
  {I}mpact on {T}ropical {P}acific {D}ecadal {V}ariability. \textit{J.
  Climate}, \textbf{26}, 7280--7297, \doi{10.1175/JCLI-D-12-00415.1}.

\bibitem[{Packard et~al.(1980)}]{PackardEtAl80}
Packard, N.~H., and Coauthors, 1980: Geometry from a time series. \textit{Phys.
  Rev. Lett.}, \textbf{45}, 712--716, \doi{10.1103/physrevlett.45.712}.

\bibitem[{Pepler et~al.(2014)Pepler, Timbal, Rakich,, and
  Coutts-Smith}]{PeplerEtAl14}
Pepler, A., B.~Timbal, C.~Rakich, and A.~Coutts-Smith, 2014: Indian {O}cean
  {D}ipole {O}verrides {ENSO}s {I}nfluence on {C}ool {S}eason {R}ainfall across
  the {E}astern {S}eaboard of {A}ustralia. \textit{J. Climate}, \textbf{26},
  7280--7297, \doi{http://dx.doi.org/10.1175/JCLI-D-13-00554.1}.

\bibitem[{Philander(1990)}]{Philander90}
Philander, G.~S., 1990: \textit{El {N}i\~no, {L}a {N}i\~na, and the {S}outhern
  {O}scillation}, International Geophysics, Vol.~46. Academic Press, San Diego.

\bibitem[{Picaut et~al.(1997)Picaut, Masia,, and du~Penhoat}]{PicautEtAl97}
Picaut, J., F.~Masia, and Y.~du~Penhoat, 1997: An advective-reflective
  conceptual model for the oscillatory nature of the {ENSO}. \textit{Science},
  \textbf{227~(5326)}, 663--666, \doi{10.1126/science.277.5326.663}.

\bibitem[{Power et~al.(1999)Power, Casey, Folland, Colman,, and
  Mehta}]{PowerEtAl99}
Power, S., C.~Casey, C.~Folland, A.~Colman, and V.~Mehta, 1999: Inter-decadal
  modulation of the impact of {ENSO} on {A}ustralia. \textit{Climate Dyn.},
  \textbf{15}, 319--324, \doi{10.1007/s003820050284}.

\bibitem[{Rasmusson and Carpenter(1982)Rasmusson, and
  Carpenter}]{RasmussonCarpenter82}
Rasmusson, R.~M., and T.~H. Carpenter, 1982: Variations in tropical sea surface
  temperature and surface wind fields associated with the {S}outhern
  {O}scillation/{E}l {N}i\~no. \textit{Mon. Wea. Rev.}, \textbf{110}, 354--384,
  \doi{10.1175/1520-0493(1982)110<0354:VITSST>2.0.CO;2}.

\bibitem[{Rayner et~al.(2003)}]{RaynerEtAl03}
Rayner, N.~A., and Coauthors, 2003: Global analyses of sea surface temperature,
  sea ice, and night marine air temperature since the late nineteenth century.
  \textit{J. Geophys. Res.}, \textbf{108~(D14)}, \doi{10.1029/2002jd002670}.

\bibitem[{Ren et~al.(2016)Ren, Zuo, Jin,, and Stuecker}]{RenEtAl16}
Ren, H.-L., J.~Zuo, F.-F. Jin, and M.~F. Stuecker, 2016: {ENSO} and annual
  cycle interaction: the combination mode representation in {CMIP5} models.
  \textit{Clim. Dyn.}, \textbf{46}, 3753--3765,
  \doi{10.1007/s00382-015-2802-z}.

\bibitem[{Saji et~al.(1999)Saji, Goswami, Vinayachandran,, and
  Yamagata}]{SajiEtAl99}
Saji, H.~N., B.~N. Goswami, P.~N. Vinayachandran, and T.~Yamagata, 1999: A
  dipole mode in the tropical {I}ndian {O}cean. \textit{Nature}, \textbf{401},
  360--363, \doi{10.1038/43854}.

\bibitem[{Sarachik and Cane(2010)Sarachik, and Cane}]{SarachikCane10}
Sarachik, E.~S., and M.~A. Cane, 2010: \textit{The El Niño-Southern
  Oscillation Phenomenon}. Cambridge University Press, Cambridge.

\bibitem[{Sauer et~al.(1991)Sauer, Yorke,, and Casdagli}]{SauerEtAl91}
Sauer, T., J.~A. Yorke, and M.~Casdagli, 1991: Embedology. \textit{J. Stat.
  Phys.}, \textbf{65~(3--4)}, 579--616, \doi{10.1007/bf01053745}.

\bibitem[{Seager et~al.(2015)Seager, Hoerling, Schubert, Wang, Lyon, Kumar,
  Nakamura,, and Henderson}]{SeagerEtAl15}
Seager, R., M.~Hoerling, S.~Schubert, H.~Wang, B.~Lyon, A.~Kumar, J.~Nakamura,
  and N.~Henderson, 2015: Causes of the 2011-14 {C}alifornia {D}rought.
  \textit{J. Climate}, \textbf{28~(18)}, 6997--7024, \doi{doi:
  10.1175/JCLI-D-14-00860.1}.

\bibitem[{Singer(2006)}]{Singer06}
Singer, A., 2006: From graph to manifold {L}aplacian: {T}he convergence rate.
  \textit{J. Appl. Comput. Harmon. Anal.}, \textbf{21}, 128--134,
  \doi{10.1016/j.acha.2006.03.004}.

\bibitem[{Solomon and Newman(2012)Solomon, and Newman}]{SolomonNewman12}
Solomon, A., and M.~Newman, 2012: Reconciling disparate twentieth-century
  {I}ndo-{P}acific ocean temperature trends in the instrumental record.
  \textit{Nature Clim. Change}, \textbf{2~(9)}, 691--699,
  \doi{10.1038/nclimate1591}.

\bibitem[{Stein et~al.(2011)Stein, Timmermann,, and Schneider}]{SteinEtAl11}
Stein, K., A.~Timmermann, and N.~Schneider, 2011: Phase synchronization of the
  {E}l {N}i\~no-{S}outhern {O}scillation with the {A}nnual {C}ycle.
  \textit{Phys. Rev. Lett.}, \textbf{107}, 128\,501,
  \doi{10.1103/PhysRevLett.107.128501}.

\bibitem[{Stein et~al.(2014)Stein, Timmermann, Schneider, Jin,, and
  Stuecker}]{SteinEtAl14}
Stein, K., A.~Timmermann, N.~Schneider, F.-F. Jin, and F.-F. Stuecker, 2014:
  {ENSO} {S}easonal {S}ynchronization {T}heory. \textit{J .Cilmate},
  \textbf{27}, 5286--5310, \doi{10.1175/JCLI-D-13-00525.1}.

\bibitem[{Stuecker et~al.(2015{\natexlab{a}})Stuecker, Jin,, and
  Timmermann}]{StueckerEtAl15}
Stuecker, M.~F., F.~Jin, and A.~Timmermann, 2015{\natexlab{a}}: El
  {Ni\~no}--{S}outhern {O}scillation frequency cascade. \textit{Proc. Natl.
  Acad. Sci.}, \textbf{112~(44)}, 13\,490--13\,495,
  \doi{10.1073/pnas.1508622112}.

\bibitem[{Stuecker et~al.(2015{\natexlab{b}})Stuecker, Timmermann,, and
  Jin}]{StueckerEtAl15b}
Stuecker, M.~F., A.~Timmermann, and F.-F. Jin, 2015{\natexlab{b}}: Tropospheric
  biennial oscillation (tbo) indistinguishable from white noise.
  \textit{Geophys. Res. Lett.}, \textbf{42}, 7785--7791,
  \doi{10.1002/2015GL065878}.

\bibitem[{Stuecker et~al.(2013)Stuecker, Timmermann, Jin, McGregor,, and
  Ren}]{StueckerEtAl13}
Stuecker, M.~F., A.~Timmermann, F.-F. Jin, S.~McGregor, and H.-L. Ren, 2013: A
  combination mode of the annual cycle and the {E}l {N}i\~no/{S}outhern
  {O}scillation. \textit{Nat. Geosci.}, \textbf{6}, 540--544,
  \doi{10.1038/NGEO1826}.

\bibitem[{Suarez and Schopf(1988)Suarez, and Schopf}]{SuarezSchopf88}
Suarez, M.~J., and P.~S. Schopf, 1988: A delayed action oscillator for {ENSO}.
  \textit{J. Atmos. Sci.}, \textbf{45~(21)}, 3283--3287,
  \doi{10.1175/1520-0469(1988)045<3283:ADAOFE>2.0.CO;2}.

\bibitem[{Sun and Yu(2009)Sun, and Yu}]{SunYu09}
Sun, F., and J.-Y. Yu, 2009: A 10--15-yr modulation cycle of {ENSO} intensity.
  \textit{J. Climate}, \textbf{22}, 1718--1735, \doi{10.1175/2008JCLI2285.1}.

\bibitem[{Takens(1981)}]{Takens81}
Takens, F., 1981: Detecting strange attractors in turbulence. \textit{Dynamical
  Systems and Turbulence, Warwick 1980}, Lecture Notes in Mathematics, Vol.
  898, Springer, Berlin, 366--381, \doi{10.1007/bfb0091924}.

\bibitem[{Thompson and Battisti(2001)Thompson, and
  Battisti}]{ThompsonBattisti01}
Thompson, C.~J., and D.~S. Battisti, 2001: A linear stochastic dynamical model
  of {ENSO}. {P}art {II}: {A}nalysis. \textit{J. Climate}, \textbf{14},
  445--466, \doi{10.1175/1520-0442(2001)014<0445:ALSDMO>2.0.CO;2}.

\bibitem[{Thomson(1982)}]{Thomson82}
Thomson, D.~J., 1982: Spectrum estimation and harmonic analysis. \textit{Proc.
  IEEE}, \textbf{70}, 1055--1096, \doi{10.1109/proc.1982.12433}.

\bibitem[{Thual et~al.(2016)Thual, Majda, Chen,, and Stechmann}]{ThualEtAl16}
Thual, S., A.~J. Majda, N.~Chen, and S.~N. Stechmann, 2016: Simple stochastic
  model for {E}l {N}i\~no with westerly wind bursts. \textit{Proc. Natl. Acad.
  Sci.}, \doi{10.1073/pnas.1612002113}, early online edition.

\bibitem[{Timmermann(2003)}]{Timmermann03}
Timmermann, A., 2003: Decadal {ENSO} amplitude modulations: {A} nonlinear
  paradigm. \textit{Global Planet. Change}, \textbf{37}, 135--156,
  \doi{10.1016/S0921-8181(02)00194-7}.

\bibitem[{Timmermann et~al.(2003)Timmermann, Jin,, and
  Abshagen}]{TimmermannEtAl03}
Timmermann, A., F.-F. Jin, and J.~Abshagen, 2003: A nonlinear theory for {E}l
  {N}i\~no bursting. \textit{J. Atmos. Sci.}, \textbf{60}, 152--165,
  \doi{10.1175/1520-0469(2003)060<0152:ANTFEN>2.0.CO;2}.

\bibitem[{Trenberth(2013)}]{Trenberth13}
Trenberth, K.~E., 2013: El ni\~no-{S}outhern {O}scillation ({ENSO}).
  \textit{Reference Module in Earth Systems and Environmental Sciences},
  S.~Elias, Ed., Elsevier, \doi{10.1016/B978-0-12-409548-9.04082-3}.

\bibitem[{Vecchi(2006)}]{Vecchi06}
Vecchi, G., 2006: The termination of the 1997-–98 {E}l {N}i\~no. {P}art {II}:
  {M}echanisms of {A}tmospheric {C}hange. \textit{J. Climate}, \textbf{19},
  2647--2664, \doi{10.1175/JCLI3780.1}.

\bibitem[{von Luxburg et~al.(2008)von Luxburg, Belkin,, and
  Bousquet}]{VonLuxburgEtAl08}
von Luxburg, U., M.~Belkin, and O.~Bousquet, 2008: Consistency of spectral
  clustering. \textit{Ann. Stat.}, \textbf{26~(2)}, 555--586,
  \doi{10.1214/009053607000000640}.

\bibitem[{Wang(2001)}]{Wang01}
Wang, C., 2001: A unified oscillator model for the {E}l {N}i\~no--{S}outhern
  {O}scillation. \textit{J. Climate}, \textbf{14}, 98--115,
  \doi{10.1175/1520-0442(2001)014<0098:AUOMFT>2.0.CO;2}.

\bibitem[{Wang et~al.(2017)Wang, Deser, Yu, DiNezio,, and Clement}]{WangEtAl17}
Wang, C., C.~Deser, J.-Y. Yu, P.~DiNezio, and A.~Clement, 2017: El {N}i{\~{n}}o
  and {S}outhern {O}scillation ({ENSO}): {A} review. \textit{Coral Reefs of the
  Eastern Tropical Pacific: Persistence and Loss in a Dynamic Environment},
  P.~W. Glynn, D.~P. Manzello, and I.~C. Enoch, Eds., Coral Reefs of the World,
  Vol.~8, Springer Netherlands, Dordrecht, 85--106,
  \doi{10.1007/978-94-017-7499-4_4}.

\bibitem[{Wang and Picaut(2004)Wang, and Picaut}]{WangPicaut04}
Wang, C., and J.~Picaut, 2004: Understanding {ENSO} physics---a review.
  \textit{Earth's Climate: The Ocean-Atmosphere Interaction}, C.~Wang, S.-P.
  Xie, and J.~A. Carton, Eds., Geophysical Monograph Series, Vol. 147, American
  Geophysical Union, Washington, 21, \doi{10.1029/147GM02}.

\bibitem[{Webster and Hoyos(2010)Webster, and Hoyos}]{WebsterHoyos10}
Webster, P.~J., and C.~D. Hoyos, 2010: Beyond the spring barrier? \textit{Nat.
  Geosci.}, \textbf{3~(3)}, 152--153, \doi{10.1038/ngeo800}.

\bibitem[{Webster et~al.(1999)Webster, Moore, Loschnigg,, and
  Leban}]{WebsterEtAl99}
Webster, P.~J., A.~Moore, J.~Loschnigg, and M.~Leban, 1999: Coupled
  ocean-atmosphere dynamics in the {I}ndian {O}cean during 1997-98.
  \textit{Nature}, \textbf{401}, 356--360, \doi{10.1038/43848}.

\bibitem[{Weisberg and Wang(1997)Weisberg, and Wang}]{WeisbergWang97}
Weisberg, R.~H., and C.~Wang, 1997: A western {P}acific oscillator paradigm for
  the {E}l {N}i\~no- {S}outhern {O}scillation. \textit{Geophys. Res. Lett.},
  \textbf{24}, 779--782, \doi{10.1029/97GL00689}.

\bibitem[{Weller and Cai(2013)Weller, and Cai}]{WellerCai13}
Weller, E., and W.~Cai, 2013: Asymmetry in the {IOD} and {ENSO} teleconnection
  in a {CMIP5} model ensemble and its relevance to regional rainfall.
  \textit{J. Climate}, \textbf{26}, 5139--5149,
  \doi{10.1175/JCLI-D-12-00789.1}.

\bibitem[{Wu and Kirtman(2004)Wu, and Kirtman}]{WuKirtman04}
Wu, R., and B.~P. Kirtman, 2004: The tropospheric biennial oscillation of the
  {M}onsoon–{ENSO} system in an interactive ensemble coupled {GCM}.
  \textit{J. Climate}, \textbf{17}, 1623--1640,
  \doi{10.1175/1520-0442(2004)017<1623:TTBOOT>2.0.CO;2}.

\bibitem[{Wyrtki(1975)}]{Wyrtki75}
Wyrtki, K., 1975: El {N}i\~no---{T}he dynamic response of the equatorial
  pacific ocean to atmospheric forcing. \textit{J. Phys. Oceanogr.},
  \textbf{5}, 572--584, \doi{10.1175/1520-0485(1975)005<0572:ENTDRO>2.0.CO;2}.

\bibitem[{Yamagata et~al.(2003)Yamagata, Behera, Rao, Guan, Ashok,, and
  Saji}]{YamagataEtAl03}
Yamagata, T., S.~Behera, S.~A. Rao, Z.~Guan, K.~Ashok, and N.~H. Saji, 2003:
  Comments on ``{D}ipoles, {T}emperature {G}radients, and {T}ropical {C}limate
  {A}nomalies''. \textit{Bull. Amer. Meteor. Soc.}, \textbf{84}, 1418--1422,
  \doi{10.1175/BAMS-84-10-1418}.

\bibitem[{Yeh and Kirtman(2004)Yeh, and Kirtman}]{YehKirtman04}
Yeh, S.-W., and B.~P. Kirtman, 2004: Tropical {P}acific decadal variability and
  {ENSO} amplitude modulation in a {CGCM}. \textit{J. Geophys. Res.},
  \textbf{109}, C11\,009, \doi{10.1029/2004JC002442}.

\bibitem[{Yuan et~al.(2013)Yuan, Zhou,, and Zhao}]{YuanEtAl13}
Yuan, D., H.~Zhou, and X.~Zhao, 2013: Interannual {C}limate {V}ariability over
  the {T}ropical {P}acific {O}cean {I}nduced by the {I}ndian {O}cean {D}ipole
  through the {I}ndonesian {T}hroughflow. \textit{J. Climate}, \textbf{26},
  2845--2861, \doi{10.1175/JCLI-D-12-00117.1}.

\bibitem[{Yuan et~al.(2011)}]{YuanEtAl11}
Yuan, D., and Coauthors, 2011: Forcing of the {I}ndian {O}cean {D}ipole on the
  {I}nterannual {V}ariations of the {T}ropical {P}acific {O}cean: {R}oles of
  the {I}ndonesian {T}hroughflow. \textit{J. Climate}, \textbf{24}, 3593--3608,
  \doi{10.1175/2011JCLI3649.1}.

\bibitem[{Zebiak and Cane(1987)Zebiak, and Cane}]{ZebiakCane87}
Zebiak, S.~E., and M.~A. Cane, 1987: A model {E}l {N}i\~no--{S}outhern
  {O}scillation. \textit{Mon. Wea. Rev.}, \textbf{115}, 2262--2278,
  \doi{10.1175/1520-0493(1987)115<2262:AMENO>2.0.CO;2}.

\bibitem[{Zhao and Nigam(2015)Zhao, and Nigam}]{ZhaoNigam15}
Zhao, Y., and S.~Nigam, 2015: The {I}ndian {O}cean {D}ipole: {A} {M}onopole in
  {SST}. \textit{J. Climate}, \textbf{28}, 3--19,
  \doi{10.1175/JCLI-D-14-00047.1}.

\end{thebibliography}

\begin{figure}[t]
  \noindent\includegraphics[scale=.5]{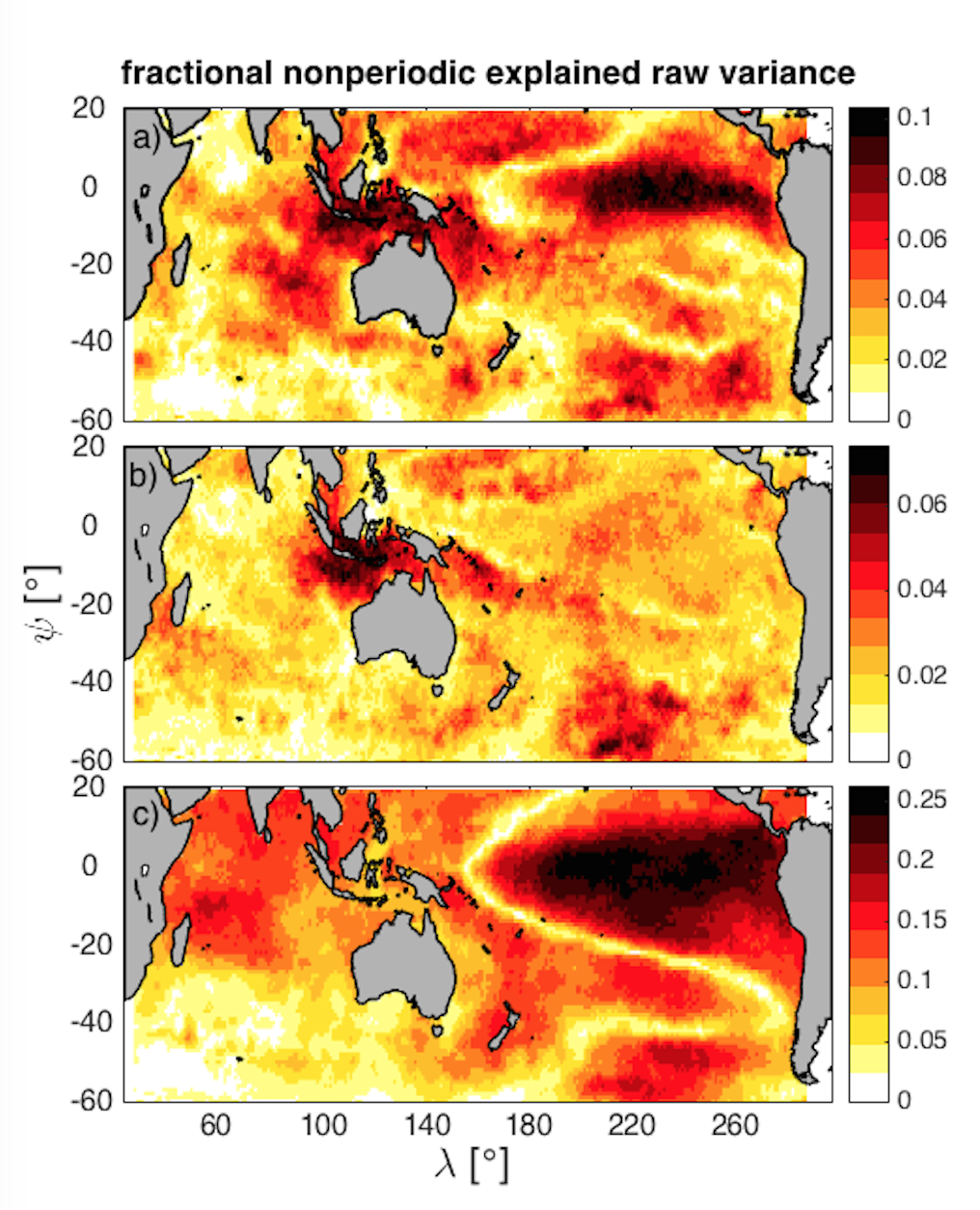}\\
  \caption{Explained variance relative to the raw SST data after removal of the seasonal cycle for (a) ENSO-A1, (b) ENSO-A2, (c) and ENSO modes as extracted via NLSA from 140 years of HadISST data.}\label{figVarRawWithoutPerHadISST}
\end{figure}

\begin{figure}[t]
\noindent\includegraphics[width=\linewidth]{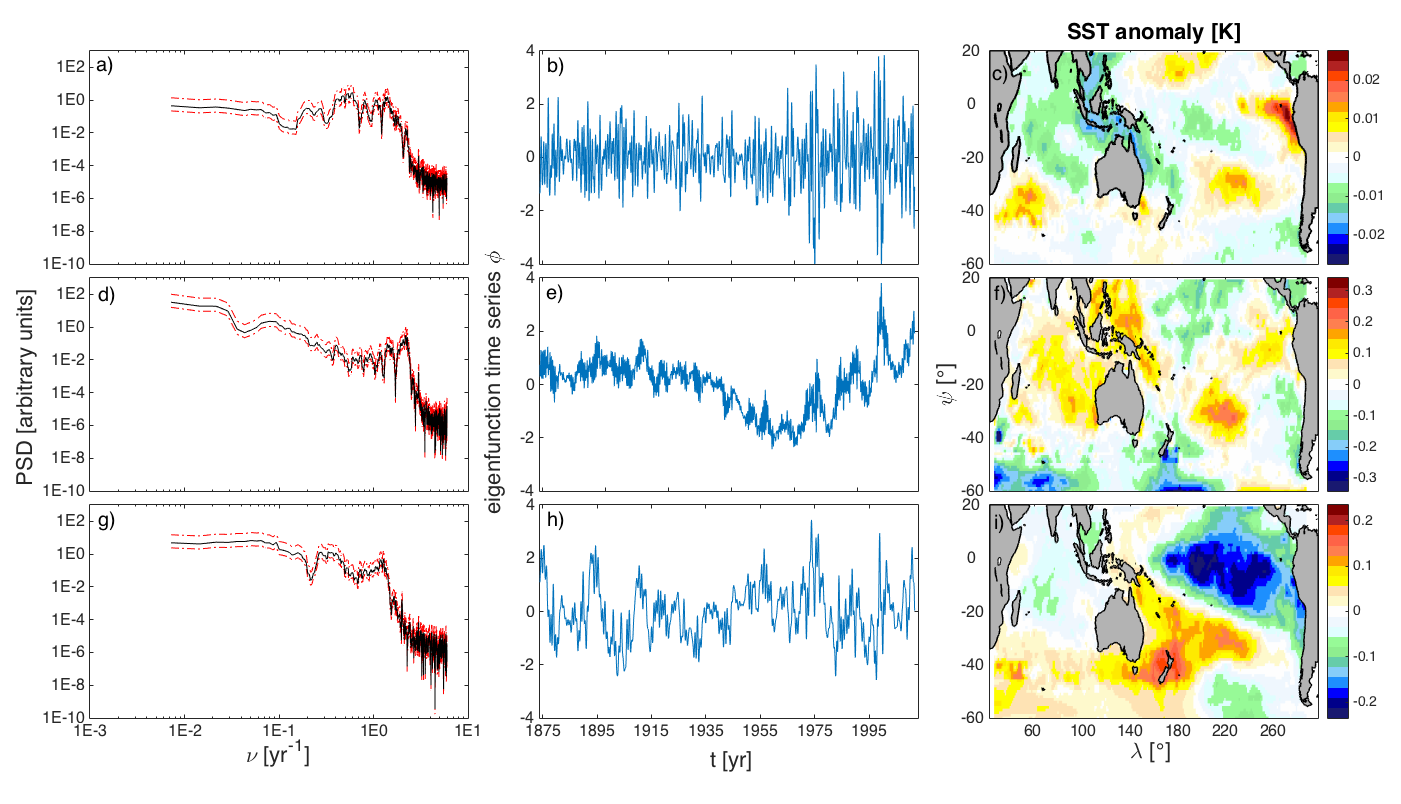}\\
  \caption{Same as Fig.~\ref{figDec}, but for the TBO (a--c), WPMM (d--f), and IPO-like (g--i) modes extracted via NLSA from the industrial-era HadISST dataset. The power spectral densities were estimated in the same manner as in Fig.~\ref{figObsPeriodic}.}
  \label{figObsDecadal}
\end{figure}

\begin{figure}[t]
\noindent\includegraphics[width=\linewidth]{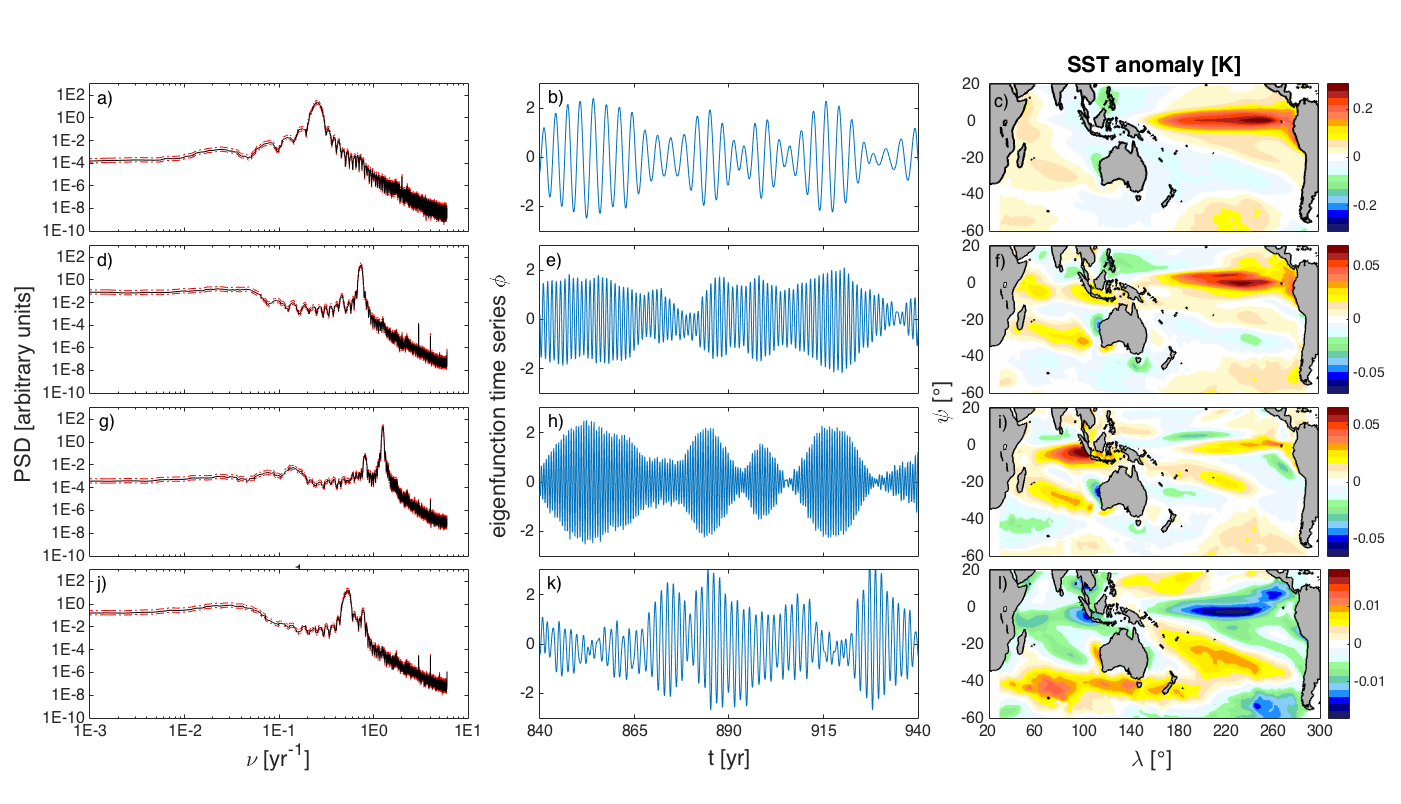}\\
  \caption{Power spectral densities (a, d, g, j), temporal patterns (b, e, h, k), and phase composites obtained from CCSM4 Indo-Pacific SST data via SSA for interannual modes. The modes shown here represent the fundamental component of ENSO (a--c), the ENSO-A1 combination mode (d--f), THE ENSO-A2 combination mode (g--i), and the fundamental component of the TBO (j--l). The power spectral densities and phase composites were computed in the same manner as their NLSA-based counterparts in Figs.~\ref{figInterannual} and~\ref{figBienn}.} \label{figSSAa}
\end{figure}

\begin{figure}[t]
\noindent\includegraphics[width=\linewidth]{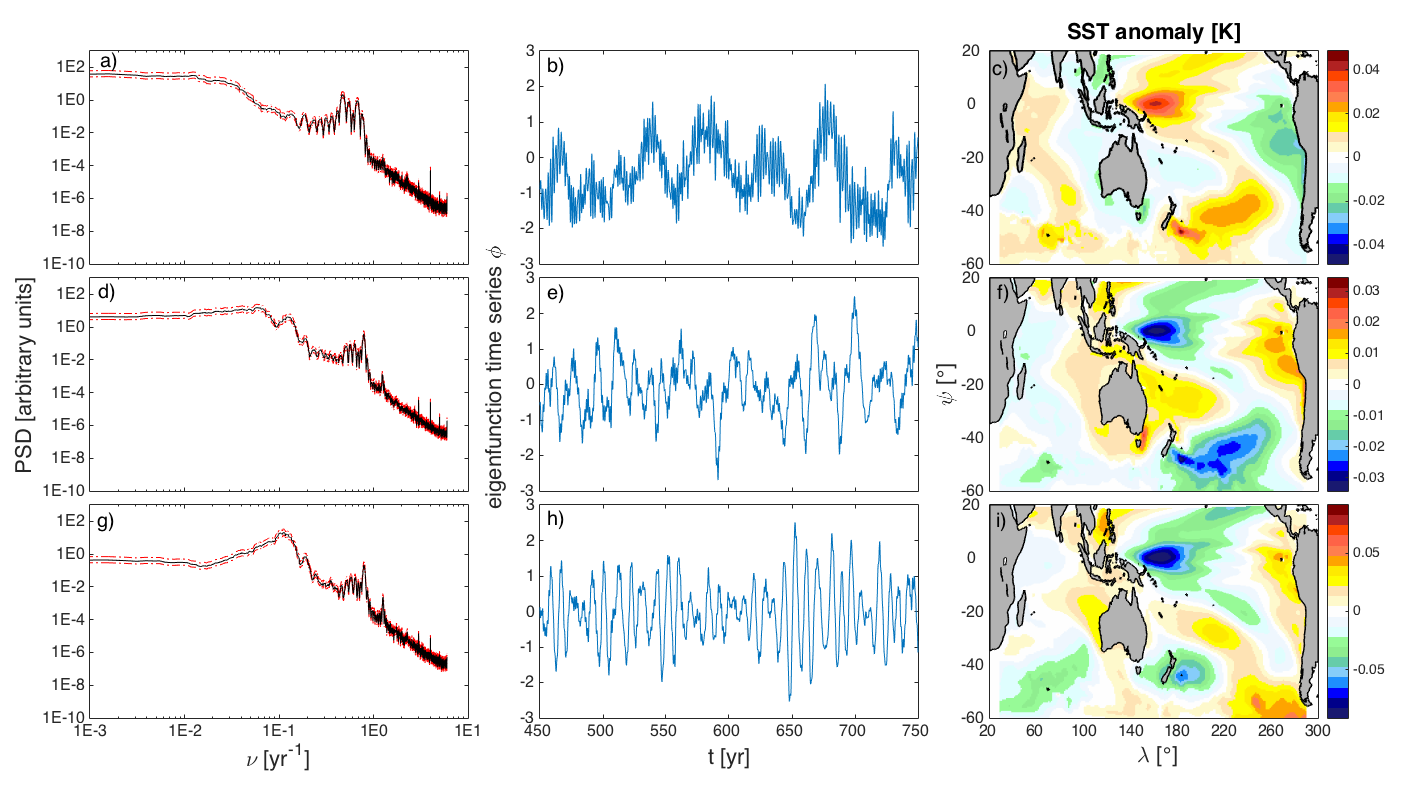}\\
  \caption{Same as Fig.~\ref{figSSAa} but for representative decadal SSA modes.} \label{figSSAb}
\end{figure}

\end{document}